\documentclass[%
reprint,
superscriptaddress,
amsmath,amssymb,
aps,
prc,
floatfix,
twocolumn,
twoside,
showkeys,
showpacs,
notitlepage
]{revtex4-2}

\usepackage[T1]{fontenc}
\usepackage{graphicx}
\usepackage{siunitx}
\usepackage{newtxtext,newtxmath}
\usepackage{hyperref}
\usepackage{fancyhdr}
\usepackage{microtype}
\usepackage{capt-of}
\usepackage{etoolbox}
\usepackage{float}
\usepackage{multirow}
\usepackage[bb=dsserif]{mathalpha}
\usepackage{bm}

\hypersetup{
    colorlinks=true,
    linkcolor=blue,
    citecolor=blue,
    urlcolor=blue,
    filecolor=blue
}

\fancypagestyle{plain}{%
  \fancyhf{}
  \fancyfoot[C]{\thepage}
  
}

\DeclareMathAlphabet{\pazocal}{OMS}{zplm}{m}{n}
\setcitestyle{numbers,square,comma,sort&compress}

\newcommand{\bra}[1]{\langle{#1\,}|}
\newcommand{\ket}[1]{|{#1\,}\rangle}
\newcommand{\braket}[2]{\langle{#1}\,|{#2\,}\rangle}
\newcommand{\T}{\textstyle}

\newcommand{\cgk}[6]{ \left( \!\! \left.\begin{array}{*{2}{c}} #1 &  #2  \\
                                               #4  &  #5 \end{array}
\right| \!\begin{array}{*{1}{c}} #3 \\ #6 \end{array} \right) }

\newcommand{\ms}{\mkern0.5mu} 

\makeatletter
\patchcmd{\frontmatter@abstract@produce}
  {\vskip200\p@\@plus1fil
   \penalty-200\relax
   \vskip-200\p@\@plus-1fil}
  {}
  {}
  {\GenericError{}{Failed to patch \string\frontmatter@abstract@produce}{}{}}
\makeatother

\begin{document}

\title{Numerical study of the three-boson bound-state problem in partial-wave
and vector-variable formulations}
\author{Wolfgang Schadow}
\email{\!wolfgang.schadow@caribou3d.com}
\affiliation{
Caribou3D Research \& Development,
Kirchplatz 1, 53424 Remagen, Germany
}

\author{M. R. Hadizadeh}
\email{\!mhadizadeh@centralstate.edu}
\affiliation{
College of Engineering, Science, Technology, and Agriculture,
Central State University, Wilberforce, OH 45384, USA
}
\affiliation{
Department of Physics and Astronomy,
Ohio University, Athens, OH 45701, USA
}
\date{November 1, 2026}

\begin{abstract}
We present a systematic benchmark of the three-boson bound-state problem in
momentum space, comparing one-dimensional (1D) spectator-amplitude,
two-dimensional (2D) partial-wave, and three-dimensional (3D) vector-variable
formulations. The benchmark controls the interaction representation by
embedding the same finite partial-wave interaction space in each formulation,
so that discrepancies reflect discretization, interpolation, and quadrature
errors. This enables direct 1D--2D--3D comparisons for separable interactions,
controlled 2D--3D tests for local interactions, and comparison with the full
local interaction in the 3D vector-variable formulation. Binding energies
agree at the $10^{-6}$~MeV level for the separable interactions. For the local
interactions at $l_{\rm max}=12$, the 2D--3D differences are
$2.2\times10^{-6}$~MeV for MT-V and $5.3\times10^{-5}$~MeV for MT-IV. The 2D and 3D
equations are also solved in both $t$-matrix-driven and bare-potential-driven
forms, whose agreement validates the permutation geometry, quadrature, and
interpolation. Fourier transforms to coordinate space yield consistent norm
decompositions and spatial observables, providing an independent check of the
momentum-space solutions.
\end{abstract}


\maketitle
\thispagestyle{plain}

\renewcommand{\arraystretch}{1.02}

\section{Introduction}
\label{sec:intro}

The quantitative description of quantum systems with three or more interacting
particles remains a central problem in few-body physics. The Faddeev and
Faddeev--Yakubovsky equations provide a systematic framework for treating
such systems in nuclear and atomic physics
\cite{Faddeev60a,Yakubovsky1967,Elster98a,Elster98b,Schadow2000d,
Polasek2000,Shertzer2001,Caia2004,Kessler2004,Kadyrov2005,Liu2005a,Ramalho2006,
Liu2007a,Hadizadeh2007a,RodriguezGallardo2008,Hadizadeh2008a,Bayegan2008b,
Harzchi2010a,Gloeckle2010a,Golak2010a,Golak2013a,Shalchi2012a,Veerasamy2013,
Harzchi2014a,Kuruoglu2016}. Applications of Faddeev methods also include
exotic three-body models, such as $N N N^{*}$ systems studied in connection
with experimental searches for $\eta$-mesic ${}^{3}\mathrm{He}$
\cite{Kelkar2019a}. The nontrivial integral kernels and permutation
operators make the numerical solution demanding, motivating internally
consistent benchmarks  for different numerical realizations.

A common strategy for constructing such benchmarks is to consider simpler
systems for which high-precision or partially analytical solutions are
available. For three-body bound states, partial-wave decomposition provides
an efficient and well-established framework, leading to coupled
two-dimensional (2D) integral equations in the magnitudes of the Jacobi
momenta. For separable interactions, these equations can be reduced further
to a set of coupled one-dimensional (1D) equations.

This one-dimensional reduction is not restricted to interactions that are
separable by construction. Nonseparable two-body interactions can also be
treated through separable representations, such as those generated by the
Ernst--Shakin--Thaler (EST) method. Koike et al.~\cite{Koike87a} employed a
separable representation of the Paris nucleon--nucleon potential in
nucleon--deuteron scattering calculations. Schadow et al.~\cite{Schadow97b}
compared triton binding energies and wave functions obtained with EST and
$W$-matrix representations of the Paris, Bonn~A, and Bonn~B interactions
against direct two-dimensional calculations.

Complementary to these partial-wave formulations, three-dimensional
(3D) approaches based directly on vector momentum variables retain
the full angular dependence of the wave function without expanding
it in partial waves. Their numerical implementation requires the
evaluation of multidimensional integrals involving shifted permutation
arguments, making quadrature and interpolation central to the
accuracy of the solution. Elster et al.~\cite{Elster98b} already
demonstrated close agreement between vector-variable and partial-wave
calculations of three-body binding energies and wave functions for
Malfliet--Tjon-type interactions.

Building on that comparison, the present work controls the
interaction representation explicitly: the same finite partial-wave
interaction space is embedded in the different formulations, allowing
numerical discrepancies to be examined without conflating them with
differences in interaction content. For Yamaguchi-type potentials
\cite{Yamaguchi54a}, this gives a
direct 1D--2D--3D comparison. Since these interactions are separable by
construction, the comparison introduces no additional separable-expansion
approximation. For local Malfliet--Tjon (MT) potentials
\cite{Malfliet69}, the same strategy provides a controlled 2D--3D comparison,
supplemented by a comparison with the full local interaction in the
3D vector-variable formulation.

Comparisons between $t$-matrix-driven and potential-driven
Faddeev equations also have precedents. Golak et al.
\cite{Golak2013a} developed a 3D operator formulation of the
three-nucleon bound state with spin and isospin degrees of
freedom and chiral two- and three-nucleon forces. For the
two-nucleon-force case, they compared numerical schemes
using either the two-body $t$-matrix or the interaction
directly. Here, we implement both forms in the 2D and 3D
spinless-boson calculations without three-body forces.
Their comparison provides a complementary test of the
permutation geometry, quadrature, and interpolation within
the same controlled interaction spaces, rather than
introducing a new potential-driven formulation.

The comparison is extended to configuration space by Fourier transforming
the momentum-space wave functions. These transformations provide access to
spatial information but require careful treatment of highly oscillatory
integrals. We employ a hybrid Filon-type quadrature
\cite{Filon1930,Press2007}, which treats the rapid trigonometric
oscillations analytically while interpolating the smooth momentum envelope.
This semi-analytical treatment enables a direct coordinate-space comparison
between the partial-wave and vector-variable representations,
complementing the momentum-space checks.

The paper is organized as follows. Section~\ref{sec:threebodyformalism}
summarizes the general three-body formalism.
Sections~\ref{sec:1dapproach} and \ref{sec:2dapproach} present the
1D and 2D formulations, respectively, followed by the 3D vector-variable
approach in Sec.~\ref{sec:3dapproach}. Numerical details, including the
treatment of the oscillatory spatial transformations, are given in
Sec.~\ref{sec:numerical_method}. Results and discussion are presented in
Sec.~\ref{sec:results}, followed by a concluding summary in
Sec.~\ref{sec:outlooksummary}.

\section{The Three-Body Bound State}
\label{sec:threebodyformalism}

The full bound-state wave function \(|\Psi\rangle\) of three identical
particles interacting via pairwise forces \(V_i\equiv V_{jk}\), where
\((i,j,k)\) denotes a cyclic permutation of \((1,2,3)\), obeys the
Schr\"odinger equation, which reads in integral form
\begin{equation}
|\Psi \rangle = G_0(E_3) \sum_{i=1}^3 V_i \,| \Psi \rangle .
\label{eq:schroedinger_integral}
\end{equation}
Here, \(G_0(E_3)=(E_3-H_0)^{-1}\) is the free three-body resolvent at the
bound-state energy \(E_3<0\), and \(H_0\) is the free Hamiltonian.
We define the individual Faddeev components by
\(|\psi_i\rangle = G_0(E_3) V_i |\Psi\rangle\).
The full three-body wave function is given by their sum,
\(|\Psi\rangle = \sum_{i=1}^3 |\psi_i\rangle\).
Using the embedded two-body transition operators
\(t_i = V_i + V_i G_0(E_3)t_i\), we obtain the standard set
of coupled Faddeev equations
\begin{equation}
| \psi_i \rangle = G_0(E_3) \, t_i \, \sum_{j\neq i} |\psi_j \rangle ,
\label{eq:faddeev_components}
\end{equation}
where \(t_i\) denotes the two-body \(t\)-matrix in subsystem \(jk\).

For a system of identical spinless bosons, the full three-particle wave
function \(|\Psi \rangle\) must be totally symmetric. The three Faddeev
components are generated from one another by particle permutations and
therefore have the same functional form in their respective Jacobi
coordinate systems. It is thus sufficient to consider a single arrangement.
We choose arrangement \(i=1\), with particle 1 as spectator and particles 2
and 3 forming the interacting pair. The corresponding Faddeev
component satisfies
\begin{equation}
|\psi_1\rangle = G_0(E_3) \, t_1 \, P \, |\psi_1\rangle ,
\label{eq:faddeev}
\end{equation}
where \(P=P_{12}P_{23}+P_{13}P_{23}\) is the sum of the cyclic
and anticyclic permutations, and \(P_{ij}\) exchanges particles
\(i\) and \(j\). The full three-boson wave function is recovered via
\begin{equation}
|\Psi\rangle = (1 + P) \, |\psi_1\rangle .
\label{eq:faddeevt}
\end{equation}

For the 2D and 3D formulations, it is useful to work with an equivalent
equation for the same Faddeev component, driven directly by the bare
potential. Starting from the definition
\(|\psi_1\rangle = G_0(E_3) V_1 |\Psi\rangle\) and substituting the sum
of components yields
\begin{equation}
    |\psi_1\rangle = G_0(E_3) \, V_1
    \Big( |\psi_1\rangle + |\psi_2\rangle + |\psi_3\rangle \Big).
\end{equation}
For identical bosons, \(|\psi_2\rangle\) and \(|\psi_3\rangle\) are
generated from \(|\psi_1\rangle\) by cyclic and anticyclic permutations,
such that
\((|\psi_2\rangle + |\psi_3\rangle) = P|\psi_1\rangle\), with \(P\)
acting on the component in arrangement 1. From this point onward, we use
\(|\psi\rangle\equiv|\psi_1\rangle\), \(t\equiv t_1\), and
\(V\equiv V_1\).
The \(V\)-driven equation then reads
\begin{equation}
    |\psi\rangle = G_0(E_3) \ms V (1 + P) \, |\psi\rangle .
    \label{eq:faddeev_tfree}
\end{equation}
This formulation avoids the explicit construction of the fully off-shell
two-body \(t\)-matrix. Consequently, the bare interaction appears directly
in the multidimensional kernel, including the direct
\(G_0(E_3) \ms V|\psi\rangle\) contribution.

To solve these abstract operator equations numerically, they must be
projected onto a well-defined momentum-space basis. The resulting integral
equations take different forms depending on the interaction type and the
chosen coordinate representation. In the following sections, we construct
these projections and define the corresponding basis states for three
distinct numerical approaches: a 1D spectator-amplitude formulation for
separable interactions, a 2D partial-wave formulation, and a 3D
vector-variable formulation in which the wave function is represented
directly as a function of Jacobi momentum vectors.

\section{One-Dimensional Approach}
\label{sec:1dapproach}

For separable interactions, the one-dimensional approach used here is the
spectator-amplitude form of the homogeneous Alt--Grassberger--Sandhas (AGS)
equations evaluated at the bound-state pole \cite{Alt67}. We derive this
equation directly from the Faddeev operator equation [Eq.~(\ref{eq:faddeev})]
at the three-body binding energy $E_3 < 0$.

To solve the Faddeev equations numerically, we must represent the abstract
operators in a well-defined momentum basis. Exploiting identical-particle
symmetry, we define the basis states in the Jacobi coordinate system of a
single arrangement. We introduce the Jacobi momenta $p$ (the relative momentum
within the interacting two-body subsystem) and $q$ (the momentum of the
spectator particle relative to the center of mass of the interacting pair).
With the mass convention used throughout this work, the free three-body
kinetic energy is $H_0=p^2+\frac{3}{4}q^2$. The conversion to MeV is performed
only when numerical results are reported.

For the partial-wave decomposition, we define the orbital angular momentum
$l$ of the two-body subsystem and the orbital angular momentum $L$ of the
spectator particle. To avoid confusion with the spectator orbital angular
momentum $L$, the total orbital angular momentum of the three-body system
is denoted by $\pazocal{L}$, with its $z$-projection $\pazocal{M}$.

The radial basis states associated with the pair and spectator Jacobi momenta
are independently delta-normalized,
\begin{align}
\langle p|p'\rangle=\frac{\delta(p-p')}{p^2},
\qquad
\langle q|q'\rangle=\frac{\delta(q-q')}{q^2}.
\end{align}
The three-body basis is constructed from the tensor product of these radial
states and the orthonormal coupled angular states. The partial-wave projected
states $\ket{p \, q \, (l L) \pazocal L \pazocal M}$ form a complete basis
for this designated arrangement,
\begin{align}
\mathbb{1} &= \sum_{\substack{l L  {\pazocal L} {\pazocal M}}} \int\limits_{0}^{\infty}
\! dp \, p^2  \int\limits_{0}^{\infty} \! dq \, q^2 \, \nonumber \\
& \qquad \times \, \ket{ p \, q \, (l L ) {\pazocal L} {\pazocal M}} \,
\bra{ p \, q \, (l L) {\pazocal L} {\pazocal M}},
\label{eq:3bodybosoncomplete}
\end{align}
and are normalized according to
\begin{align}
\langle p \, q \, (l L) & {\pazocal L}{\pazocal M}|{ p' \, q'
  \, (l' L\!') {\pazocal L'} {\pazocal M'}} \rangle \label{eq:3bstatealphaalpha} \\
& = \frac{ \delta(  p  - p\, ' ) }{ p^2} \, \frac{ \delta(  q  - q\, ' ) }{ q^2}
\, \delta_{l l'} \, \delta_{L L'} \, \delta_{{\pazocal L}
{\pazocal L}'} \, \delta_{{\pazocal M} {\pazocal M}'}. \nonumber
\end{align}

For the three-boson ground state considered here, the total orbital angular
momentum is $\pazocal{L}=0$, which requires $L=l$. Consequently, we drop the
redundant indices $L$, $\pazocal{L}$, and $\pazocal{M}$, denoting the basis
states compactly as $|p \, q \, l\rangle$.

The original Yamaguchi potential \cite{Yamaguchi54a} provides the prototype
for the separable interactions considered here. The generalized form factors
used for higher partial waves are specified in Sec.~\ref{sec:interactions}.
In each active partial wave, the two-body potential is written as
\begin{equation}
    V_l(p,p') = -4\pi\lambda_l \, g_l(p)g_l(p') ,
    \label{eq:separable_potential}
\end{equation}
where $g_l(p)$ is the momentum-space form factor. Let $E$ denote the
two-body energy measured from the free-pair threshold. The corresponding
two-body $t$-matrix has the same separable structure,
\begin{equation}
    t_l(p,p';E) = g_l(p)\,\Delta_l(E)\,g_l(p') ,
    \label{eq:separable_tmatrix}
\end{equation}
where the reduced propagator $\Delta_l(E)$ contains the full two-body
energy dependence.

To distinguish the two-body propagator entering the Lippmann--Schwinger
equation from the free three-body resolvent $G_0(E_3)$ introduced above, we
denote the free two-body resolvent by
$G_0^{(2)}(E)=(E-h_0)^{-1}$, with $h_0=p^2$ in the mass convention used here.
With the convention of Eq.~(\ref{eq:separable_potential}) and the
Lippmann--Schwinger equation
$t(E)=V+V G_0^{(2)}(E)\,t(E)$, the reduced propagator is
\begin{equation}
\Delta_l(E) =
-\left[
(4\pi\lambda_l)^{-1}
+
\langle g_l | G_0^{(2)}(E) | g_l \rangle
\right]^{-1}.
\label{eq:propagator}
\end{equation}
For the negative two-body energies relevant to the present bound-state
problem, the matrix element appearing here is explicitly
\begin{align}
\langle g_l | G_0^{(2)}(E) | g_l \rangle
=
\int\limits_0^\infty \! dk\,k^2\,
\frac{|g_l(k)|^2}{E-k^2},
\qquad E<0.
\end{align}
The dimensions of $\lambda_l$ are fixed by the chosen form-factor
normalization.

In the three-body problem, the energy argument $E$ of
$t_l(p,p';E)$ and $\Delta_l(E)$ is determined by the spectator momentum.
After projection onto a state with spectator momentum $q$, it is evaluated
at the subsystem energy
\begin{equation}
    E_{\rm sub}(q) = E_3 - \frac{3}{4}q^2 .
    \label{eq:esub}
\end{equation}
Here $E_3<0$ is the three-body bound-state energy measured from the
three-particle threshold, while $E_2<0$ denotes the dimer bound-state
energy measured from the free-pair threshold. Since the three-body bound
states considered here lie below the particle--dimer threshold,
$E_3<E_2<0$, it follows that
$E_{\rm sub}(q)\le E_3<E_2$ for all real spectator momenta $q$.
Thus the subsystem energy remains below the two-body bound-state pole,
and no singularity of $\Delta_l(E_{\rm sub})$ is encountered on the
integration contour.

Projecting Eq.~(\ref{eq:faddeev}) onto the momentum basis states
$|p \, q \, l\rangle$ and utilizing the separable form of
Eq.~(\ref{eq:separable_tmatrix}), we obtain the functional form of the
Faddeev component
\begin{align}
    \psi_l(p,q) &= \frac{g_l(p)}{E_3 - p^2 - \T\frac{3}{4}q^2} \;
    \Delta_l\!\left(E_3 - \T\frac{3}{4}q^2\right)  \nonumber \\
    & \quad \times  \int\limits_0^\infty \! dp' \,
    p'^2 \, g_l(p') \, \langle p' \ms q \, l | P | \psi \rangle.
\end{align}

We define the spectator amplitude $F_l(q)$ as the integral over the internal
subsystem momentum,
\begin{equation}
    F_l(q) = \int\limits_0^\infty \! dp' \, p'^2 \, g_l(p') \,
    \langle p' \ms q \, l | P | \psi \rangle.
    \label{eq:fl_def}
\end{equation}
Because the internal momentum $p'$ has been integrated out, this amplitude
depends only on the spectator momentum $q$. For a given spectator amplitude
$F_l(q)$, the Faddeev component is then given by
\begin{equation}
    \psi_l(p,q) =
    \frac{g_l(p)}
    {E_3 - p^2 - \T\frac{3}{4}q^2} \;
    \Delta_l\!\left(E_3 - \T\frac{3}{4}q^2\right) F_l(q).
    \label{eq:psi_from_F}
\end{equation}

To find the integral equation governing $F_l(q)$, we insert a completeness
relation in the partial-wave basis into the definition of $F_l(q)$,
\begin{align}
    F_l(q) &= \sum_{l'} \int\limits_0^\infty \! dp' \, p'^2
    \int\limits_0^\infty \! dp'' \, p''^2 \int\limits_0^\infty \! dq' \, q'^2 \,
    \nonumber \\
& \quad \times \,  g_l(p') \, \langle p' \, q \, l | P | p'' \, q' \, l' \rangle \,
    \psi_{l'}(p'', q').
\end{align}
Substituting the functional form of $\psi_{l'}$ from Eq.~(\ref{eq:psi_from_F})
into this expression yields the one-dimensional integral equation
\begin{align}
    F_l(q) &= \sum_{l'} \int\limits_0^\infty \! dq' \, q'^2 \,
    \mathcal{V}_{ll'}(q, q', E_3) \nonumber \\
    & \quad \times \, \Delta_{l'}\!\left(E_3 - \T\frac{3}{4}q'^2\right) F_{l'}(q'),
    \label{eq:formfactor_1d}
\end{align}
where the effective spectator kernel $\mathcal{V}_{ll'}$ is defined by the
matrix element
\begin{align}
    \mathcal{V}_{ll'}(q, q', E_3) &= \int\limits_0^\infty \! dp \, p^2
    \int\limits_0^\infty \! dp' \, p'^2 \, g_l(p) \nonumber \\
    & \quad \times \, \frac{ \langle p \, q \, l | P | p' \, q' \, l' \rangle }
    {E_3 - p'^2 - \T\frac{3}{4}q'^2} \, g_{l'}(p') \,.
    \label{eq:v_eff_def}
\end{align}

The permutation operator connects the different Jacobi coordinate systems.
With the completeness and normalization conventions of
Eqs.~(\ref{eq:3bodybosoncomplete}) and (\ref{eq:3bstatealphaalpha}), the
standard partial-wave projected permutation matrix element for identical
spinless bosons is  \cite{Elster98b,Gloeckle83}
\begin{align}
    \langle  p \, q \, l | P | p' \, q' \, l' \rangle
    &= \int\limits_{-1}^1 \! dx \,
    \frac{\delta(p-\pi_1)}{p^{l+2}} \nonumber \\
    & \quad \times \, \frac{\delta(p'-\pi_2)}{p'^{l'+2}}
    \, G_{ll'}(q,q',x),
  \label{eq:matrixperm}
\end{align}
where the shifted momentum arguments are dictated by the kinematics
\begin{align}
    \pi_1 &= \sqrt{\T\frac{1}{4}q^2 + q'^2 + q \ms q' \ms x} \nonumber \\
    \pi_2 &= \sqrt{q^2 + \T\frac{1}{4}q'^2 + q \ms q' \ms x} \,.
\label{eq:pi12shift}
\end{align}
The geometric factor $G_{ll'}(q,q',x)$ is given in
Appendix~\ref{sec:appendixperm}, which documents the conventional
recoupling representation used in this work. Alternative
partial-wave representations evaluate the angular functions
directly at the permuted Jacobi momenta, without further expanding
their angular dependence into recoupling sums. Such implementations
have been reported to improve numerical efficiency and stability,
particularly at large orbital angular momenta
\cite{Hebeler2015a,Miller2022a}.

Inserting Eq.~(\ref{eq:matrixperm}) into the effective kernel, the Dirac
delta distributions collapse the integrations over the subsystem momenta,
setting $p=\pi_1$ and $p'=\pi_2$. The radial measure $p^2dp$ cancels the
measure-related $p^2$ part of $p^{l+2}$, leaving $g_l(\pi_1)/\pi_1^l$;
analogously, the $p'$-integration yields $g_{l'}(\pi_2)/\pi_2^{l'}$. This
scaling is numerically stable because the form factors used here behave as
$g_l(p) \propto p^l$ near the origin.

After evaluating the free resolvent denominator at $p' = \pi_2$, the remaining
integration is purely angular. The reduced spectator kernel thus reduces to a
single integral over the angle $x$
\begin{align}
    \mathcal{V}_{ll'}(q, q', E_3) &= \int\limits_{-1}^1 \! dx \;
    \frac{g_l(\pi_1)}{\pi_1^l}  \, \frac{g_{l'}(\pi_2)}{\pi_2^{l'}} \nonumber \\
    & \quad \times \, \frac{G_{ll'}(q,q',x)}{E_3 - q^2 - q'^2 - qq'x} \,,
    \label{eq:v_eff_reduced}
\end{align}
where the denominator follows from $\pi_2^2+\frac{3}{4}q'^2=q^2+q'^2+qq'x$.
Equation~(\ref{eq:formfactor_1d}), driven by this reduced spectator kernel,
can now be discretized into a linear eigenvalue problem, reducing the
three-body kernel to a set of coupled one-dimensional integral equations in
the spectator momentum. The energy $E_3$ is varied until the corresponding
eigenvalue equals unity.

Once the spectator amplitudes $F_l(q)$ are found, the partial-wave Faddeev
components $\psi_{l}(p,q)$ are constructed directly via
Eq.~(\ref{eq:psi_from_F}). The full three-body wave function
$\Psi_l(p,q)$ is then generated via Eq.~(\ref{eq:faddeevt})
\begin{align}
\Psi_l(p,q) &= \psi_l(p,q)  \nonumber \\
&  \quad + \sum_{l'} \int\limits_{-1}^1 \! dx \;
 \tilde G_{ll'} (p,q,x)\, \frac{ \psi_{l'}(\tilde \pi_1,\tilde \pi_2)}
 {\tilde \pi_1^{l'} \,\tilde \pi_2^{l'}} \,.
\label{eq:partialwave}
\end{align}
The shifted arguments are
\begin{align}
\tilde \pi_1 &= \sqrt{\textstyle{\frac{1}{4}}p^2+
 \textstyle{\frac{9}{16}}q^2 +\textstyle{\frac{3}{4}}qpx},  \nonumber \\
\tilde \pi_2 &= \sqrt{p^2 +\textstyle{\frac{1}{4}}q^2 -qpx} \,.
\label{eq:shiftedarguments}
\end{align}
The quantity $\tilde G_{ll'} (p,q,x)$, which emerges from applying the
permutation operator to construct the full wave function components, can be
found in Ref.~\cite{Hueber93}. The explicit expression for the bosonic case
considered here is given in Appendix~\ref{sec:appendixperm}.

Enforcing the wave function normalization and utilizing the symmetry
of the Faddeev components, the norm is evaluated directly as
\begin{align}
    3 \, \langle\psi | \Psi \rangle
    &= 3 \sum_{l} \int\limits_0^\infty \! dq \, q^2
    \int\limits_0^\infty \! dp \, p^2 \nonumber \\
    & \quad \times \, \psi_{l}^*(p,q) \, \Psi_{l}(p,q) = 1 \, .
    \label{eq:norm_pwd}
\end{align}
A crucial distinction arises here between the partial-wave content of the
interaction and that of the full wave function. Because the Faddeev component
$\psi$ is driven directly by the two-body interaction, $\psi_l(p,q)$ is
strictly non-zero only for the active partial waves $l \le l_{\rm max}$
included in the potential. Consequently, the sum in the mixed overlap
$3\langle\psi|\Psi\rangle$ terminates exactly at $l_{\rm max}$. In contrast,
the permutation operator in Eq.~(\ref{eq:partialwave}) can generate components
outside the active interaction space in the reconstructed full wave function.
Evaluating the direct norm $\langle\Psi | \Psi\rangle$ in a partial-wave
basis would therefore require summing these induced components, whereas
the mixed Faddeev overlap terminates at the active $l_{\rm max}$.

The full Hamiltonian expectation value is evaluated as
\begin{equation}
\langle H\rangle =
3\,\langle \psi |H_0| \Psi \rangle
+
3\,\langle \Psi | V_1 | \Psi \rangle \, .
\label{eq:energy_expectation_pwd}
\end{equation}
The first factor of three follows from the identical-particle relation
$\langle\Psi|\Psi\rangle=3\langle\psi|\Psi\rangle$, while the second
accounts for the three identical pair interactions. The reduced
kinetic-energy matrix element is
\begin{align}
\langle \psi |H_0| \Psi \rangle
&= \sum_l \int\limits_0^{\infty} \! dp \, p^2
  \int\limits_0^{\infty} \! dq \, q^2
  \left ( p^2 + \frac{3}{4}q^2 \right) \nonumber \\
& \quad \times \psi_l^*(p,q) \, \Psi_l(p,q) \,,
\label{eq:h0expect_pwd}
\end{align}
and the matrix element of one pair interaction is
\begin{align}
\langle \Psi | V_1 | \Psi \rangle
&= \sum_l \int\limits_0^{\infty} \! dp \, p^2
  \int\limits_0^{\infty} \! dq \, q^2
  \int\limits_0^{\infty} \! dp' \, p'^2 \nonumber \\
& \quad \times \Psi_l^*(p,q) \, V_l(p,p') \, \Psi_l(p',q) \,.
\label{eq:vexpect_pwd}
\end{align}
In the numerical results below, the consistency of $\langle H\rangle$ with
the eigenvalue energy $E_3$ provides an independent check of the
reconstructed full wave function.

\section{Two-Dimensional Approach}
\label{sec:2dapproach}

For general non-separable interactions, such as the local Malfliet--Tjon
potential used below, the two-body $t$-matrix cannot be factored into a
product of functions depending on single momenta. Consequently, the internal
subsystem momentum $p$ cannot be eliminated using the separable factorization
employed in the one-dimensional approach and remains explicitly coupled to
the spectator momentum $q$. This leads to a
two-dimensional (2D) partial-wave formulation.

We begin by projecting the standard Faddeev operator equation
[Eq.~(\ref{eq:faddeev})] onto the partial-wave momentum basis
$|p \, q \, (l L) \pazocal{L} \pazocal{M} \rangle$. To transition from the
abstract operators to the integral equations, we require the matrix elements
of the free propagator, the two-body $t$-matrix, and the permutation operator.

For central, spin-independent interactions, the free propagator and the
two-body $t$-matrix are diagonal in the spectator momentum $q$, as well
as in the angular momenta $l$, $L$, $\pazocal{L}$, and $\pazocal{M}$

\begin{align}
\langle p \, q \, (l L) \pazocal{L} \pazocal{M} &| G_0(E_3) | p' \, q' \, (l' L')
\pazocal{L'} \pazocal{M'} \rangle \nonumber \\
&= \frac{\delta(p-p')}{p^2} \, \frac{\delta(q-q')}{q^2} \nonumber \\
& \quad \times \, \delta_{ll'} \, \delta_{LL'} \,
\delta_{\pazocal{L}\pazocal{L'}} \, \delta_{\pazocal{M}\pazocal{M'}} \,
\frac{1}{E_3 - p^2 - \frac{3}{4}q^2} \,,
\label{eq:g0_matrix_pwd} \\
\langle p \, q \, (l L) \pazocal{L} \pazocal{M} &| t(E_3) | p' \, q' \, (l' L')
\pazocal{L'} \pazocal{M'} \rangle \nonumber \\
&= \frac{\delta(q-q')}{q^2} \, \delta_{ll'} \, \delta_{LL'} \,
\delta_{\pazocal{L}\pazocal{L'}} \, \delta_{\pazocal{M}\pazocal{M'}}
\nonumber \\
& \quad \times \, t_l\!\left(p, p'; E_3 - \textstyle{\frac{3}{4}}q^2\right).
\label{eq:t_matrix_pwd}
\end{align}

Because the two-body $t$-matrix and the bare potential $V$ operate only within
the interacting pair, their dynamical matrix elements depend exclusively on the
subsystem angular momentum $l$. The matrix element of the permutation operator
$P$ has the same form as in the one-dimensional case, Eq.~(\ref{eq:matrixperm}),
and the shifted momenta $\pi_1$ and $\pi_2$ are those defined in
Eq.~(\ref{eq:pi12shift}).

For the bound state considered here, the total angular momentum is
$\pazocal{L}=0$, which requires $L=l$. We consequently drop the redundant
indices $L$, $\pazocal{L}$, and $\pazocal{M}$ from the basis states and the
Faddeev components. Because the two-body $t$-matrix is diagonal in the
spectator momentum, it is evaluated at the subsystem energy
$E_3-\frac{3}{4} q^2$. Inserting completeness relations into the standard
Faddeev Eq.~(\ref{eq:faddeev}) and integrating over the delta functions
yields the coupled two-dimensional integral equations
\begin{align}
\psi_l(p,q) &= \frac{1}{ E_3- p^2 -\frac{3}{4} q^2} \nonumber \\
& \quad \times \sum_{l'} \int\limits_0^{\infty} \! dq' \, q'^2
  \int\limits_{-1}^1 \! dx' \;
  \frac{t_l\!\left(p,\pi_1;E_3-\textstyle{\frac{3}{4}} q^2\right)}
       {\pi_1^l} \nonumber \\
& \quad \times \, G_{ll'}(q,q',x') \;
  \frac{\psi_{l'}\!\ms(\pi_2,q')}{\pi_2^{l'}} \, .
\label{eq:faddeev_2d_t}
\end{align}
The factors $\pi_1^{-l}$ and $\pi_2^{-l'}$ originate from the normalization
convention used in the partial-wave permutation matrix element,
Eq.~(\ref{eq:matrixperm}). For identical spinless bosons with central
interactions, only even values of $l$ contribute. In numerical calculations,
this infinite set is truncated at a sufficiently high value of $l_{\rm max}$.

The $t$-driven equation contains only the exchange term
$G_0 t P |\psi\rangle$, because the multiple scattering inside the
interacting pair is already summed into $t_l$. In contrast, the
potential-driven equation contains both the direct term
$G_0 V |\psi\rangle$ and the exchange term $G_0 V P |\psi\rangle$.
At the operator level, the equality of the resulting bound-state solutions
follows directly from the Faddeev definition
$|\psi\rangle = G_0 V (1+P) |\psi\rangle$, whereas the $t$-driven form is
obtained after resumming repeated pair interactions into the two-body
transition operator. The two-body potential shares the same
spectator-diagonal structure as the $t$-matrix,
\begin{equation}
\langle p \, q \, l | V | p' \, q' \, l' \rangle
= \frac{\delta(q-q')}{q^2} \, \delta_{ll'} \, V_l(p, p') \,.
\label{eq:v_matrix_pwd}
\end{equation}

Evaluating the direct term gives an interaction that changes only the
subsystem momentum, leaving the spectator momentum $q$ unchanged.
In the exchange term, the integration over the intermediate subsystem
momentum evaluates the product of $V_l(p,p'')$ and the permutation matrix
element $\langle p'' \, q \, l | P | p' \, q' \, l' \rangle$. The Dirac delta
distributions in the permutation matrix element collapse the subsystem momentum
integrals, setting $p''=\pi_1$ and $p'=\pi_2$.

Consequently, the exchange part of the $V$-driven equation has the same
permutation structure as Eq.~(\ref{eq:faddeev_2d_t}), with $t_l$ replaced by
$V_l$. The full $V$-driven equation additionally contains the direct potential
term
\begin{align}
\psi_l(p,q) &= \frac{1}{E_3 - p^2 - \frac{3}{4}q^2} \bigg[
\int\limits_0^\infty \! dp' \, p'^2 \, V_l(p,p') \, \psi_l(p',q) \nonumber \\
& \quad + \sum_{l'} \int\limits_0^\infty \! dq' \, q'^2
\int\limits_{-1}^1 \! dx' \, \frac{V_l(p, \pi_1)}{\pi_1^l} \,
G_{ll'}(q,q',x') \nonumber \\
& \quad \times  \; \frac{\psi_{l'}(\pi_2, q')}{\pi_2^{l'}} \bigg] .
\label{eq:faddeev_2d_v}
\end{align}

This formulation shows that the bare potential can drive the three-body
system using the same permutation geometry as the $t$-driven approach,
while avoiding the explicit pre-calculation and interpolation of the
fully off-shell two-body $t$-matrix. Depending on the grid sizes and
interpolation strategy, this formulation may also reduce the computational
cost \cite{Mohammadzadeh2024}. Whether determined via the $t$-driven
Eq.~(\ref{eq:faddeev_2d_t}) or the potential-driven
Eq.~(\ref{eq:faddeev_2d_v}), the resulting Faddeev components
$\psi_l(p,q)$ determine the bound state within the chosen partial-wave
model space. The full wave function $\Psi_l(p,q)$ is constructed from
the Faddeev component as in Eq.~(\ref{eq:partialwave}). The normalization
condition $3\langle\psi|\Psi\rangle=1$ and the Hamiltonian expectation
value, Eqs.~(\ref{eq:energy_expectation_pwd})--(\ref{eq:vexpect_pwd}), are
then evaluated in the same way as in Sec.~\ref{sec:1dapproach}.

\section{Three-Dimensional Approach}
\label{sec:3dapproach}

In the three-dimensional (3D) approach, the Faddeev equations are projected
directly onto the Jacobi vector momentum states $\ket{\mathbf{p} \, \mathbf{q}}$.
Without decomposing the wave function into partial waves, this approach
retains the continuous angular dependence between the Jacobi momentum
vectors.

To treat the equations numerically in this vector-variable representation, we
define the basis states such that they satisfy the completeness relation
\begin{equation}
\mathbb{1} = \int \! d^3 p  \int \! d^3 q \;
\ket{ {\mathbf p} \, {\mathbf q} } \, \bra{ {\mathbf  p} \,
 {\mathbf q} } ,
\label{eq:complete3bpq}
\end{equation}
and are normalized according to
\begin{align}
  \braket{ {\mathbf p} \, {\mathbf q}} { {\mathbf  p}' \,
    {\mathbf  q}'} &=
  \delta( {\mathbf p}  - {\mathbf p}^\prime) \, \delta( {\mathbf q} -
  {\mathbf q}^\prime ) \nonumber \\
  & = \frac{\delta(p-p')}{p^2}\, \delta(\hat{\mathbf{p}}-\hat{\mathbf{p}}')
  \, \frac{\delta(q-q')}{q^2}\, \delta(\hat{\mathbf{q}}-\hat{\mathbf{q}}') .
\label{eq:normalization3bpq}
\end{align}
Here $\hat{\mathbf{p}}$ and $\hat{\mathbf{q}}$ denote the unit vectors specified
by the solid angles $\Omega_{p,q}=(\theta,\varphi)$. Detailed geometric
derivations for projecting operators onto this vector basis can be found in
Ref.~\cite{Elster98b}.

Before constructing the three-body equations, it is instructive to define the
two-body transition operator in this 3D vector-variable basis. The two-body
transition operator satisfies the Lippmann--Schwinger equation
$t(E)=V+V\ms G_0^{(2)}(E)\ms t(E)$, where $G_0^{(2)}(E)$ is the free two-body
 resolvent defined in Sec.~\ref{sec:1dapproach}. Restricting to
nonrelativistic, local potentials and spinless particles, the matrix elements
depend only on the magnitudes of the initial and final momenta and the angle
between them.

Specifically, we define the variables $x' = \hat{\mathbf{p}}' \cdot \hat{\mathbf{p}}$,
$x'' = \hat{\mathbf{p}}'' \cdot \hat{\mathbf{p}}$, and
$y = \hat{\mathbf{p}}' \cdot \hat{\mathbf{p}}''$. The explicit integral
equation reads
\begin{align}
    t(p', p, x'; E) &= V(p', p, x') + \int\limits_0^\infty \!dp'' \, p''^2
    \int\limits_{-1}^1 \! dx'' \int\limits_0^{2\pi} \! d\varphi'' \nonumber \\
    & \quad \times \, \frac{V(p', p'', y)}{E - p''^2} \; t(p'', p, x''; E),
    \label{eq:ls_3d_explicit}
\end{align}
where the internal angle $y$ is expressed via the azimuthal angle $\varphi''$ as
$y = x' \ms x'' + \sqrt{1-x'^2} \, \sqrt{1-x''^2} \, \cos\varphi''$.

By defining the azimuthally integrated potential
\begin{align}
    v(p', p'',&\, x', x'') \equiv \int\limits_0^{2\pi} \! d\varphi'' \label{eq:vppxx} \\
    & \times V\Big(p', p'', x' x'' + \sqrt{1-x'^2} \,\sqrt{1-x''^2} \,
     \cos\varphi''\Big) , \nonumber
\end{align}
Eq.~(\ref{eq:ls_3d_explicit}) reduces to a 2D integral equation in $p'$ and $x'$,
driven by the azimuthally integrated potential, with the inhomogeneous term
evaluated at $p'' = p$ and $x'' = 1$
\begin{align}
    t(p', p, x';E) &= \frac{1}{2\pi} \; v(p', p, x', 1) +
    \int\limits_0^\infty \! dp'' \, p''^2 \int\limits_{-1}^1 \! dx'' \nonumber \\
    & \quad \times \, \frac{v(p', p'', x', x'')}{E - p''^2} \; t(p'', p, x'';E).
\label{eq:ls_3d_reduced}
\end{align}
This isolates the azimuthal geometry entirely within the azimuthally integrated
potential $v$, a technique we now apply directly to the three-body equations.

Projecting the standard $t$-driven Faddeev Eq.~(\ref{eq:faddeev}) onto the
vector basis and choosing the spectator momentum $\mathbf{q}$ along the
$z$-axis, the Faddeev component depends on $p$, $q$, and
$x=\hat{\mathbf{p}}\cdot\hat{\mathbf{q}}$. In the following integral,
$x'=\hat{\mathbf{q}}'\cdot\hat{\mathbf{q}}$ and $\varphi'$ is the azimuthal
angle of $\mathbf{q}'$ around $\mathbf{q}$. These three-body angular variables
should not be confused with the variables used above in the two-body
Lippmann--Schwinger equation. This projection yields the 3D integral equation
\begin{align}
\psi(p,q,x)&= \frac{1}{ E_3 - p^2 -\frac{3}{4} q^2}
 \int\limits_0^{\infty} \! dq' \, q'^2 \int\limits_{-1}^1 \! dx'
 \int\limits_0^{2\pi} \! d\varphi' \nonumber \\
 & \quad \times \, t_{\rm s} \!\left(p,\pi_1,
 \frac{\frac{1}{2}q \ms x+q' \ms y} {\pi_1}; E_3-\textstyle{\frac{3}{4}} q^2 \right)
 \nonumber \\
 & \quad \times \, \psi\! \left(\pi_2, q', \frac{q \ms x'+\frac{1}{2}q'}{\pi_2}
 \right),
 \label{eq:faddeev3d}
\end{align}
where $t_{\rm s}$ denotes the symmetrized two-body $t$-matrix

\begin{equation}
 \langle \mathbf{p} | t_{\rm s}(E) | \mathbf{q} \rangle =  \langle
  \mathbf{p} | t(E) | \mathbf{q} \rangle +  \langle \mathbf{p}
  | t(E) | -\mathbf{q} \rangle
\label{eq:tsym_def}
\end{equation}

which implements the exchange symmetry of the two particles in the interacting
pair.

The shifted momentum magnitudes $\pi_1$ and $\pi_2$ naturally emerge from the
vector addition, taking the same form defined for the 1D permutation in
Eq.~(\ref{eq:pi12shift}), and the internal angular coupling is given by
$y = x \ms x' + \sqrt{1-x^2} \, \sqrt{1-x'^2} \, \cos\varphi'$.

Alternatively, projecting the potential-driven Eq.~(\ref{eq:faddeev_tfree})
onto the same vector basis produces a structural form that explicitly separates
the direct potential interaction from the geometric permutation. Because the
potential $V$ in the direct term acts only on the subsystem momentum $p$, the
spectator momentum $q$ is strictly conserved across this interaction.

As in the 2D formulation, the bare potential is diagonal in the spectator
momentum. In the direct term this leaves $q$ unchanged, while in the exchange
term the shifted Jacobi momenta enter through the same permutation geometry as
in the $t$-driven equation. In the direct term, the integration over the
azimuthal angle of the intermediate subsystem momentum is already contained
in the azimuthally integrated potential $v$, so that no additional azimuthal
integration appears. Utilizing $v$ defined in Eq.~(\ref{eq:vppxx}) for the
direct term, and replacing the $t$-matrix with the symmetrized bare potential
$V_{\rm s}$ in the permutation term, the $V$-driven 3D equation becomes
\begin{widetext}
\begin{align}
\psi(p,q,x) &= \frac{1}{E_3 - p^2 - \frac{3}{4}q^2} \: \Bigg[
\int\limits_0^\infty \! dp' \, p'^2 \int\limits_{-1}^1 \! dx' \, v(p,p',x,x') \, \psi(p',q,x') \nonumber \\
&\qquad \qquad  \qquad  \qquad
 + \int\limits_0^\infty \! dq' \, q'^2 \int\limits_{-1}^1 \! dx' \int\limits_0^{2\pi} \! d\varphi' \,
V_{\rm s}\!\left(p, \pi_1, \frac{\frac{1}{2}q \ms x + q' \ms y}{\pi_1} \right) \,
 \psi\!\left(\pi_2, q',
\frac{q \ms x'+\frac{1}{2}q'}{\pi_2}\right) \Bigg].
\label{eq:faddeev3d_v}
\end{align}
\end{widetext}
This formulation mirrors the structure of the 2D potential-driven
Eq.~(\ref{eq:faddeev_2d_v}). The bare potential drives the three-body system
through the same permutation geometry as in the $t$-driven equation, while
avoiding the explicit pre-calculation and interpolation of the fully off-shell
two-body $t$-matrix. Depending on the grid sizes and
interpolation strategy, this formulation may also reduce the computational
cost \cite{Mohammadzadeh2024}.

While Eq.~(\ref{eq:vppxx}) allows the use of the full bare interaction directly,
validating the 3D approach against the 1D and 2D methods requires a strict
one-to-one comparison. With the partial-wave normalization used in
Secs.~\ref{sec:1dapproach} and \ref{sec:2dapproach}, the interaction can be
expanded as
\begin{equation}
V(p,p',\hat{\mathbf{p}}\cdot\hat{\mathbf{p}}')
= \sum_l \frac{2l+1}{4\pi}\,
P_l(\hat{\mathbf{p}}\cdot\hat{\mathbf{p}}')\, V_l(p,p') \,.
\end{equation}
Inserting this expansion into the azimuthal integration and using the standard
azimuthal projection identity yields
\begin{align}
v(p,p',x,x')
&= \sum_l \frac{2l+1}{4\pi}\, V_l(p,p')
\int\limits_0^{2\pi} \!d\varphi'\, P_l(y) \nonumber \\
&= \sum_l \frac{2l+1}{2}\, V_l(p,p')\, P_l(x)\, P_l(x') \,,
\label{eq:vppxxpart}
\end{align}
which allows the same finite partial-wave interaction space to be embedded
directly into the 3D vector formulation.

The full three-boson wave function is constructed via Eq.~(\ref{eq:faddeevt}).
For the spinless bosonic state considered here,
$\Psi(\mathbf{p},\mathbf{q}) = \Psi(-\mathbf{p},\mathbf{q})$. Mapping the
internal vector angles, the full wave function can be compactly written as
\begin{align}
\Psi(p,q,x) &= \psi(p,q,x)  \nonumber \\
& \quad + \psi(\tilde \pi_1^+, \tilde \pi_2^+, x^+) +
\psi(\tilde \pi_1^-, \tilde \pi_2^-, x^-),
\label{eq:fullwf_3d}
\end{align}
where the shifted momentum magnitudes for the cyclic ($+$) and anticyclic ($-$)
permutations are given by
\begin{align}
\tilde \pi_1^\pm &= \sqrt{\textstyle{\frac{1}{4}}p^2 +
\textstyle{\frac{9}{16}}q^2 \pm \textstyle{\frac{3}{4}}p \ms q  \ms x}
\nonumber \\
\tilde \pi_2^\pm &= \sqrt{p^2 + \textstyle{\frac{1}{4}}q^2 \mp p \ms q \ms x}.
\label{eq:shiftedarguments_3d_pm}
\end{align}
Notice that the positive branches, $\tilde \pi_1^+$ and $\tilde \pi_2^+$, are
precisely the shifted variables $\tilde \pi_1$ and $\tilde \pi_2$ defined in
Eq.~(\ref{eq:shiftedarguments}) for the partial-wave recoupling.

The corresponding internal angular arguments are determined by the scalar
products of the shifted vectors
\begin{equation}
x^\pm = \frac{ \pm \left( \frac{3}{8}q^2 - \frac{1}{2}p^2 \right) -
\frac{1}{2}p \ms q \ms x } {\tilde \pi_1^\pm \, \tilde \pi_2^\pm}.
\label{eq:xpm_3d}
\end{equation}
This compact notation shows that the permutation geometry used in the 3D
continuous-variable representation is the same geometry that underlies the
shifted arguments in the coupled partial-wave representation.

The full three-body wave function is normalized according to
\begin{align}
1 &= \langle\Psi| \Psi \rangle \nonumber \\
   &= 8 \pi^2 \int\limits_0^{\infty} \! dp \, p^2
        \int\limits_0^{\infty} \! dq \, q^2
        \int\limits_{-1}^1  \! dx \; |\Psi(p,q,x)|^2 \label{eq:3d_norm1} \\
  &= 3 \langle\psi | \Psi \rangle \nonumber \\
   &= 24 \pi^2 \int\limits_0^{\infty} \! dp \, p^2
        \int\limits_0^{\infty} \! dq \, q^2
        \int\limits_{-1}^1  \! dx \; \psi^*(p,q,x) \, \Psi(p,q,x) \,.
        \label{eq:3dpsinormexplicit}
\end{align}
In the exact continuous limit, both integral forms must yield identical results.
In our discretized 3D vector approach, computing both forms provides a
stringent numerical consistency check.

The full Hamiltonian expectation value is evaluated as
\begin{equation}
\langle H\rangle =
3\langle \psi |H_0| \Psi \rangle
+
3\langle \Psi | V_1 | \Psi \rangle .
\label{eq:energy_expectation_3d}
\end{equation}
The reduced kinetic-energy matrix element is
\begin{align}
\langle \psi |H_0| \Psi \rangle &= 8 \pi^2 \int\limits_0^{\infty}
\! dp \, p^2 \int\limits_0^{\infty} \! dq \, q^2
\left (
  p^2 + \frac{3}{4}q^2
\right) \nonumber \\
 & \quad \times \, \int\limits_{-1}^1 \! dx  \,
 \psi^*\!(p,q,x) \, \Psi(p,q,x),
\label{eq:h0expect_3d}
\end{align}
and the matrix element of one pair interaction is
\begin{align}
\langle \Psi | V_1 | \Psi \rangle &= 8 \pi^2 \int\limits_0^{\infty}
  \! dp \, p^2 \int\limits_0^{\infty} \! dq \, q^2 \int\limits_0^{\infty}
  \! dp' \, p'^2 \int\limits_{-1}^1 \! dx \int\limits_{-1}^1 \! dx' \nonumber \\
  & \quad \times \, \Psi^*\!(p,q,x) \, v(p,p',x,x') \,
  \Psi(p',q,x') \,,
\label{eq:vexpect_3d}
\end{align}
where the azimuthal dependence has already been absorbed into the azimuthally
integrated potential $v$ defined in Eq.~(\ref{eq:vppxx}), or into its
truncated partial-wave representation defined in Eq.~(\ref{eq:vppxxpart}).

The full wave function $\Psi(\mathbf{p},\mathbf{q})$ can be expanded in terms
of partial-wave components $\Psi_l(p,q)$
\begin{align}
\Psi(\mathbf{p},\mathbf{q}) = \Psi(p,q,x) = \sum_l \frac{\sqrt{2l+1}}{4\pi} \,
P_l(x) \, \Psi_l(p,q).
\label{eq:pwe_expansion}
\end{align}
Conversely, the partial-wave components are obtained by projecting the
3D solutions onto the Legendre polynomials
\begin{align}
\Psi_l(p,q) = 2\pi \sqrt{2l+1} \int\limits_{-1}^1 \! dx \, P_l(x) \,
\Psi(p,q,x).
\label{eq:pwe_projection}
\end{align}
These relations allow for a direct numerical mapping between the 3D
vector-variable results and the converged partial-wave expansions used in
the 1D and 2D approaches, facilitating the benchmark comparisons presented
in Sec.~\ref{sec:results}.

\section{Observables and Wave Function Analysis}
\label{sec:observables}

In physical applications, the multidimensional wave function is usually
analyzed through expectation values and reduced distributions.
For instance, the spectator momentum distribution
$n_{\mathrm{spec}}(q)$ represents the probability of finding a boson with
momentum $q$ relative to the center of mass of the interacting pair.
Complementarily, the subsystem momentum distribution $n_{\mathrm{sub}}(p)$
represents the internal momentum of that pair. Using the full 3D wave
function and the partial-wave components, these distributions are evaluated as
\begin{align}
  n_{\mathrm{spec}}(q) &= 2\pi \ms q^2 \int\limits_0^{\infty} \! dp \, p^2
  \int\limits_{-1}^1 \! dx \, |\Psi(p,q,x)|^2  \nonumber \\
       &= \frac{q^2}{4\pi} \sum_l  \int\limits_0^{\infty} \! dp \, p^2 \,
                 |\Psi_l(p,q)|^2, \label{eq:mom_dist_q} \\[8pt]
  n_{\mathrm{sub}}(p) &= 2\pi \ms p^2 \int\limits_0^{\infty} \! dq \, q^2
  \int\limits_{-1}^1 \! dx \, |\Psi(p,q,x)|^2  \nonumber \\
  &= \frac{p^2}{4\pi} \sum_l  \int\limits_0^{\infty} \! dq \, q^2 \,
     |\Psi_l(p,q)|^2, \label{eq:mom_dist_p}
\end{align}
where $x$ generally denotes the cosine of the angle between the relevant
Jacobi vectors. With this convention, the remaining solid-angle factor is kept
outside the definition of the radial distributions, so that
$4\pi \int_0^\infty \! dq \, n_{\mathrm{spec}}(q) = 1$ and
$4\pi \int_0^\infty \! dp \, n_{\mathrm{sub}}(p) = 1$. By comparing the
partial-wave sums to the corresponding 3D vector integrals, we can quantify
the angular momentum content required to saturate the state within the chosen
numerical accuracy.

\subsection{Coordinate-Space Transformation: 3D Vector Approach}

To investigate the spatial structure of the bound state, we transform the
momentum-space wave function $\Psi(\mathbf{p},\mathbf{q})$ into
configuration space. We define the conjugate spatial Jacobi coordinates
$\mathbf{r}$ (the two-body subsystem distance) and $\boldsymbol{\rho}$
(the distance of the third particle to the pair's center of mass).

For the 3D vector-variable approach, we perform the unitary transformation
without introducing a partial-wave expansion. The full double Fourier
integral is given by
\begin{equation}
\tilde \Psi(\mathbf{r},\boldsymbol{\rho}) =
\frac{1}{(2\pi)^3}
\int \! d^3p \int \! d^3q \,
\Psi(\mathbf{p},\mathbf{q}) \,
e^{i\mathbf{p}\cdot\mathbf{r}} \,
e^{i\mathbf{q}\cdot\boldsymbol{\rho}} .
\label{eq:fourier_full}
\end{equation}

We evaluate this in two steps to preserve numerical stability \cite{Liu2003a}.
First, we transform the spectator momentum $\mathbf{q}$ to $\boldsymbol{\rho}$
\begin{equation}
I_q(\mathbf{p},\boldsymbol{\rho}) =
\frac{1}{(2\pi)^{3/2}}
\int \! d^3q \,
\Psi(\mathbf{p},\mathbf{q}) \,
e^{i\mathbf{q}\cdot\boldsymbol{\rho}} .
\label{eq:iq_def}
\end{equation}
Choosing $\mathbf{p}\parallel\hat{z}$, we define the internal angles
$x_q = \hat{\mathbf{q}}\cdot\hat{\mathbf{p}}$ and
$\mu = \hat{\boldsymbol{\rho}}\cdot\hat{\mathbf{p}}$. The azimuthal integration
over $\phi_q$ reduces the angular dependence to a zeroth-order cylindrical
Bessel function $J_0$. For the real bound-state wave functions considered
here, the imaginary part of the Fourier integral vanishes by symmetry. We
therefore retain only the real part, which after the azimuthal integration is
\begin{align}
I_q(p,\rho,\mu)
&=
\frac{1}{\sqrt{2\pi}}
\int\limits_0^\infty \! dq \, q^2
\int\limits_{-1}^{1} \! dx_q  \, \cos(q \ms \rho \ms x_q  \ms \mu) \nonumber \\
& \quad \times \, J_0\!\left(q \ms \rho \ms \sqrt{1-x_q^2} \,
\sqrt{1-\mu^2}\right) \, \Psi(p,q,x_q).
\label{eq:iq_final}
\end{align}
Parseval's theorem for this $q\to\rho$ transformation yields the partial
unitarity identity
\begin{equation}
8\pi^2
\int\limits_0^\infty \! dp \, p^2
\int\limits_0^\infty \! d\rho \, \rho^2
\int\limits_{-1}^{1} \! d\mu \;
|I_q(p,\rho,\mu)|^2
= 1 ,
\label{eq:reduced_unitarity_step1}
\end{equation}
which provides a stringent numerical consistency check for the intermediate
projection prior to the final spatial integration.

The second step completes the transformation into full coordinate space
\begin{equation}
\tilde  \Psi(\mathbf{r},\boldsymbol{\rho}) =
\frac{1}{(2\pi)^{3/2}}
\int \! d^3p \,
I_q(\mathbf{p},\boldsymbol{\rho}) \,
e^{i\mathbf{p}\cdot\mathbf{r}} .
\label{eq:second_step}
\end{equation}
Choosing $\mathbf{r}\parallel\hat{z}$ and defining the spatial angles
$x_p = \hat{\mathbf{p}}\cdot\hat{\mathbf{r}}$ and
$x_r = \hat{\boldsymbol{\rho}}\cdot\hat{\mathbf{r}}$, the internal angular
coupling becomes $\mu = x_p \ms x_r + \sqrt{1-x_p^2} \, \sqrt{1-x_r^2} \,
\cos\phi_p$.

An angular reduction of the final \(p\)-integration to a spherical Bessel
factor would be valid only if the remaining integrand were independent of the
direction of \(\mathbf p\). In the present two-step transformation this
condition is not satisfied. After the \(q\to\rho\) transformation, the
intermediate amplitude \(I_q\) depends on the angle
\(\mu=\hat{\boldsymbol\rho}\cdot\hat{\mathbf p}\). During the subsequent
integration over the direction of \(\mathbf p\), this angle varies according to
\begin{equation}
\mu =
x_p \ms x_r+\sqrt{1-x_p^2}\, \sqrt{1-x_r^2}\,\cos\phi_p .
\end{equation}
Thus \(I_q(p,\rho,\mu)\) cannot be taken outside the angular integration over
\(\mathbf p\), and the full azimuthal integral over \(\phi_p\) must be
retained. For the same
symmetry reasons noted above, only the real part of the final Fourier integral
contributes
\begin{align}
\tilde \Psi(r,\rho,x_r)
&=
\frac{1}{(2\pi)^{3/2}}
\int\limits_0^\infty  \! dp \, p^2
\int\limits_{-1}^{1} \! dx_p
\int\limits_0^{2\pi} \! d\phi_p \nonumber \\
& \quad \times
\cos(p \ms r \ms x_p) \,
I_q(p,\rho,\mu).
\label{eq:final_unitary}
\end{align}
In the continuous limit, this formulation exactly preserves the norm of
Eq.~(\ref{eq:3d_norm1}); computational deviations from this identity provide a
useful measure of the numerical accuracy of the spatial transformation.

\subsection{Coordinate-Space Transformation: Partial-Wave Approach}

For a bound state with zero total angular momentum ($\pazocal{L}=0$), we
insert the Legendre expansion of the momentum-space wave function, given
in Eq.~(\ref{eq:pwe_expansion}), into the unitary transformation of
Eq.~(\ref{eq:fourier_full}). Expanding the plane waves into spherical
harmonics and spherical Bessel functions $j_l(z)$, and applying the
spherical harmonic addition theorem, the angular integrations decouple
completely.

Using the orthogonality of the spherical harmonics and collecting the phases,
the spatial wave function takes the partial-wave form
\begin{equation}
 \tilde  \Psi(\mathbf{r},\boldsymbol{\rho}) = \sum_{l}
  \frac{\sqrt{2l+1}}{4\pi} \, P_l(\hat{\mathbf{r}}\cdot\hat{\boldsymbol{\rho}})
  \, \tilde  \Psi_l(r,\rho),
\label{eq:coord_pwe_expansion}
\end{equation}
where the radial components $\tilde \Psi_l(r,\rho)$ are given by the double
spherical Bessel transform
\begin{align}
 \tilde  \Psi_l(r,\rho) &= \frac{2}{\pi} (-1)^l  \nonumber \\
 & \quad \times  \int\limits_0^\infty \! dp \, p^2 j_l(p \ms r)
  \int\limits_0^\infty \!dq \, q^2 j_l(q  \ms \rho) \, \Psi_l(p,q).
\label{eq:coord_bessel_transform}
\end{align}
Because the angular integrations have been performed analytically, this
representation avoids the azimuthal coupling inherent to the direct 3D
transform, requiring only the evaluation of the oscillatory radial integrals.

\subsection{Spatial Geometry and Observables}

Using this spatial representation, the radial correlation function
$c_{\mathrm{sub}}(r)$ describes the distribution of separations $r$ between
the two subsystem bosons. Similarly, $c_{\mathrm{spec}}(\rho)$ describes the
distribution of distances $\rho$ between the third boson and the subsystem
center of mass. In terms of the full 3D state and the partial-wave
components, these are evaluated as
\begin{align}
c_{\mathrm{sub}}(r) &= 2\pi \ms r^2 \int\limits_0^\infty \! d\rho \, \rho^2
  \int\limits_{-1}^1 \! dx_r \, | \tilde \Psi(r, \rho ,x_r)|^2 \nonumber \\
  &= \frac{r^2}{4\pi} \sum_l  \int\limits_0^\infty \! d\rho \,
\rho^2 \, |\tilde \Psi_l(r,\rho)|^2, \label{eq:corr_function_combined} \\[8pt]
c_{\mathrm{spec}}(\rho) &= 2\pi \ms \rho^2 \int\limits_0^\infty \! dr \, r^2
 \int\limits_{-1}^1 \! dx_r\, |\tilde \Psi(r, \rho ,x_r)|^2 \nonumber \\
 &= \frac{\rho^2}{4\pi} \sum_l
\int\limits_0^\infty \! dr \, r^2 \, |\tilde \Psi_l(r,\rho)|^2.
\label{eq:corr_spectator_combined}
\end{align}
As for the momentum distributions, the remaining solid-angle factor is kept
outside the definition of the radial distributions. Thus
$4\pi \int_0^\infty \! dr\,c_{\mathrm{sub}}(r)=1$ and
$4\pi \int_0^\infty \! d\rho\,c_{\mathrm{spec}}(\rho)=1$.

A symmetric spatial configuration of three identical bosons can be geometrically
compared to an equilateral triangle. The expectation values of the spatial
Jacobi coordinates represent the base length $\langle r \rangle$ and the
height $\langle \rho \rangle$ of this triangle
\begin{align}
\langle r \rangle &= 8\pi^2 \int\limits_0^\infty \! d\rho \, \rho^2
\int\limits_0^\infty \! dr \, r^3 \int\limits_{-1}^1 \! dx_r \,
|\tilde  \Psi(r,\rho,x_r)|^2 \nonumber \\
  &= \sum_{l} \int\limits_0^\infty \! d\rho \, \rho^2
  \int\limits_0^\infty \! dr \, r^3 \: |\tilde \Psi_l(r,\rho)|^2,
  \label{eq:exp_r} \\[10pt]
\langle \rho \rangle &= 8\pi^2 \int\limits_0^\infty \! d\rho \, \rho^3
\int\limits_0^\infty \! dr \, r^2 \int\limits_{-1}^1 \! dx_r \,
|\tilde \Psi(r,\rho,x_r)|^2 \nonumber \\
  &= \sum_{l} \int\limits_0^\infty \!dr \, r^2
  \int\limits_0^\infty \! d\rho \, \rho^3 \, |\tilde \Psi_l(r,\rho)|^2.
  \label{eq:exp_rho}
\end{align}
Equivalently, the same expectation values follow from the radial
correlation functions,
\begin{align}
\langle r\rangle &=
4\pi\int\limits_0^\infty \! dr\, r\, c_{\mathrm{sub}}(r), \nonumber \\
\langle\rho\rangle &=
4\pi\int\limits_0^\infty \! d\rho\, \rho\, c_{\mathrm{spec}}(\rho).
\label{eq:exp_from_corr}
\end{align}

For a single equilateral configuration, the ratio between the base and height
is $r/\rho=2/\sqrt{3}$. The quantum bound state does not impose this relation
pointwise on the probability distribution. Nevertheless, the ratio
$\langle r\rangle/\langle\rho\rangle$ provides a useful diagnostic of the
average spatial correlations. We therefore define the deviation from the
equilateral reference value as
\begin{equation}
\delta = \frac{\langle r \rangle / \langle \rho \rangle - 2/\sqrt{3}}
{2/\sqrt{3}} \times 100 .
\label{eq:triangle_deviation}
\end{equation}
Evaluating these spatial integrals provides a compact measure of how the
average geometry depends on the underlying interaction.

\section{Two-Body Interactions}
\label{sec:interactions}

For the separable interactions introduced in Sec.~\ref{sec:1dapproach},
we employ generalized Yamaguchi form factors defined for all partial waves as
\begin{equation}
g_l(p) = \frac{p^l}{(p^2+\beta^2)^{l+1}} \,.
\label{eq:yamaguchiform}
\end{equation}
For $l=0$, this reduces to the original Yamaguchi form factor
\cite{Yamaguchi54a}. The range parameter $\beta$ is taken to be the same
in all retained partial waves. With momenta expressed in fm$^{-1}$,
$g_l(p)$ has units fm$^{l+2}$. Consequently, with the potential convention
of Eq.~(\ref{eq:separable_potential}), the partial-wave strength has units
\begin{equation}
[\lambda_l]={\rm fm}^{-(2l+3)} .
\end{equation}

In the calculations reported here, the same numerical coefficient is used
for the strength in every retained even partial wave. We therefore introduce
the dimensionless numerical coefficient $\bar\lambda$ and define
\begin{equation}
\lambda_l =
\bar\lambda\,{\rm fm}^{-(2l+3)},
\qquad l=0,2,4,\ldots,l_{\rm max}.
\label{eq:yamaguchi_strengths}
\end{equation}
The values of $\bar\lambda$ are listed in
Table~\ref{tab:yamaguchiparams}. Thus, for example, $\lambda_0$,
$\lambda_2$, and $\lambda_4$ have units fm$^{-3}$, fm$^{-7}$,
and fm$^{-11}$, respectively. This prescription defines the generalized
higher-partial-wave Yamaguchi interaction used in the present benchmark;
the original Yamaguchi form corresponds to the $s$ wave.
With the Fourier--Bessel convention used in this work,
the corresponding $l=0$ coordinate-space form factor is
\begin{equation}
\tilde  g_0(r) = \sqrt{\frac{\pi}{2}} \, \frac{e^{-\beta \ms r}}{r} \,.
\label{eq:yamaguchiform_r}
\end{equation}

Although the calculations are performed in momentum space, the
coordinate-space representation provides useful checks for spatial
observables and potential-energy expectation values. Exact analytical
spatial representations can also be derived for higher partial waves
($l>0$). In the literature, several parameter sets are used for the
Yamaguchi potential, often together with slightly different conversion
conventions. Table~\ref{tab:yamaguchiparams} lists the parameter sets
considered here, the conversion factors employed, and the exact $s$-wave
two-body binding energy $E_2$ obtained analytically \cite{Schadow2026b}.

\begin{table*}[!tbp]
\caption{
Parameters and conversion factors for the Yamaguchi interactions.
The final column identifies the source of each parameter set.
The dimensionless coefficient $\bar\lambda$ defines the partial-wave
strengths through
$\lambda_l=\bar\lambda\,{\rm fm}^{-(2l+3)}$ for all retained even
partial waves. The listed $s$-wave two-body binding energies $E_2$
are taken from Ref.~\cite{Schadow2026b}, where they were calculated
both analytically and by numerical solution of the two-body
Schr\"odinger equation. The listed parameters and conversion
factors are those used in the present calculations.
}
\label{tab:yamaguchiparams}
\setlength{\tabcolsep}{0pt}
\begin{ruledtabular}
\begin{tabular*}{\textwidth}{
    @{\extracolsep{\fill}}
    l
    S[table-format=1.7, group-digits=false]
    S[table-format=1.7, group-digits=false]
    S[table-format=2.8, group-digits=false]
    S[table-format=3.7, group-digits=false]
    S[table-format=-1.12, group-digits=false]
    c
}
Potential &
\multicolumn{1}{c}{$\beta$} &
\multicolumn{1}{c}{$\bar\lambda$} &
\multicolumn{1}{c}{${(\hbar c)^2}/{m_N}$} &
\multicolumn{1}{c}{$\hbar c$} &
\multicolumn{1}{c}{$E_2$} &
Ref. \\
 & {[fm$^{-1}$]} & {} & {[MeV fm$^2$]} & {[MeV fm]} & {[MeV]} & \\
\hline\\[-8pt]
YAMA-23  & 1.3905818 & 0.3707654 & 41.47103997 & 197.3269804 & -2.22456720 &
\cite{Schadow2026b}\\
YAMA-IV  & 1.15      & 0.179     & 41.47       & 197.3286    & -0.33178133 &
\cite{Hadizadeh2007a}\\
\end{tabular*}
\end{ruledtabular}
\end{table*}

\begin{table*}[!tbp]
\caption{
Malfliet--Tjon interaction parameters and conversion factors
adopted from Ref.~\cite{Elster98b}. The two energy columns give
the two-body binding energies calculated in
Ref.~\cite{Schadow2026b} and the earlier values reported in
Ref.~\cite{Elster98b}, respectively. The earlier energies are
reproduced at their original published precision. The listed
parameters and conversion factors are those used in the
present calculations.
}
\label{tab:mtparams}
\begin{ruledtabular}
\setlength{\tabcolsep}{0pt}
\sisetup{exponent-product = \cdot}
\begin{tabular*}{\textwidth}{
    @{\extracolsep{\fill}}
    l
    S[table-format=-3.4, group-digits=false]
    S[table-format=1.3, group-digits=false]
    S[table-format=4.4, group-digits=false]
    S[table-format=1.2, group-digits=false]
    S[table-format=2.2, group-digits=false]
    S[table-format=3.1, group-digits=false]
    S[table-format=-1.8, group-digits=false]
    S[table-format=-1.4, group-digits=false]
}
Potential &
\multicolumn{1}{c}{$V_a$} &
\multicolumn{1}{c}{$\mu_a$} &
\multicolumn{1}{c}{$V_r$} &
\multicolumn{1}{c}{$\mu_r$} &
\multicolumn{1}{c}{${(\hbar c)^2}/{m_N}$} &
\multicolumn{1}{c}{$\hbar c$} &
\multicolumn{1}{c}{$E_2$ \cite{Schadow2026b}} &
\multicolumn{1}{c}{$E_2$ \cite{Elster98b}} \\
 & {[MeV fm]} & {[fm$^{-1}$]} &
   {[MeV fm]} & {[fm$^{-1}$]} &
   {[MeV fm$^2$]} & {[MeV fm]} &
   {[MeV]} & {[MeV]} \\
\hline\\[-8pt]
MT-IV & -65.109   & 0.633 & {--}      & {--} &
        41.47 & 197.3 & -2.20862918 & -2.2087 \\
MT-V  & -570.3316 & 1.55  & 1438.4812 & 3.11 &
        41.47 & 197.3 & -0.35000049 & -0.3500 \\
\end{tabular*}
\end{ruledtabular}
\end{table*}

As a second class of interaction models, we employ local Malfliet--Tjon
(MT) potentials of Yukawa type \cite{Yukawa35a,Malfliet69}. While these
central interactions do not include tensor components or spin dependence,
they provide useful benchmark potentials with simple analytic momentum-space
representations. Depending on the parameter set, they also include a finite
short-range repulsive component and therefore test the convergence of the
high-momentum part of the wave function. In coordinate space, they are given by
\begin{equation}
\tilde V(r) = V_a \, \frac{e^{-\mu_a \ms r}}{r} + V_r \,
\frac{e^{-\mu_r \ms r}}{r} \,,
\label{eq:malfliet-tjon}
\end{equation}
where the long-range attractive part is characterized by $V_a$ and $\mu_a$,
and the short-range repulsive part by $V_r$ and $\mu_r$. In the parameter
sets used below, $V_a < 0$ and $V_r \ge 0$.

With the momentum-space normalization used in this work, the partial-wave
matrix elements take the closed form
\begin{equation}
V_l(p,p') = \frac{1}{\pi p \ms p'}\left[ V_a \, Q_l(z_a)
+ V_r \, Q_l(z_r) \right] \,,
\label{eq:mt_partial_wave}
\end{equation}
with
\begin{equation}
z_{a/r} = \frac{p^2 + p'^2 + \mu_{a/r}^2}{2 \ms p \ms p'} \,,
\label{eq:mt_z}
\end{equation}
where $Q_l(z)$ denotes the Legendre function of the second kind. For regular
central potentials in this normalization, the partial-wave matrix elements
with $l>0$ vanish when either external momentum approaches zero. For numerical
use near the grid boundaries, the $l=0$ matrix element is evaluated using
the limiting form
\begin{align}
&V_0(p,p') \bigg|_{p\approx0\ {\rm or}\ p'\approx0} \nonumber \\
& \qquad \qquad\approx \frac{2}{\pi} \left( \frac{V_a}{p^2 + p'^2 + \mu_a^2}
 + \frac{V_r}{p^2 + p'^2 + \mu_r^2} \right) .
 \label{eq:mt_l0_boundary}
\end{align}

With this normalization, the full vector-variable matrix element is related
to the partial-wave matrix elements through the expansion introduced in
Sec.~\ref{sec:3dapproach}. It is given by
\begin{equation}
V(\mathbf{p},\mathbf{p}') =
\frac{1}{2\pi^2}
\left[
  \frac{V_a}{(\mathbf{p}-\mathbf{p}')^2+\mu_a^2}
  +
  \frac{V_r}{(\mathbf{p}-\mathbf{p}')^2+\mu_r^2}
\right] \,.
\label{eq:mt_vector_potential}
\end{equation}
With the definition of $v$ as the full azimuthal integral in
Eq.~(\ref{eq:vppxx}), this integration can be performed analytically.
Writing $x$ and $x'$ as the cosines of the polar angles entering
Eq.~(\ref{eq:vppxx}), the resulting azimuthally integrated interaction
$v(p,p',x,x')$, used directly in the 3D integral kernels, is explicitly
given by
\begin{widetext}
\begin{align}
v(p,p',x,x') &= \frac{1}{\pi}\left[
\frac{V_a}{\sqrt{
\left(p^2+p'^2-2 \,p \, p' \, x \, x' + \mu_a^2\right)^2
-4 \, p^2 \,p'^2(1-x^2)(1-x'^2)}}
\right. \nonumber \\
&\qquad\qquad\qquad\qquad \left.
+ \frac{V_r}{\sqrt{
\left(p^2+p'^2-2 \, p \, p' \, x \, x' + \mu_r^2\right)^2
-4 \, p^2 \,p'^2(1-x^2)(1-x'^2)}}
\right].
\label{eq:vppxx_analytic}
\end{align}
\end{widetext}

Table~\ref{tab:mtparams} collects the parameter sets adopted here and the
corresponding conversion factors.

\section{Numerical Methods}
\label{sec:numerical_method}

The numerical strategy closely follows the approach established in the
two-body study \cite{Schadow2026b}, extended here to the higher-dimensional
three-body kernels. The determination of the three-body bound-state energy
and wave function in momentum space requires the solution of the homogeneous
Faddeev equations at negative energies. For a fixed trial energy $E$, the
discretized equation is written as an eigenvalue problem for the kernel
$K(E)$
\begin{equation}
K(E) |\psi\rangle = \eta(E) \, |\psi\rangle .
\label{eq:kernel_eigenproblem}
\end{equation}
The physical three-body binding energy $E_3$ is determined by the condition
\begin{equation}
\eta(E_3)=1 .
\label{eq:binding_condition}
\end{equation}

For each trial energy, the two-body input is constructed either as a fully
off-shell $t$-matrix or directly from the potential matrix elements,
depending on whether the $t$-driven or $V$-driven formulation is used.
The resulting three-body kernel is applied to a trial Faddeev component on
the discretized momentum and angular grids. A Krylov subspace method
\cite{Arnoldi51,Stadler91a} is then used to determine the eigenvalue
associated with the physical bound-state branch. The total energy $E$ is
varied until the condition $\eta(E_3)=1$ is fulfilled. After convergence,
the Faddeev component is used to reconstruct the full wave function, normalize
the state, evaluate expectation values, and, where required, transform the
momentum-space wave function into coordinate space.

\subsection{Discretization and Iteration}

All continuous momentum and angular variables are discretized using
Gauss--Legendre quadrature rules. The semi-infinite momentum intervals
are truncated at finite cutoffs $p_{\rm max}$ and $q_{\rm max}$.
As detailed in the convergence studies below, evaluating the
Hamiltonian expectation values requires significantly larger cutoffs
than determining the binding energies alone. For the production
calculations reported below, we use a spectator momentum cutoff of
$q_{\rm max}=400~\mathrm{fm}^{-1}$ for the separable Yamaguchi
interactions. To accommodate the shifted momentum arguments in the
permutation kernel, we choose the corresponding pair-momentum cutoff
as $p_{\rm max}=\frac{3}{2}q_{\rm max}=600~\mathrm{fm}^{-1}$.
For the local Malfliet--Tjon potentials,
$q_{\rm max}=200~\mathrm{fm}^{-1}$
($p_{\rm max}=300~\mathrm{fm}^{-1}$) is sufficient at the quoted
accuracy.

To obtain a dense grid in the physically dominant low-momentum
region, the momentum intervals are divided into subintervals,
$(0,p_0)\cup(p_0,p_{\rm max})$ and
$(0,q_0)\cup(q_0,q_{\rm max})$, with representative boundary values
$p_0=9~\mathrm{fm}^{-1}$ and $q_0=10~\mathrm{fm}^{-1}$.
The mappings described in Ref.~\cite{Gloeckle91} are used to
distribute the quadrature points efficiently. The momentum-grid
sizes are denoted by $N_p$ and $N_q$.
The required mesh sizes depend on the interaction, formulation,
and observable. As illustrated by the convergence studies below,
the binding energy can stabilize on a smaller grid than that
required for accurate evaluation of the Hamiltonian expectation
values.

In the 3D vector-variable formulation, the angular grid for
$x=\hat{\mathbf p}\cdot\hat{\mathbf q}$ contains
$N_x=34$--$50$ points, including the endpoints $x=\pm1$.
Gauss--Legendre quadrature is used for the interior points.
The azimuthal integration uses $N_\varphi=32$ quadrature points
for YAMA-23 and $N_\varphi=48$ for the Malfliet--Tjon interactions.
Test calculations with larger values of $N_\varphi$ showed no
changes within the numerical accuracy quoted in this work.
The boundary points $p=0$, $q=0$, and $x=\pm1$ are included
with zero quadrature weight. They do not contribute to the
quadrature sums but are available for interpolation and function
evaluations at kinematic boundaries, avoiding extrapolation
at these points.

The permutation operator introduces shifted momenta, such as
$(\pi_1,\pi_2)$, which generally do not coincide with the
quadrature points. Evaluating the integral kernels therefore
requires interpolation of the wave functions and, in the
$t$-driven formulations, of the fully off-shell two-body
$t$-matrix. We employ local cubic Hermite splines
\cite{Hueber93} for interpolation in the continuous momentum
and angular variables. This provides a stable representation
of the off-shell $t$-matrix and Faddeev amplitudes while avoiding
the Runge oscillations associated with high-order global
polynomial interpolation \cite{Press2007}.

With $N_l$ retained partial-wave channels, the dimensions of the
discretized state vectors are
\[
D_{\rm 1D}=N_q \times N_l, \quad
D_{\rm 2D}=N_p \times N_q \times N_l, \quad
D_{\rm 3D}=N_p \times N_q \times N_x.
\]
The corresponding kernels formally have dimensions $D\times D$.
The 1D separable formulation leads to a matrix eigenvalue problem
of moderate size that can be solved by standard direct methods.
For the much larger 2D and 3D problems, explicit construction
and storage of the full kernel matrix $K(E)$ are impractical.
Instead, the equations are solved iteratively by computing only
the kernel action,
\begin{equation}
|v'\rangle = K(E) \,|v\rangle .
\end{equation}
This matrix-free operation is used to construct the Krylov
subspace from which the relevant eigenvalues are extracted.

As a check on numerical robustness, two independent eigensolvers
are implemented. The first is a custom subspace projection
method based on Ref.~\cite{Stadler91a}. It constructs an
orthogonal basis by repeated kernel applications, starting from
a smooth analytic trial function,
$\psi(p,q,x)\sim[(0.5+p^2)(0.5+q^2)]^{-1}$, and diagonalizes
the resulting small Rayleigh--Ritz matrix.

The second, primary solver uses the implicitly restarted
Arnoldi method \cite{Arnoldi51} provided by the ARPACK library.
Through its reverse-communication interface, the method builds
the Krylov subspace using matrix-vector products without
requiring a prescribed analytic starting vector.
Wave-function overlap tracking (root homing) is used to
identify the physical bound state and follow it across
successive trial energies. For all calculations reported below,
the eigensolver iterations are terminated when the relative
change of the eigenvalue falls below $10^{-10}$.

Since the kernel depends nonlinearly on the total energy $E$,
the eigenvalue problem must be solved repeatedly to determine
the physical energy $E_3$. An outer root-finding procedure,
initialized with two trial energies, uses a stabilized secant
iteration to update $E$. The search is stopped when both
$|\eta(E)-1|$ and the change in $E$ fall below the prescribed
tolerances, typically $10^{-10}$ and $10^{-8}$~MeV, respectively.

The computationally demanding multidimensional integrations
are parallelized using the Message Passing Interface (MPI).
The spectator momentum grid $q$ is distributed across MPI
ranks, and the resulting partial integrals are combined through
global reductions at each iteration.

Table~\ref{tab:runtimes_yama23} summarizes representative mesh
sizes, kernel orders, and production costs for YAMA-23 with
identical interaction content in all three formulations.
The partial-wave calculations retain the seven even channels
$l=0,2,\ldots,12$, so that $N_l=7$. In the 3D calculation,
the same interaction content is embedded in the kernel, while
the unknown is represented directly on a
$N_p\times N_q\times N_x$ grid.
The mesh columns describe this state representation rather
than all auxiliary integration grids. In particular,
$N_\varphi$ affects the cost of applying the 3D kernel but
does not enter its order $D$.

The 1D timing is divided into two stages: initialization and
the spectator-amplitude solution, including the binding-energy
search, followed by full-wave-function reconstruction,
normalization, and momentum-space expectation values on a
separate two-momentum grid. The reconstruction does not
involve another eigenvalue problem, so no kernel order is
assigned to this row. The complete 2D and 3D timings include
both stages and refer to the $t$-matrix-driven formulation.
Coordinate-space transformations are excluded throughout.

The 1D programs use one core and one thread. The parallel
calculations use one single-threaded MPI rank per core,
with the node counts based on 32 ranks per node.
To distinguish elapsed runtime from computational resource
cost, the table also reports allocated core-hours, calculated
as the number of cores multiplied by the wall-clock time
in hours.

\begin{table*}[!tbp]
\caption{
Mesh sizes, kernel orders $D$, and representative production
costs for YAMA-23 with identical interaction content through
$l_{\rm max}=12$. Timings cover the momentum-space stages
defined in the text and exclude coordinate-space transformations.
Dashes indicate quantities not applicable to the corresponding
representation or stage.
}
\label{tab:runtimes_yama23}
\centering
\setlength{\tabcolsep}{0pt}
\sisetup{group-digits=false}
\begin{ruledtabular}
\begin{tabular*}{\textwidth}{
    @{\extracolsep{\fill}}
    l
    S[table-format=3.0]
    S[table-format=3.0]
    S[table-format=1.0]
    S[table-format=2.0]
    S[table-format=7.0]
    S[table-format=3.0]
    S[table-format=2.0]
    S[table-format=3.1]
    S[table-format=3.2]
}
Calculation &
\multicolumn{1}{c}{$N_p$} &
\multicolumn{1}{c}{$N_q$} &
\multicolumn{1}{c}{$N_l$} &
\multicolumn{1}{c}{$N_x$} &
\multicolumn{1}{c}{$D$} &
\multicolumn{1}{c}{Cores} &
\multicolumn{1}{c}{Nodes} &
\multicolumn{1}{c}{$T_{\rm wall}$ [s]} &
\multicolumn{1}{c}{Core-hours} \\
\hline\\[-8pt]
1D solution
& {--} & 767 & 7 & {--}
& 5369 & 1 & 1 & 193.9 & 0.05 \\
1D reconstruction
& 768 & 768 & 7 & {--}
& {--} & 1 & 1 & 702.5 & 0.20 \\
\hline\\[-8pt]
2D complete
& 768 & 768 & 7 & {--}
& 4128768 & 256 & 8 & 104.2 & 7.41 \\
3D complete
& 512 & 512 & {--} & 34
& 8912896 & 512 & 16 & 995.8 & 141.63 \\
\end{tabular*}
\end{ruledtabular}
\end{table*}

\begin{table*}[htbp]
\setlength{\tabcolsep}{0pt}
\caption{
\linespread{1.0}\selectfont
Convergence of the three-body binding energy $E_3$ obtained directly from the
1D spectator amplitude equation for the YAMA-23 potential at
$l_{\rm max} = 12$. Here $N_q$ denotes the number of spectator-momentum
grid points.}
\label{tab:form_convergence_YAMA23}
\begin{ruledtabular}
\begin{tabular*}{\columnwidth}{
    @{\extracolsep{\fill}}
    S[table-format=3.0]
    S[table-format=-2.8, group-digits=false]
    S[table-format=3.0]
}
\multicolumn{1}{c}{$q_{\rm cut}$ [fm$^{-1}$]} &
\multicolumn{1}{c}{$E_3$ [MeV]} &
\multicolumn{1}{c}{$N_q$} \\
\hline\\[-8pt]
 10 & -24.53848452 & 48 \\
 20 & -24.53856898 & 48 \\
 30 & -24.53856914 & 48 \\
 40 & -24.53856914 & 48 \\
 50 & -24.53856914 & 48 \\
100 & -24.53856914 & 48 \\
\end{tabular*}
\end{ruledtabular}
\end{table*}

The separate 1D timings illustrate that reconstructing the
full wave function and evaluating its expectation values
can be more expensive than solving the spectator-amplitude
equation itself. Nevertheless, the 1D reduction remains the
most economical in allocated core time for this separable
benchmark. The shorter elapsed time of the 2D calculation
comes with a substantially larger core allocation.
These production timings therefore characterize the stated
meshes and computational settings, rather than providing
an equal-resource performance comparison.

The greater computational cost of the 3D formulation reflects
the solution of a three-dimensional integral equation for
$\psi(p,q,x)$, with additional angular quadrature and
interpolation requirements. Its purpose is not to reduce
runtime for the present bound-state benchmarks, but to avoid
an explicit partial-wave expansion of the wave function and
the associated growth of the coupled-channel space.
This distinction concerns the representation of the unknown
amplitude; the finite partial-wave content of the interaction
is still matched across formulations in the present comparison.

\subsection{Numerical Evaluation of Spatial Observables}

The unitary transformation of the momentum-space wave functions into
coordinate space introduces significant numerical challenges due to
oscillatory integrands and finite-domain truncation errors. To transform the
3D vector-variable wave function into coordinate space, we employ two
complementary procedures. In the first, the 3D momentum-space wave function
is projected onto partial waves and transformed using the double spherical
Bessel transform, mimicking the native 2D formulation. In the second, the
vector-variable Fourier transform is evaluated directly after the analytical
angular reductions described in Sec.~\ref{sec:observables}, without expanding
the wave function in partial waves. Agreement between the two procedures
provides a sensitive internal consistency check.

The partial-wave transformation relies on integrating the momentum-space
wave functions against spherical Bessel functions $j_l(p \ms\ms r)$ and
$j_l(q \ms \rho)$. Because these functions oscillate rapidly at large spatial
distances and high momenta, the momentum grids must resolve the relevant
oscillation scales. In the calculations reported here, the sums are carried
out up to $l_{\rm max}=12$. For these angular momenta, standard upward
recurrence relations for the spherical Bessel functions remain numerically
stable.

For the direct vector-variable transformation, the intermediate
$q\to\rho$ projection contains the cylindrical Bessel function $J_0$.
Evaluating this integral directly on the original angular grid can lead to
aliasing errors in the oscillatory kernel. To suppress this error, the wave
function $\Psi(p,q,x_q)$ is first interpolated onto a dense 256-point
Gauss--Legendre mesh in $x_q$, which provides a more accurate resolution of
the Bessel oscillations.

For the final $p\to r$ projection, standard Gaussian quadrature is again
inefficient for the highly oscillatory $\cos(p \ms r \ms x_p)$ kernel. We
therefore employ a hybrid Filon quadrature scheme \cite{Filon1930,Press2007}.
The smooth momentum-space envelope is spline-interpolated, while the
trigonometric factor is integrated analytically over fine subintervals. For
$pr<0.1$, where the analytical Filon weights are susceptible to
floating-point cancellation, the algorithm automatically switches to a stabilized
high-density trapezoidal rule using a locally refined mesh. The accuracy of
the spatial transformations is assessed by comparing norms and expectation values
in momentum and coordinate space, as discussed in Sec.~\ref{sec:res_wavefunctions}.

To evaluate the kinetic-energy expectation value in coordinate space, we use
the mixed Faddeev overlap for the full kinetic energy,
\begin{equation}
\langle H_0\rangle = 3\,\langle\psi|H_0|\Psi\rangle .
\end{equation}
With the Jacobi convention used here, the coordinate-space kinetic operator is
$H_0=-\nabla_r^2-\frac{3}{4}\nabla_\rho^2$. Applying the second-derivative
Laplacian operator directly to numerically transformed coordinate-space wave
functions is prone to instability, as finite momentum cutoffs produce
oscillatory ringing near the spatial origin. We avoid this by utilizing
integration by parts to recast the kinetic-energy matrix element entirely in
terms of first spatial derivatives.

Rather than computing these gradients via finite differences on the spatial
mesh, the radial derivatives $\partial/\partial r$ and $\partial/\partial \rho$
are evaluated analytically during the momentum-space transformations. By
substituting the spherical Bessel functions $j_l(k \ms x)$ with their exact
analytical derivatives $k j'_l(k \ms x)$ within the integration kernels,
first-derivative wave functions based on the analytical kernel derivatives
are generated concurrently with the standard spatial wave functions. This
approach eliminates numerical differentiation artifacts and yields a highly
stable coordinate-space evaluation of $\langle H_0\rangle$.

Finally, the evaluation of the potential-energy expectation value in
coordinate space depends on the nature of the interaction. For local
interactions, such as the Malfliet--Tjon potential, $\langle V\rangle$ is
evaluated by integrating the coordinate-space potential $\tilde V(r)$
over the coordinate-space probability density. Conversely, for separable
interactions such as the Yamaguchi potential, the coordinate-space potential
is represented through the nonlocal separable kernel constructed from the
spatial form factors $\tilde g_l(r)$. Because the corresponding momentum-space
form factors $g_l(p)$ decay slowly at high momenta, their Fourier transforms
are highly susceptible to truncation artifacts. To suppress these artifacts,
the spatial form factors $\tilde g_l(r)$ for $l>0$ are evaluated using Filon
quadrature. This semi-analytical integration scheme explicitly accounts for
the rapid oscillations of the spherical Bessel functions, allowing us to
accurately integrate over an independent, high-density momentum grid extending
to $p=5000~\mathrm{fm}^{-1}$. For the $s$-wave, numerical integration is
bypassed entirely in favor of the exact analytical representation of
Eq.~(\ref{eq:yamaguchiform_r}).

\section{Results and Discussion}
\label{sec:results}

\subsection{Baseline Validation: Separable Interactions}
\label{sec:res_yamaguchi}

As a first benchmark, we consider three identical bosons interacting through
 separable Yamaguchi potentials. In this case, the two-body
$t$-matrix is known analytically, and the Faddeev equations reduce to
coupled one-dimensional integral equations for the spectator amplitudes
$F_l(q)$. The 1D calculation therefore provides a high-precision reference
against which the higher-dimensional formulations are tested.

\begin{figure*}[htbp]
  \includegraphics[width=0.49\textwidth, trim=0cm 0cm 0cm 0cm, clip]
  {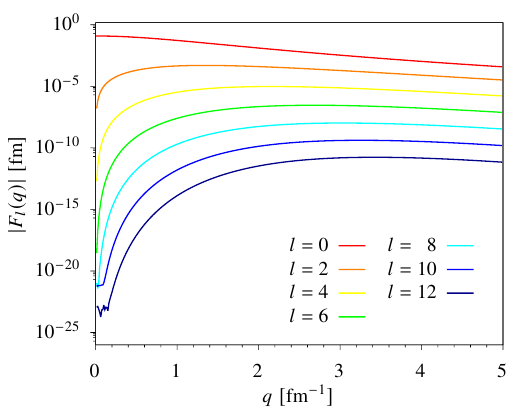}
  \hfill
  \includegraphics[width=0.49\textwidth, trim=0cm 0cm 0cm 0cm, clip]
  {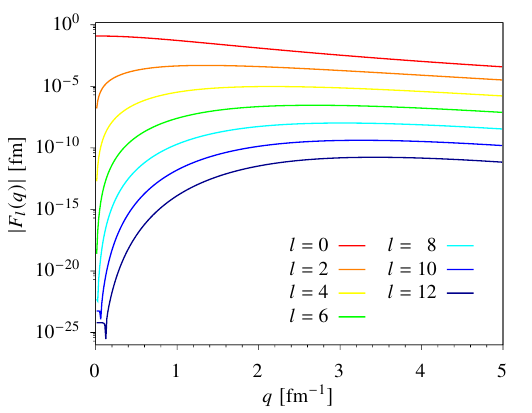}
  \caption{
  Three-boson spectator amplitudes $|F_l(q)|$ for the YAMA-23
  potential with partial waves $l=0,2,4,6,8,10,12$. The left panel
  shows the double-precision calculation, while the right panel shows the
  corresponding quadruple-precision calculation.
  }
  \label{fig:formfactor}
\end{figure*}

Figure~\ref{fig:formfactor} shows the corresponding spectator amplitudes
$|F_l(q)|$ for the YAMA-23 potential evaluated in both double and quadruple
precision. While the dominant partial waves are perfectly stable, the
double-precision calculation exhibits visible numerical noise near the origin
for the highly suppressed $l=10$ and $l=12$ channels, where the amplitudes
drop into the $10^{-20}$ to $10^{-25}$ range. As shown in the right panel,
elevating the calculation to quadruple precision
removes the visible round-off noise, restoring the smooth behavior of the
amplitudes down to the lowest momenta, while also providing a slight correction
to the \(l=8\) channel. This comparison indicates that the underlying integral
equations remain well-conditioned, and that the observed noise is a
finite-precision artifact.

\begin{table*}[htp]
\caption{
Convergence of the three-body expectation values for the YAMA-23 potential
within the 1D approach at $l_{\rm max}=12$ as a function of the momentum cutoff
$q_{\rm cut}$. The wave function has been calculated with $N_p = N_q = 768$
mesh points. }
\label{tab:3b_expect_convergence_YAMA23}
\begin{ruledtabular}
\setlength{\tabcolsep}{0pt}
\begin{tabular*}{\textwidth}{
    @{\extracolsep{\fill}}
    S[table-format=3.0]
    S[table-format=2.7, group-digits=false]
    S[table-format=-2.7, group-digits=false]
    S[table-format=-2.7, group-digits=false]
    S[table-format=-2.7, group-digits=false]
    S[table-format=1.7, group-digits=false]
}
\multicolumn{1}{c}{$q_{\rm cut}$} &
\multicolumn{1}{c}{$\langle H_0 \rangle$} &
\multicolumn{1}{c}{$\langle V \rangle$} &
\multicolumn{1}{c}{$\langle H \rangle$} &
\multicolumn{1}{c}{$E_3$} &
\multicolumn{1}{c}{$|E_{3}-\langle H\rangle|$} \\
\multicolumn{1}{c}{[fm$^{-1}$]} &
\multicolumn{1}{c}{[MeV]} &
\multicolumn{1}{c}{[MeV]} &
\multicolumn{1}{c}{[MeV]} &
\multicolumn{1}{c}{[MeV]} &
\multicolumn{1}{c}{[MeV]} \\
\hline\\[-8pt]
 40 & 67.1486391 & -91.6865494 & -24.5379103 & -24.5385691 & 0.0006588 \\
 60 & 67.1491022 & -91.6874758 & -24.5383737 & -24.5385691 & 0.0001955 \\
 80 & 67.1492150 & -91.6877015 & -24.5384865 & -24.5385691 & 0.0000826 \\
100 & 67.1492552 & -91.6877819 & -24.5385267 & -24.5385691 & 0.0000424 \\
150 & 67.1492849 & -91.6878413 & -24.5385564 & -24.5385691 & 0.0000128 \\
200 & 67.1492921 & -91.6878557 & -24.5385636 & -24.5385691 & 0.0000055 \\
300 & 67.1492958 & -91.6878631 & -24.5385673 & -24.5385691 & 0.0000018 \\
400 & 67.1492877 & -91.6878559 & -24.5385682 & -24.5385691 & 0.0000009 \\
600 & 67.1492972 & -91.6878659 & -24.5385687 & -24.5385691 & 0.0000005 \\
\end{tabular*}
\end{ruledtabular}
\end{table*}

\begin{table*}[!htbp]
\caption{
Mesh convergence of the three-body expectation values for the YAMA-23 potential
within the 1D approach at $l_{\rm max}=12$, evaluated at a fixed momentum cutoff
of $q_{\rm cut} = 400$~fm$^{-1}$. Here $N=N_p=N_q$ denotes the number of grid
points in each momentum variable. }
\label{tab:mesh_convergence_YAMA23}
\begin{ruledtabular}
\setlength{\tabcolsep}{0pt}
\begin{tabular*}{\textwidth}{
    @{\extracolsep{\fill}}
    S[table-format=3.0]
    S[table-format=2.7, group-digits=false]
    S[table-format=-2.7, group-digits=false]
    S[table-format=-2.7, group-digits=false]
    S[table-format=-2.7, group-digits=false]
    S[table-format=1.7, group-digits=false]
}
\multicolumn{1}{c}{$N$} &
\multicolumn{1}{c}{$\langle H_0 \rangle$} &
\multicolumn{1}{c}{$\langle V \rangle$} &
\multicolumn{1}{c}{$\langle H \rangle$} &
\multicolumn{1}{c}{$E_3$} &
\multicolumn{1}{c}{$|E_{3}-\langle H\rangle|$} \\
 &
\multicolumn{1}{c}{[MeV]} &
\multicolumn{1}{c}{[MeV]} &
\multicolumn{1}{c}{[MeV]} &
\multicolumn{1}{c}{[MeV]} &
\multicolumn{1}{c}{[MeV]} \\
\hline\\[-8pt]
 64 & 67.1461233 & -91.6795865 & -24.5334632 & -24.5385691 & 0.0051059 \\
 96 & 67.1486518 & -91.6862016 & -24.5375499 & -24.5385691 & 0.0010193 \\
128 & 67.1490834 & -91.6873256 & -24.5382422 & -24.5385691 & 0.0003269 \\
192 & 67.1492489 & -91.6877522 & -24.5385032 & -24.5385691 & 0.0000659 \\
256 & 67.1492753 & -91.6878231 & -24.5385478 & -24.5385691 & 0.0000214 \\
384 & 67.1492852 & -91.6878496 & -24.5385643 & -24.5385691 & 0.0000048 \\
512 & 67.1492870 & -91.6878542 & -24.5385672 & -24.5385691 & 0.0000020 \\
768 & 67.1492877 & -91.6878559 & -24.5385682 & -24.5385691 & 0.0000009 \\
\end{tabular*}
\end{ruledtabular}
\end{table*}

A notable numerical feature of the one-dimensional approach is the large
difference between the effort required to determine the eigenvalue energy and
that required to evaluate expectation values from the reconstructed full wave
function. As can be seen in Table~\ref{tab:form_convergence_YAMA23},
the eigenvalue energy obtained from Eq.~(\ref{eq:formfactor_1d}) converges
very rapidly. For the YAMA-23 potential at $l_{\rm max}=12$, a cutoff of
$q_{\rm cut}=30~\mathrm{fm}^{-1}$ with $N_q=48$ Gauss--Legendre points is
already sufficient to stabilize $E_3$ at the displayed precision.

The expectation values are considerably more demanding. Their evaluation
requires the reconstructed wave function $\Psi_l(p,q)$ over the
two-dimensional momentum domain. In particular, the kinetic-energy operator
$p^2+\frac34q^2$, together with the integration measure $p^2dp\,q^2dq$,
strongly enhances the high-momentum tail. Consequently, a much larger
cutoff is needed to obtain the same internal consistency in $\langle H\rangle$.
Table~\ref{tab:3b_expect_convergence_YAMA23} shows that
$\Delta E=|E_3-\langle H\rangle|$ falls below $10^{-5}$~MeV only once the
cutoff reaches about $q_{\rm cut}=200~\mathrm{fm}^{-1}$, and is reduced to
$9\times10^{-7}$~MeV at $q_{\rm cut}=400~\mathrm{fm}^{-1}$.

Having established the necessary momentum boundary, we next determine the
required grid density. Table~\ref{tab:mesh_convergence_YAMA23}
shows the convergence of the expectation values with respect to the number of
grid points $N$, evaluated at a fixed, sufficient cutoff of
$q_{\rm cut} = 400$~fm$^{-1}$. While $N=128$ already gives a useful estimate,
suppressing the internal residual $\Delta E$ below $10^{-5}$~MeV requires dense
integration grids of $N \ge 384$. This stiffness illustrates the numerical
demands imposed by the three-body geometry and foreshadows the more demanding
convergence behavior encountered for the local Malfliet--Tjon potentials.

A strict comparison between the 1D, 2D, and 3D calculations requires that
all formulations be evaluated in the same truncated partial-wave interaction
space. For this reason, the truncated partial-wave interaction is embedded
directly into the 2D and 3D kernels using Eq.~(\ref{eq:vppxxpart}). Thus,
differences between the calculated binding energies can be attributed to
discretization, interpolation, and permutation geometry, rather than to
different model spaces.

Table~\ref{tab:yama_1d_convergence} shows the convergence of the three-body
binding energies and expectation values for the YAMA-23 and YAMA-IV
potentials as a function of the included two-body angular momentum
$l_{\rm max}$, establishing the final 1D baseline. The partial-wave expansion
converges rapidly. At the displayed precision, increasing $l_{\rm max}$ from
4 to 6 changes the binding energy by at most $10^{-7}$~MeV, and the
binding energies remain unchanged between $l_{\rm max}=6$ and 8.
To ensure a strict comparison with the 2D and 3D calculations
presented in subsequent sections, the 1D momentum grid is chosen
to match the high-density 2D grids
($q_{\rm cut} = 400$~fm$^{-1}$, $p_{\rm cut} = 600$~fm$^{-1}$,
$N_p = N_q = 768$).

\begin{table*}[htp]
\caption{
\linespread{1.0}\selectfont
Expectation values $\langle H\rangle$, $\langle H_0\rangle$, $\langle V\rangle$,
and three-body binding energy $E_{3}$ for the Yamaguchi potentials within the
1D approach, shown as a function of the maximum two-body angular momentum
$l_{\rm max}$, with $q_{\rm cut} = 400$~fm$^{-1}$ ($p_{\rm cut} = 600$~fm$^{-1}$) and
$N_p = N_q = 768$ points.}
\label{tab:yama_1d_convergence}
\begin{ruledtabular}
  \setlength{\tabcolsep}{0pt}
\begin{tabular*}{\textwidth}{
    @{\extracolsep{\fill}}
    l
    S[table-format=2.0, group-digits=false]
    S[table-format=2.7, group-digits=false]
    S[table-format=-2.7, group-digits=false]
    S[table-format=-2.7, group-digits=false]
    S[table-format=-2.7, group-digits=false]
    S[table-format=1.7, group-digits=false]
}
\multicolumn{1}{c}{Potential} &
\multicolumn{1}{c}{$l_{\rm max}$} &
\multicolumn{1}{c}{$\langle H_0\rangle$} &
\multicolumn{1}{c}{$\langle V\rangle$} &
\multicolumn{1}{c}{$\langle H\rangle$} &
\multicolumn{1}{c}{$E_{3}$} &
\multicolumn{1}{c}{$|E_{3}-\langle H\rangle|$} \\
& &
\multicolumn{1}{c}{[MeV]} &
\multicolumn{1}{c}{[MeV]} &
\multicolumn{1}{c}{[MeV]} &
\multicolumn{1}{c}{[MeV]} &
\multicolumn{1}{c}{[MeV]} \\
\hline\\[-8pt]
YAMA-23    & 0  & 67.1444332 & -91.6723094 & -24.5278762 & -24.5278771 & 0.0000009 \\
YAMA-23    & 2  & 67.1493091 & -91.6878677 & -24.5385586 & -24.5385595 & 0.0000009 \\
YAMA-23    & 4  & 67.1492968 & -91.6878650 & -24.5385682 & -24.5385691 & 0.0000009 \\
YAMA-23    & 6  & 67.1492967 & -91.6878649 & -24.5385682 & -24.5385691 & 0.0000009 \\
YAMA-23    & 8  & 67.1492967 & -91.6878649 & -24.5385682 & -24.5385691 & 0.0000009 \\
\hline\\[-8pt]
YAMA-IV    & 0  & 33.8706511 & -42.3795985 & -8.5089475  & -8.5089478  & 0.0000003 \\
YAMA-IV    & 2  & 33.8831066 & -42.4040312 & -8.5209246  & -8.5209249  & 0.0000003 \\
YAMA-IV    & 4  & 33.8830798 & -42.4040352 & -8.5209554  & -8.5209557  & 0.0000003 \\
YAMA-IV    & 6  & 33.8830794 & -42.4040350 & -8.5209555  & -8.5209558  & 0.0000003 \\
YAMA-IV    & 8  & 33.8830794 & -42.4040350 & -8.5209555  & -8.5209558  & 0.0000003 \\
\end{tabular*}
\end{ruledtabular}
\end{table*}

\begin{table*}[htp]
\caption{
\linespread{1.0}\selectfont
Expectation values $\langle H\rangle$, $\langle H_0\rangle$, $\langle V\rangle$,
and three-body binding energy $E_{3}$ for the Yamaguchi potentials within the
2D approach, shown as a function of the maximum two-body angular momentum
$l_{\rm max}$, with $q_{\rm cut} = 400$~fm$^{-1}$ ($p_{\rm cut} = 600$~fm$^{-1}$) and
$N_p = N_q = 768$ points.
}
\label{tab:yama_2d_convergence}
\begin{ruledtabular}
\setlength{\tabcolsep}{0pt}
\begin{tabular*}{\textwidth}{
    @{\extracolsep{\fill}}
    l
    S[table-format=2.0, group-digits=false]
    S[table-format=2.7, group-digits=false]
    S[table-format=-3.7, group-digits=false]
    S[table-format=-2.7, group-digits=false]
    S[table-format=-2.7, group-digits=false]
    S[table-format=1.7, group-digits=false]
}
\multicolumn{1}{c}{Potential} &
\multicolumn{1}{c}{$l_{\rm max}$} &
\multicolumn{1}{c}{$\langle H_0\rangle$} &
\multicolumn{1}{c}{$\langle V\rangle$} &
\multicolumn{1}{c}{$\langle H\rangle$} &
\multicolumn{1}{c}{$E_{3}$} &
\multicolumn{1}{c}{$|E_{3}-\langle H\rangle|$} \\
& &
\multicolumn{1}{c}{[MeV]} &
\multicolumn{1}{c}{[MeV]} &
\multicolumn{1}{c}{[MeV]} &
\multicolumn{1}{c}{[MeV]} &
\multicolumn{1}{c}{[MeV]} \\
\hline\\[-8pt]
YAMA-23     & 0  & 67.1444152 &    -91.6722913 &    -24.5278761 &    -24.5278776 &      0.0000015 \\
YAMA-23     & 2  & 67.1492911 &    -91.6878496 &    -24.5385585 &    -24.5385600 &      0.0000015 \\
YAMA-23     & 4  & 67.1492787 &    -91.6878468 &    -24.5385681 &    -24.5385697 &      0.0000015 \\
YAMA-23     & 6  & 67.1492787 &    -91.6878468 &    -24.5385681 &    -24.5385697 &      0.0000015 \\
YAMA-23     & 8  & 67.1492787 &    -91.6878468 &    -24.5385681 &    -24.5385697 &      0.0000015 \\
\hline\\[-8pt]
YAMA-IV    & 0  & 33.8706473 &    -42.3795948 &     -8.5089474 &     -8.5089480 &      0.0000006 \\
YAMA-IV    & 2  & 33.8831029 &    -42.4040275 &     -8.5209246 &     -8.5209252 &      0.0000006 \\
YAMA-IV    & 4  & 33.8830761 &    -42.4040315 &     -8.5209554 &     -8.5209559 &      0.0000006 \\
YAMA-IV    & 6  & 33.8830757 &    -42.4040312 &     -8.5209555 &     -8.5209561 &      0.0000006 \\
YAMA-IV    & 8  & 33.8830757 &    -42.4040312 &     -8.5209555 &     -8.5209561 &      0.0000006 \\
\end{tabular*}
\end{ruledtabular}
\end{table*}

\begin{table*}[htp]
\caption{
\linespread{1.0}\selectfont
Expectation values $\langle H\rangle$, $\langle H_0\rangle$,
$\langle V\rangle$, and three-body binding energy $E_{3}$ calculated for
the Yamaguchi potentials within the three-dimensional scheme (3D),
using $q_{\rm cut} = 400$~fm$^{-1}$ ($p_{\rm cut} = 600$~fm$^{-1}$), $N_p = N_q = 512$ and
$N_x = 34$ mesh points.}
\label{tab:yama_3d_convergence}
\begin{ruledtabular}
\setlength{\tabcolsep}{0pt}
\begin{tabular*}{\textwidth}{
    @{\extracolsep{\fill}}
    l
    S[table-format=2.0, group-digits=false]
    S[table-format=2.7, group-digits=false]
    S[table-format=-3.7, group-digits=false]
    S[table-format=-2.7, group-digits=false]
    S[table-format=-2.7, group-digits=false]
    S[table-format=1.8, group-digits=false]
}
\multicolumn{1}{c}{Potential} &
\multicolumn{1}{c}{$l_{\rm max}$} &
\multicolumn{1}{c}{$\langle H_0\rangle$} &
\multicolumn{1}{c}{$\langle V\rangle$} &
\multicolumn{1}{c}{$\langle H\rangle$} &
\multicolumn{1}{c}{$E_{3}$} &
\multicolumn{1}{c}{$|E_{3}-\langle H\rangle|$} \\
& &
\multicolumn{1}{c}{[MeV]} &
\multicolumn{1}{c}{[MeV]} &
\multicolumn{1}{c}{[MeV]} &
\multicolumn{1}{c}{[MeV]} &
\multicolumn{1}{c}{[MeV]} \\
\hline\\[-8pt]
YAMA-23 &  0 & 67.1443111 & -91.6721857 & -24.5278746 & -24.5278788 & 0.0000042 \\
YAMA-23 &  2 & 67.1491870 & -91.6877440 & -24.5385570 & -24.5385612 & 0.0000042 \\
YAMA-23 &  4 & 67.1491714 & -91.6877380 & -24.5385666 & -24.5385697 & 0.0000031 \\
YAMA-23 &  6 & 67.1491712 & -91.6877378 & -24.5385667 & -24.5385697 & 0.0000030 \\
YAMA-23 &  8 & 67.1491712 & -91.6877378 & -24.5385667 & -24.5385697 & 0.0000030 \\
\hline\\[-8pt]
YAMA-IV &  0 & 33.8706252 & -42.3795718 & -8.5089467 & -8.5089492 & 0.0000025 \\
YAMA-IV &  2 & 33.8830807 & -42.4040046 & -8.5209238 & -8.5209264 & 0.0000025 \\
YAMA-IV &  4 & 33.8830459 & -42.4040005 & -8.5209546 & -8.5209552 & 0.0000006 \\
YAMA-IV &  6 & 33.8830448 & -42.4039995 & -8.5209547 & -8.5209552 & 0.0000004 \\
YAMA-IV &  8 & 33.8830447 & -42.4039995 & -8.5209547 & -8.5209552 & 0.0000004 \\
\end{tabular*}
\end{ruledtabular}
\end{table*}

\begin{table*}[htp]
\caption{
\linespread{1.0}\selectfont
Comparison of partial-wave contributions $P_l$ (in \%) to the total wave
function norm for the YAMA-23 potential and
$q_{\rm cut} = 400$~fm$^{-1}$ ($p_{\rm cut} = 600$~fm$^{-1}$). The values are shown to
highlight the agreement among the 1D, 2D, and 3D numerical schemes.
The wave functions are normalized according to $3\langle\psi | \Psi\rangle = 1$.
\label{tab:yama_norm_8digit}}
\begin{ruledtabular}
\setlength{\tabcolsep}{0pt}
\begin{tabular*}{\textwidth}{
    @{\extracolsep{\fill}}
    l
  S[table-format=2.8, group-digits=false]
  S[table-format=2.8, group-digits=false]
  S[table-format=2.8, group-digits=false]}
\multicolumn{1}{c}{$l$} &
\multicolumn{1}{c}{1D [\%]} &
\multicolumn{1}{c}{2D [\%]} &
\multicolumn{1}{c}{3D [\%]} \\
\hline\\[-8pt]
 0 & 99.15395104 & 99.15395100 & 99.15395139 \\
 2 &  0.78070564 &  0.78070530 &  0.78070495 \\
 4 &  0.05445832 &  0.05445770 &  0.05445755 \\
 6 &  0.00808918 &  0.00808838 &  0.00808827 \\
 8 &  0.00185261 &  0.00185168 &  0.00185160 \\
10 &  0.00056014 &  0.00055912 &  0.00055906 \\
12 &  0.00020520 &  0.00020411 &  0.00020407 \\
\hline\\[-8pt]
$\langle\Psi | \Psi\rangle$ [\%] & 99.99982214 & 99.99981729 & 99.99981689 \\
\end{tabular*}
\end{ruledtabular}
\end{table*}

The saturated 1D results can now be compared directly with the independently
implemented 2D partial-wave formulation in Table~\ref{tab:yama_2d_convergence}
and the 3D vector-variable formulation in Table~\ref{tab:yama_3d_convergence}.
For YAMA-23, the corresponding binding energies are

\begin{align}
E_3^{\rm 1D} &= -24.5385691~{\rm MeV}, \nonumber \\
E_3^{\rm 2D} &= -24.5385697~{\rm MeV}, \nonumber \\
E_3^{\rm 3D} &= -24.5385697~{\rm MeV},
\end{align}
showing agreement at the $10^{-6}$~MeV level. The individual expectation
values $\langle H_0\rangle$ and $\langle V\rangle$ are more sensitive to
high-momentum tails, interpolation, and multidimensional quadrature, and
therefore agree at the $10^{-5}$--$10^{-4}$~MeV level. Their sum
$\langle H\rangle$, however, remains consistent with the eigenvalue energy
at the few-$10^{-6}$~MeV level. Consistent with the 1D baseline, the 2D
calculations are performed with $q_{\rm cut}=400~\mathrm{fm}^{-1}$
($p_{\rm cut}=600~\mathrm{fm}^{-1}$) using $N_p=N_q=768$ mesh points.
 The 3D calculations employ a slightly
reduced momentum grid, $N_p=N_q=512$, together with $N_x=34$, due to the
added azimuthal dimensionality. A similar precision is obtained for YAMA-IV.

The residual $|E_3-\langle H\rangle|$ provides an independent internal
check of the reconstructed wave functions. For the final Yamaguchi
calculations, the residuals are at the $10^{-6}$~MeV level in all three
formulations, with the 3D calculation reaching $3.0\times10^{-6}$~MeV for
YAMA-23 and $4\times10^{-7}$~MeV for YAMA-IV at the largest displayed
partial-wave truncation.

The angular content of the calculated wave functions is further tested by
decomposing the total norm into individual partial-wave contributions $P_l$.
As discussed in Sec.~\ref{sec:1dapproach}, while the Faddeev components
are strictly bounded by the interaction truncation $l_{\rm max}$, the full
wave function $\Psi$ contains an infinite sequence of partial waves generated
by the permutation operator. To quantify this extended angular content, we
project the full wave function $\Psi$ onto partial waves up to $l=12$.
Table~\ref{tab:yama_norm_8digit} presents a detailed comparison of these
contributions for the YAMA-23 potential across all three numerical schemes,
with the underlying interaction itself truncated at $l_{\rm max}=12$.

The results reveal excellent consistency. The dominant $l=0$ component,
which accounts for about $99.154\%$ of the total norm, agrees across the
three formulations to better than $4\times10^{-7}$ percentage points.
Comparable absolute consistency persists for the higher partial waves,
despite their rapidly decreasing magnitude. This shows that the 3D
vector-variable approach accurately reproduces the angular content of the
state when projected onto partial waves, even though no partial-wave basis
is used in solving the 3D equation.

The final row of Table~\ref{tab:yama_norm_8digit} shows that the partial-wave
sum through $l=12$ captures about $99.9998\%$ of the total norm. The
remaining deviation from $100\%$ reflects the neglected $l>12$ tail of the
full wave function generated by the permutation operator, while the small
differences among the 1D, 2D, and 3D columns reflect the different
interpolation and projection procedures used in the three formulations.
Thus the comparison demonstrates both the saturation of
the partial-wave expansion and the consistency of the 3D vector-variable
projection onto partial waves.

\begin{figure*}[!htpb]

\par\addvspace{\intextsep}
\noindent
\begin{minipage}{\textwidth}
  \vspace{-15mm}
  \includegraphics[width=0.48\textwidth]{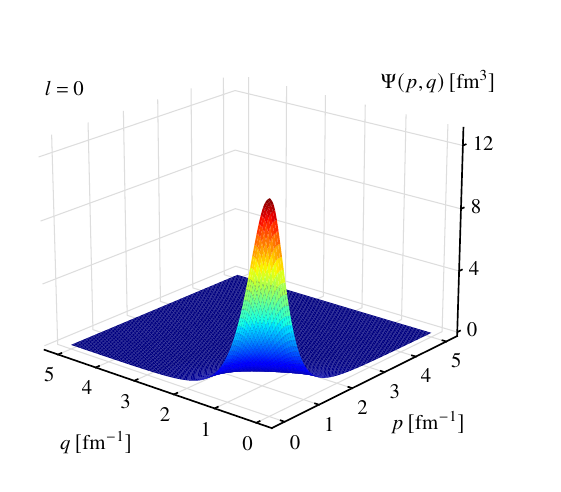}
  \hfill
  \includegraphics[width=0.48\textwidth]{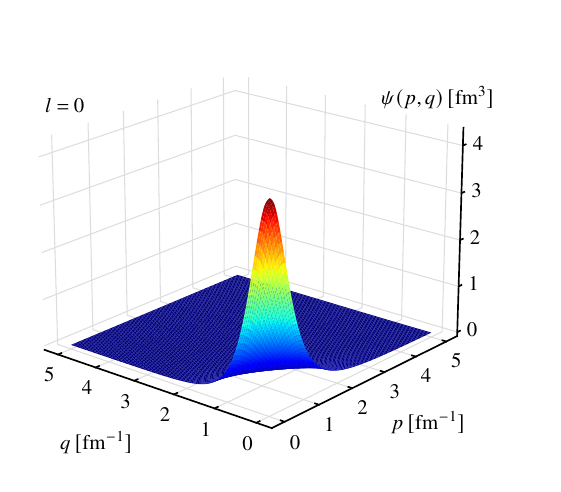}
\end{minipage}
\vspace{-4mm}

\noindent
\begin{minipage}{\textwidth}
  \includegraphics[width=0.48\textwidth]{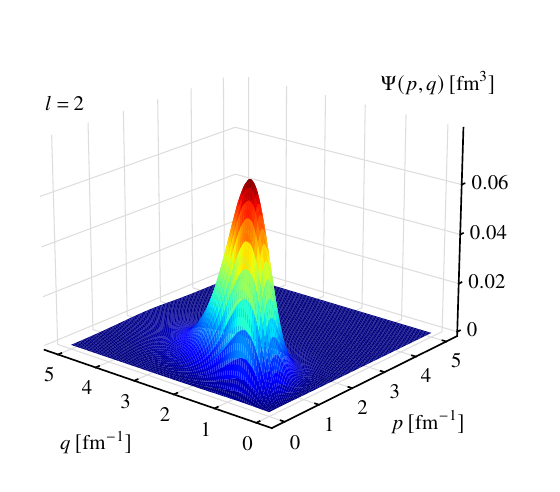}
  \hfill
  \includegraphics[width=0.48\textwidth]{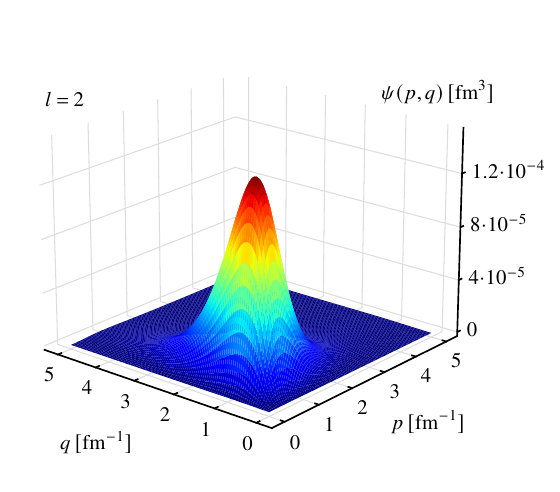}
\end{minipage}
\vspace{-4mm}

\noindent
\begin{minipage}{\textwidth}
  \includegraphics[width=0.48\textwidth]{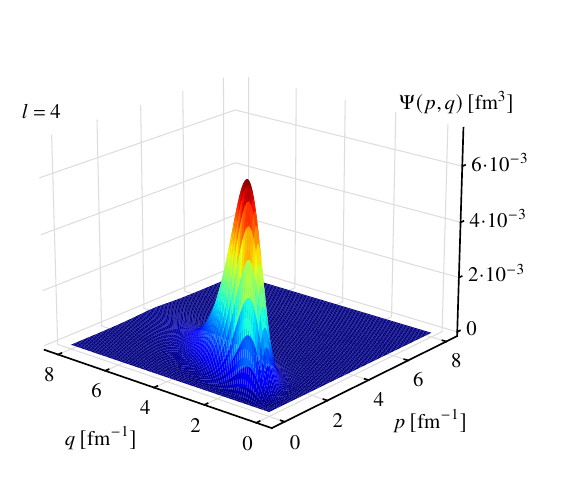}
  \hfill
  \includegraphics[width=0.48\textwidth]{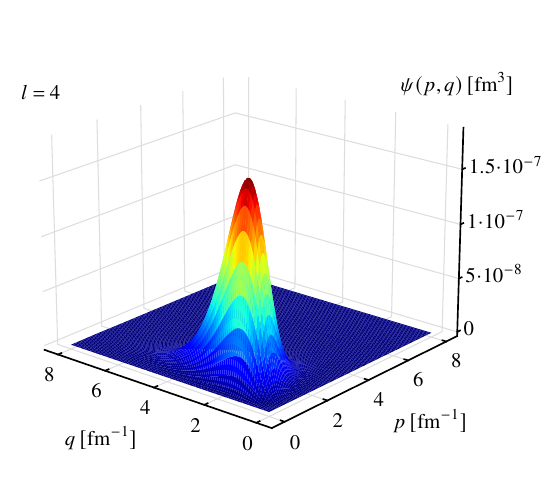}
  \captionof{figure}{\label{fig:tritwave_yama_low_l} Three-boson wave
  functions $\Psi_l(p,q)$ (left)
  and the Faddeev components $\psi_l(p,q)$ (right) with the YAMA-23 potential
  and the angular momentum $l = 0, 2, 4$ in a 2D approach.}
\end{minipage}
\end{figure*}

\begin{figure*}[!htpb]

\par\addvspace{\intextsep}
\noindent
\begin{minipage}{\textwidth}
  \vspace{-15mm}
  \includegraphics[width=0.48\textwidth]{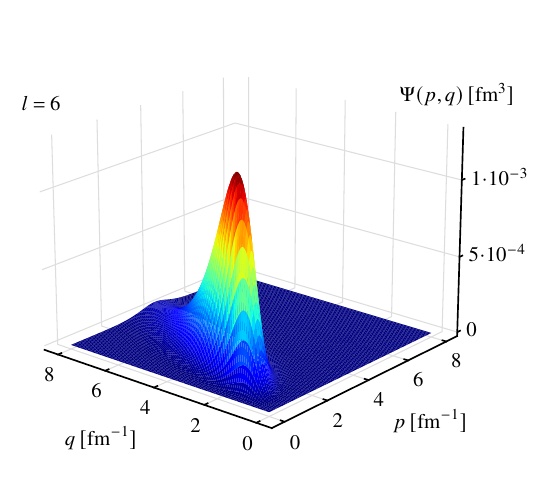}
  \hfill
  \includegraphics[width=0.48\textwidth]{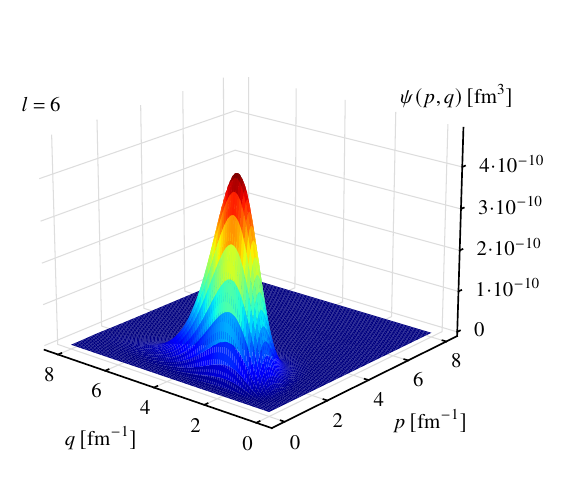}
\end{minipage}
\vspace{-4mm}

\noindent
\begin{minipage}{\textwidth}
  \includegraphics[width=0.48\textwidth]{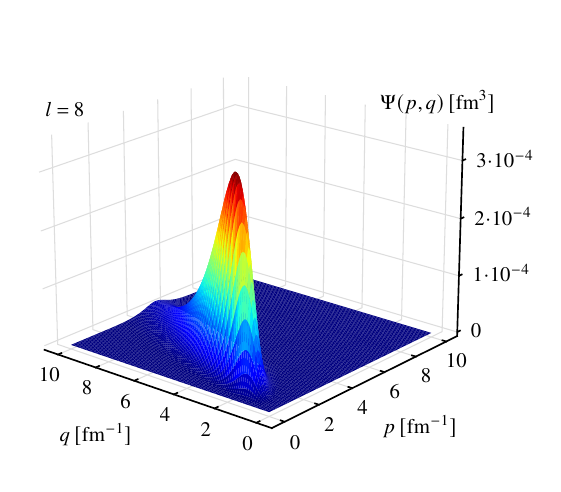}
  \hfill
  \includegraphics[width=0.48\textwidth]{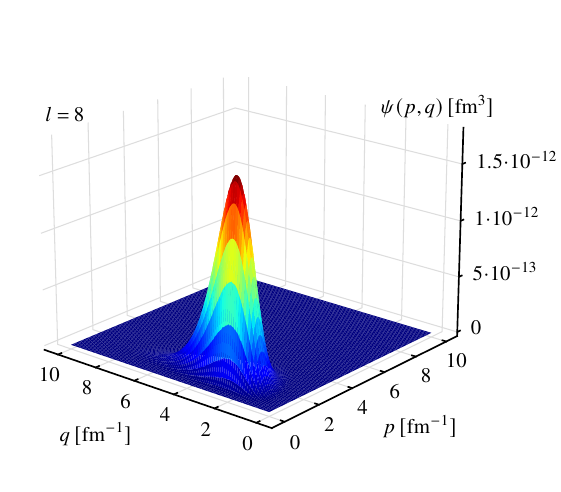}
\end{minipage}
\vspace{-4mm}

\noindent
\begin{minipage}{\textwidth}
  \includegraphics[width=0.48\textwidth]{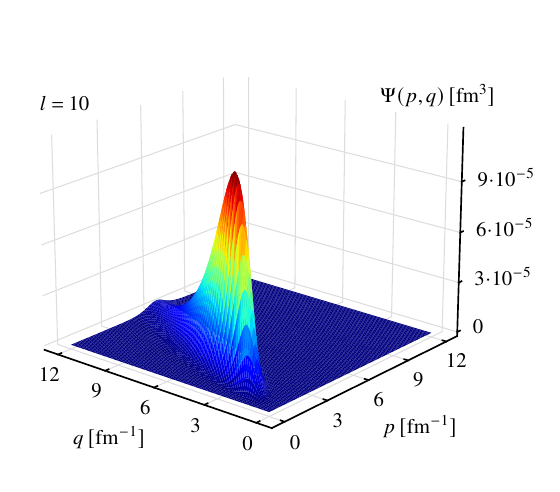}
  \hfill
  \includegraphics[width=0.48\textwidth]{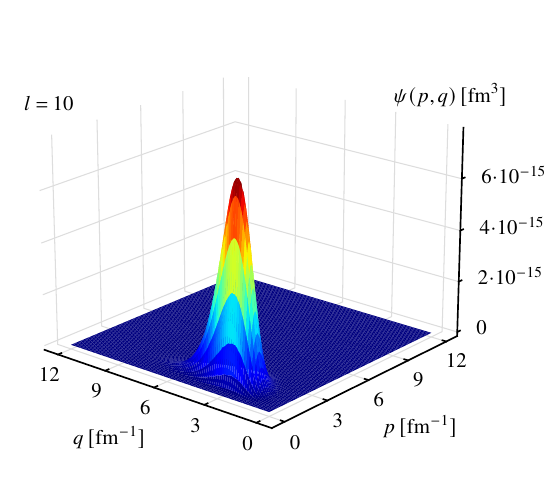}
  \captionof{figure}{\label{fig:tritwave_yama_high_l} Three-boson wave functions
  $\Psi_l(p,q)$ (left)
  and the Faddeev components $\psi_l(p,q)$ (right) with the YAMA-23 potential
   and the angular momentum $l = 6, 8, 10$ in
  a two-dimensional approach (2D).}
\end{minipage}
\end{figure*}

The smoothness of the multidimensional calculations is further illustrated by
examining the two-dimensional structures of the physical states.
Figures~\ref{fig:tritwave_yama_low_l} and \ref{fig:tritwave_yama_high_l} display
contour plots of the full three-body wave functions $\Psi_l(p,q)$ alongside
their corresponding Faddeev components $\psi_l(p,q)$ for the YAMA-23
potential, spanning partial waves from $l=0$ up to $l=10$. For the Yamaguchi
interaction, the Faddeev components reflect its separable structure, while the
full wave functions exhibit the coupled structures generated by the permutation
operator. Even in the highly suppressed $l=10$ channel, the contour lines remain
smooth and free of visible numerical artifacts, providing a visual check that
the spline interpolations and Gauss--Legendre quadratures handle the shifted
momentum arguments smoothly.

\subsection{\boldmath Two-Dimensional Formulation: $t$-Matrix vs.
Bare Potential}
\label{sec:res_2d_t_vs_v}

The two-dimensional partial-wave formulation provides a direct test of the
potential-driven equation derived in Eq.~(\ref{eq:faddeev_2d_v}). Unlike the
standard formulation, Eq.~(\ref{eq:faddeev_2d_t}), which uses the fully
off-shell two-body $t$-matrix, the $V$-driven equation acts directly with
the bare interaction and contains both the direct $G_0V|\psi\rangle$ term
and the permutation contribution $G_0VP|\psi\rangle$. Agreement between
the two formulations therefore tests not only the two-body input, but also
the implementation of the permutation geometry.

We perform this comparison for the YAMA-23 interaction in the same truncated
partial-wave space used in the benchmark calculations. The interaction
includes all even partial waves up to $l_{\rm max}=12$, and the momentum
cutoffs are fixed at $q_{\rm cut}=400~\mathrm{fm}^{-1}$ and
$p_{\rm cut}=600~\mathrm{fm}^{-1}$.

\begin{table*}[!tbp]
\caption{
Comparison of the three-body binding energies $E_3$ and internal
residuals $\Delta E=|E_3-\langle H\rangle|$ obtained with the
$t$-matrix-driven and potential-driven ($V$-driven) formulations.
The entries use matched momentum meshes within each comparison:
$N_p=N_q=512$ for 3D YAMA-23 and $N_p=N_q=384$ for the
remaining cases. The absolute difference
$\Delta E_3^{t-V}=|E_3^t-E_3^V|$ measures the agreement between
the two kernel implementations.
}
\label{tab:summary_t_vs_v}
\begin{ruledtabular}
\setlength{\tabcolsep}{0pt}
\begin{tabular*}{\textwidth}{@{\extracolsep\fill}l l c c c}
Potential & Formulation & $E_3$ [MeV] & $\Delta E$ [MeV] & $\Delta E_3^{t-V}$ [MeV] \\
\hline\\[-8pt]
YAMA-23 (2D, $l_{\rm max}=12$)  & $t$-driven & -24.5385690 & 0.0000054 & \multirow{2}{*}{$0.0000010$} \\
                        & $V$-driven & -24.5385680 & 0.0000044 & \\
\hline\\[-8pt]
YAMA-23 (3D, $l_{\rm max}=12$)  & $t$-driven & -24.5385697 & 0.0000030 & \multirow{2}{*}{$0.0000015$} \\
                        & $V$-driven & -24.5385682 & 0.0000014 & \\
\hline\\[-8pt]
MT-V (2D, $l_{\rm max}=12$)  & $t$-driven & -7.7365935  & 0.0000111 & \multirow{2}{*}{$0.0000035$} \\
                        & $V$-driven & -7.7365900  & 0.0000075 & \\
\hline\\[-8pt]
MT-V (3D, $l_{\rm max}=12$) & $t$-driven & -7.7365957  & 0.0000087 & \multirow{2}{*}{$0.0000057$} \\
                        & $V$-driven & -7.7365900  & 0.0000029 & \\
\hline\\[-8pt]
MT-V (3D, $l_{\rm max}=\infty$)  & $t$-driven & -7.7366067  & 0.0000082 & \multirow{2}{*}{$0.0000056$} \\
                        & $V$-driven & -7.7366011  & 0.0000026 & \\
\end{tabular*}
\end{ruledtabular}
\end{table*}

Table~\ref{tab:summary_t_vs_v} compares the binding energies and
internal residuals obtained with the two formulations on matched
momentum meshes. For the 2D YAMA-23 calculation at
$N_p=N_q=384$, the $t$-driven equation gives
$E_3=-24.5385690~\mathrm{MeV}$, while the $V$-driven equation
gives $E_3=-24.5385680~\mathrm{MeV}$. Their difference,
$1.0\times10^{-6}~\mathrm{MeV}$, is smaller than the
corresponding internal residuals of
$5.4\times10^{-6}~\mathrm{MeV}$ and
$4.4\times10^{-6}~\mathrm{MeV}$.
The mesh dependence is documented in
Table~\ref{tab:mesh_yama_t_vs_v_2d} in Appendix~A.
This agreement supports the consistency of the two numerical
kernels for the present benchmark, while the $V$-driven
formulation avoids the explicit construction of the fully
off-shell two-body $t$-matrix.

\subsection{\boldmath Three-Dimensional Vector Formulation: $t$-Matrix vs.
Bare Potential}
\label{sec:res_3d_comparison}

The three-dimensional (3D) vector-variable formulation provides the most
direct test of the multidimensional permutation geometry. In this approach,
the wave function is represented as a function of the Jacobi momentum
vectors, without expanding it in partial waves. The shifted permutation
arguments must therefore be handled directly in the continuous angular
variables, requiring multidimensional quadratures and interpolations of the
shifted wave-function arguments and, in the $t$-driven case, of the off-shell
two-body $t$-matrix.

As in the 2D case, we compare the standard $t$-matrix-driven equation,
Eq.~(\ref{eq:faddeev3d}), with the potential-driven equation,
Eq.~(\ref{eq:faddeev3d_v}). For a strict comparison with the 1D and 2D
benchmarks, the same truncated YAMA-23 interaction space is embedded into
the 3D kernel, including even partial waves up to $l_{\rm max}=12$. The
momentum cutoffs are fixed at $q_{\rm cut}=400~\mathrm{fm}^{-1}$ and
$p_{\rm cut}=600~\mathrm{fm}^{-1}$.

Table~\ref{tab:summary_t_vs_v} also presents the comparison of
the two 3D kernels at $N_p=N_q=512$. The $t$-driven calculation
gives $E_3=-24.5385697~\mathrm{MeV}$, while the $V$-driven
calculation gives $E_3=-24.5385682~\mathrm{MeV}$.
The difference, $1.5\times10^{-6}~\mathrm{MeV}$, is comparable
to the corresponding internal residuals of
$3.0\times10^{-6}~\mathrm{MeV}$ and
$1.4\times10^{-6}~\mathrm{MeV}$.
The mesh dependence of this comparison is documented in
Table~\ref{tab:mesh_yama_t_vs_v_3d} in Appendix~A.

This agreement shows that the $V$-driven 3D equation reproduces the same
permutation geometry as the standard $t$-driven formulation within the quoted
numerical accuracy. It also indicates that the vector-variable interpolation,
angular quadrature, and shifted-momentum mappings introduce no detectable
systematic discrepancy at this level.

\subsection{Local Interactions: 2D Partial-Wave Convergence}
\label{sec:res_mt_2d}

Having established the numerical baseline with soft separable interactions,
we next consider local Malfliet--Tjon (MT) potentials in the 2D partial-wave
formulation. Compared with the Yamaguchi interactions, these local potentials
generate stronger short-range and high-momentum structures in the wave
function. Their numerical treatment therefore requires careful control of the
momentum cutoff, dense Gauss--Legendre grids, and higher partial-wave
truncations in order to obtain stable expectation values.

Tables~\ref{tab:cutoff_convergence_MT5_N512} and
\ref{tab:mesh_convergence_MTV_q200} show the cutoff and mesh
convergence for the MT-V potential at $l_{\rm max}=12$. The internal residual
$\Delta E=|E_3-\langle H\rangle|$ is reduced to approximately
$10^{-5}~\mathrm{MeV}$ for $q_{\rm cut}=200~\mathrm{fm}^{-1}$ and
$N=512$. Increasing the cutoff to $q_{\rm cut}=300~\mathrm{fm}^{-1}$ lowers
the residual only slightly, to $9.4\times10^{-6}~\mathrm{MeV}$, indicating
that the remaining discrepancy is dominated mainly by the difficulty of the
expectation-value integration rather than by the eigenvalue determination.

Table~\ref{tab:mt_2d_convergence} presents the 2D binding
energies and expectation values for the MT-IV and MT-V
potentials as functions of the maximum two-body angular
momentum $l_{\rm max}$. Compared with the separable Yamaguchi
interactions, the local MT potentials exhibit slower
partial-wave convergence. Increasing $l_{\rm max}$ from
10 to 12 changes the binding energy by
$2.4\times10^{-5}$~MeV for MT-IV and $1.99\times10^{-5}$~MeV for MT-V.
Higher partial waves therefore remain relevant to the last
displayed digits.

\begin{table*}[htp]
\caption{
\linespread{1.0}\selectfont
Momentum cutoff convergence of the three-body expectation values for the MT-V
potential in the 2D approach at $l_{\rm max}=12$, evaluated on a fixed dense
grid of $N_p=N_q=512$ points.}
\label{tab:cutoff_convergence_MT5_N512}
\begin{ruledtabular}
\setlength{\tabcolsep}{0pt}
\begin{tabular*}{\textwidth}{
    @{\extracolsep{\fill}}
    S[table-format=3.0]
    S[table-format=2.8, group-digits=false]
    S[table-format=-2.8, group-digits=false]
    S[table-format=-2.8, group-digits=false]
    S[table-format=-2.8, group-digits=false]
    S[table-format=1.8, group-digits=false]
}
\multicolumn{1}{c}{$q_{\rm cut}$} &
\multicolumn{1}{c}{$\langle H_0 \rangle$} &
\multicolumn{1}{c}{$\langle V \rangle$} &
\multicolumn{1}{c}{$\langle H \rangle$} &
\multicolumn{1}{c}{$E_3$} &
\multicolumn{1}{c}{$|E_{3}-\langle H\rangle|$} \\
\multicolumn{1}{c}{[fm$^{-1}$]} &
\multicolumn{1}{c}{[MeV]} &
\multicolumn{1}{c}{[MeV]} &
\multicolumn{1}{c}{[MeV]} &
\multicolumn{1}{c}{[MeV]} &
\multicolumn{1}{c}{[MeV]} \\
\hline\\[-8pt]
 60 & 29.7775714 & -37.5141177 & -7.7365463 & -7.7365935 & 0.0000471 \\
 80 & 29.7775909 & -37.5141596 & -7.7365687 & -7.7365935 & 0.0000247 \\
100 & 29.7775987 & -37.5141752 & -7.7365765 & -7.7365935 & 0.0000170 \\
150 & 29.7776029 & -37.5141850 & -7.7365821 & -7.7365935 & 0.0000114 \\
200 & 29.7776046 & -37.5141880 & -7.7365834 & -7.7365935 & 0.0000100 \\
300 & 29.7776098 & -37.5141938 & -7.7365841 & -7.7365935 & 0.0000094 \\
\end{tabular*}
\end{ruledtabular}
\end{table*}

\begin{table*}[htp]
\caption{
\linespread{1.0}\selectfont
Mesh convergence of the three-body expectation values for the MT-V potential
in the 2D approach with $l_{\rm max}=12$, evaluated at a fixed momentum cutoff of
$q_{\rm cut} = 200$~fm$^{-1}$. Here $N=N_p=N_q$ denotes the number of grid points
in each momentum variable.}
\label{tab:mesh_convergence_MTV_q200}
\begin{ruledtabular}
\setlength{\tabcolsep}{0pt}
\begin{tabular*}{\textwidth}{
  @{\extracolsep{\fill}}
  S[table-format=3.0]
  S[table-format=2.7, group-digits=false]
  S[table-format=-2.7, group-digits=false]
  S[table-format=-2.7, group-digits=false]
  S[table-format=-2.7, group-digits=false]
  S[table-format=1.7, group-digits=false]
}
\multicolumn{1}{c}{$N$} &
\multicolumn{1}{c}{$\langle H_0 \rangle$} &
\multicolumn{1}{c}{$\langle V \rangle$} &
\multicolumn{1}{c}{$\langle H \rangle$} &
\multicolumn{1}{c}{$E_3$} &
\multicolumn{1}{c}{$|E_{3}-\langle H\rangle|$} \\
 &
\multicolumn{1}{c}{[MeV]} &
\multicolumn{1}{c}{[MeV]} &
\multicolumn{1}{c}{[MeV]} &
\multicolumn{1}{c}{[MeV]} &
\multicolumn{1}{c}{[MeV]} \\
\hline\\[-8pt]
 64 & 29.7763543 & -37.5110421 & -7.7346878 & -7.7366295 & 0.0019417 \\
 96 & 29.7773495 & -37.5135787 & -7.7362292 & -7.7366025 & 0.0003733 \\
128 & 29.7775183 & -37.5139908 & -7.7364725 & -7.7365960 & 0.0001235 \\
192 & 29.7775796 & -37.5141411 & -7.7365615 & -7.7365942 & 0.0000327 \\
384 & 29.7776037 & -37.5141861 & -7.7365824 & -7.7365935 & 0.0000111 \\
512 & 29.7776046 & -37.5141880 & -7.7365834 & -7.7365935 & 0.0000100 \\
\end{tabular*}
\end{ruledtabular}
\end{table*}

\begin{table*}[htp]
\caption{
\linespread{1.0}\selectfont
The expectation values $\langle H\rangle$, $\langle H_0\rangle$,
$\langle V\rangle$, and three-body binding energy $E_{3}$ calculated for
the Malfliet--Tjon potentials within the two-dimensional scheme (2D).
The calculations were performed using $N_p=N_q=512$ mesh points and a
momentum cutoff
of $q_{\rm cut} = 200$~fm$^{-1}$. The final column shows the corresponding
partial-wave binding energies from Table~3 of Ref.~\cite{Elster98b}; these
earlier values were obtained from the coupled 2D partial-wave Faddeev equations.}
\label{tab:mt_2d_convergence}
\begin{ruledtabular}
\setlength{\tabcolsep}{0pt}
\begin{tabular*}{\textwidth}{
    @{\extracolsep{\fill}}
    l
    S[table-format=2.0, group-digits=false]
    S[table-format=2.7, group-digits=false]
    S[table-format=-3.7, group-digits=false]
    S[table-format=-2.7, group-digits=false]
    S[table-format=-2.7, group-digits=false]
    S[table-format=1.7, group-digits=false]
    S[table-format=-2.5, group-digits=false]
}
\multicolumn{1}{c}{Potential} &
\multicolumn{1}{c}{$l_{\rm max}$} &
\multicolumn{1}{c}{$\langle H_0\rangle$} &
\multicolumn{1}{c}{$\langle V\rangle$} &
\multicolumn{1}{c}{$\langle H\rangle$} &
\multicolumn{1}{c}{$E_{3}$} &
\multicolumn{1}{c}{$|E_{3}-\langle H\rangle|$} &
\multicolumn{1}{c}{$E_{3}$ \cite{Elster98b}} \\
& &
\multicolumn{1}{c}{[MeV]} &
\multicolumn{1}{c}{[MeV]} &
\multicolumn{1}{c}{[MeV]} &
\multicolumn{1}{c}{[MeV]} &
\multicolumn{1}{c}{[MeV]} &
\multicolumn{1}{c}{[MeV]} \\
\hline\\[-8pt]
MT-IV &  0 & 76.820039 & -101.681869 & -24.861830 & -24.861869 & 0.000038 &
  -24.8616 \\
MT-IV &  2 & 77.245094 & -102.291805 & -25.046711 & -25.046751 & 0.000040 &
  -25.0465 \\
MT-IV &  4 & 77.265189 & -102.320617 & -25.055428 & -25.055468 & 0.000040 &
  -25.0552 \\
MT-IV &  6 & 77.267671 & -102.324183 & -25.056512 & -25.056553 & 0.000041 &
  -25.0562 \\
MT-IV &  8 & 77.268189 & -102.324928 & -25.056739 & -25.056781 & 0.000042 &
  -25.0564 \\
MT-IV & 10 & 77.268338 & -102.325142 & -25.056804 & -25.056846 & 0.000042 &
  -25.0565 \\
MT-IV & 12 & 77.268391 & -102.325218 & -25.056827 & -25.056870 & 0.000043 &
  -25.0565 \\
\hline\\[-8pt]
MT-V  &  0 & 29.0115917 & -36.5513512 & -7.5397595 & -7.5397624 & 0.0000029 &
  -7.53975 \\
MT-V  &  2 & 29.7092580 & -37.4239640 & -7.7147060 & -7.7147085 & 0.0000026 &
  -7.71470 \\
MT-V  &  4 & 29.7715490 & -37.5053881 & -7.7338391 & -7.7338466 & 0.0000075 &
  -7.73383 \\
MT-V  &  6 & 29.7769602 & -37.5130889 & -7.7361287 & -7.7361372 & 0.0000085 &
  -7.73613 \\
MT-V  &  8 & 29.7775279 & -37.5140173 & -7.7364894 & -7.7364986 & 0.0000092 &
  -7.73649 \\
MT-V  & 10 & 29.7775974 & -37.5141613 & -7.7365639 & -7.7365736 & 0.0000097 &
  -7.73656 \\
MT-V  & 12 & 29.7776046 & -37.5141880 & -7.7365834 & -7.7365935 & 0.0000100 &
  -7.73658 \\
\end{tabular*}
\end{ruledtabular}
\end{table*}

For comparison, Table~\ref{tab:mt_2d_convergence} additionally includes
the earlier partial-wave binding energies reported in Table~3 of
Ref.~\cite{Elster98b}. Those results were obtained from the coupled
two-dimensional partial-wave Faddeev equations and therefore provide
a direct historical comparison with the present 2D calculation.
At $l_{\rm max}=12$, the absolute binding-energy differences are
$3.7\times10^{-4}$~MeV for MT-IV and
$1.35\times10^{-5}$~MeV for MT-V.
The cutoff and mesh dependence of the present MT-V calculation
is documented in Tables~\ref{tab:cutoff_convergence_MT5_N512}
and \ref{tab:mesh_convergence_MTV_q200}.

Despite these more demanding convergence properties, the 2D formulation
remains stable. At $l_{\rm max}=12$, the final MT-V result is
$E_3=-7.7365935~\mathrm{MeV}$ with
$\Delta E=1.0\times10^{-5}~\mathrm{MeV}$, while MT-IV gives
$E_3=-25.056870~\mathrm{MeV}$ with
$\Delta E=4.3\times10^{-5}~\mathrm{MeV}$. These values provide the local
interaction benchmark against which the 3D vector-variable calculations are
compared below.

\begin{table*}[htp]
\caption{
\linespread{1.0}\selectfont
Comparison of partial-wave norm contributions and global expectation
values evaluated in momentum space and coordinate space for the YAMA-23
and MT-V potentials in the 2D approach. $l_{\rm max} = 12$ and
$q_{\rm cut} = 400$~fm$^{-1}$ ($p_{\rm cut} = 600$~fm$^{-1}$) for YAMA-23 and
$q_{\rm cut} = 200$~fm$^{-1}$ ($p_{\rm cut} = 300$~fm$^{-1}$) for MT-V.
The wave functions are normalized according to
$3\langle\psi | \Psi\rangle = 1$. For comparison, the final column
contains the MT-V norm contributions from Table~4 and the
partial-wave energy expectation values from Table~5 of
Ref.~\cite{Elster98b}, reproduced at their original reported
precision.
The individual norm contributions $P_l$ are given in percent.
}
\label{tab:combined_observables}
\begin{ruledtabular}
\setlength{\tabcolsep}{0pt}
\begin{tabular*}{\textwidth}{
    @{\extracolsep{\fill}}
    l
    S[table-format=-2.8, group-digits=false]
    S[table-format=-2.8, group-digits=false]
    S[table-format=-2.8, group-digits=false]
    S[table-format=-2.8, group-digits=false]
    S[table-format=-2.6, group-digits=false]
}
\multicolumn{1}{c}{} &
\multicolumn{2}{c}{YAMA-23} &
\multicolumn{3}{c}{MT-V} \\
\cline{2-3} \cline{4-6} \\[-8pt]
\multicolumn{1}{c}{$l$} &
\multicolumn{1}{c}{Momentum space} &
\multicolumn{1}{c}{Coordinate space} &
\multicolumn{1}{c}{Momentum space} &
\multicolumn{1}{c}{Coordinate space} &
\multicolumn{1}{c}{Ref.~\cite{Elster98b}} \\
\hline\\[-8pt]
\quad 0  & 99.15395100 & 99.15395431 & 99.08519532 & 99.08518248 & 99.0851 \\
\quad 2  &  0.78070530 &  0.78070529 &  0.74823192 &  0.74822946 &  0.7482 \\
\quad 4  &  0.05445770 &  0.05445766 &  0.11591648 &  0.11591595 &  0.1159 \\
\quad 6  &  0.00808838 &  0.00808835 &  0.03304501 &  0.03304576 &  0.03305 \\
\quad 8  &  0.00185168 &  0.00185167 &  0.01087725 &  0.01087857 &  0.01088 \\
\quad 10 &  0.00055912 &  0.00055911 &  0.00393585 &  0.00393723 &  0.003939 \\
\quad 12 &  0.00020411 &  0.00020410 &  0.00154116 &  0.00154233 &  0.001545 \\
\hline\\[-8pt]
$\langle\Psi | \Psi\rangle$ [\%] & 99.99981729 & 99.99982049 & 99.99874300 &
  99.99873178 & 99.998614 \\
\hline\\[-8pt]
$\langle H_0 \rangle$ [MeV]      & 67.1492787 & 67.1495505 & 29.7776046 &
  29.7776111 & 29.7776 \\
$\langle V \rangle$ [MeV]        & -91.6878468 & -91.6878500 & -37.5141880 &
  -37.5141737 & -37.5139 \\
$\langle H \rangle$ [MeV]        & -24.5385681 & -24.5382995 & -7.7365834 &
  -7.7365627 & -7.73634 \\
\end{tabular*}
\end{ruledtabular}
\end{table*}

\begin{table*}[htp]
\caption{
\linespread{1.0}\selectfont
Momentum cutoff convergence of the three-body expectation values for the
MT-V potential in the 3D approach at $l_{\rm max}=12$, evaluated on a fixed
dense grid of $N_p=N_q=384$ and $N_x=50$ points.
The results at $q_{\rm cut}=60$ and $80$~fm$^{-1}$ were obtained
with ARPACK, whereas those at $q_{\rm cut}=100$, $150$, $200$,
and $300$~fm$^{-1}$ were obtained with the custom subspace solver
described in Sec.~\ref{sec:numerical_method}.
}
\label{tab:cutoff_convergence_MT53D_N384}
\begin{ruledtabular}
\setlength{\tabcolsep}{0pt}
\begin{tabular*}{\textwidth}{
  @{\extracolsep{\fill}}
  S[table-format=3.0]
  S[table-format=2.7, group-digits=false]
  S[table-format=-2.7, group-digits=false]
  S[table-format=-2.7, group-digits=false]
  S[table-format=-2.7, group-digits=false]
  S[table-format=1.7, group-digits=false]
}
\multicolumn{1}{c}{$q_{\rm cut}$} &
\multicolumn{1}{c}{$\langle H_0 \rangle$} &
\multicolumn{1}{c}{$\langle V \rangle$} &
\multicolumn{1}{c}{$\langle H \rangle$} &
\multicolumn{1}{c}{$E_3$} &
\multicolumn{1}{c}{$|E_{3}-\langle H\rangle|$} \\
\multicolumn{1}{c}{[fm$^{-1}$]} &
\multicolumn{1}{c}{[MeV]} &
\multicolumn{1}{c}{[MeV]} &
\multicolumn{1}{c}{[MeV]} &
\multicolumn{1}{c}{[MeV]} &
\multicolumn{1}{c}{[MeV]} \\
\hline\\[-8pt]
 60 & 29.7775534 & -37.5141033 & -7.7365499 & -7.7365957 & 0.0000458 \\
 80 & 29.7775735 & -37.5141458 & -7.7365723 & -7.7365957 & 0.0000234 \\
100 & 29.7775839 & -37.5141639 & -7.7365801 & -7.7365959 & 0.0000159 \\
150 & 29.7775878 & -37.5141734 & -7.7365857 & -7.7365959 & 0.0000102 \\
200 & 29.7775885 & -37.5141755 & -7.7365870 & -7.7365959 & 0.0000089 \\
300 & 29.7775885 & -37.5141762 & -7.7365877 & -7.7365959 & 0.0000082 \\
\end{tabular*}
\end{ruledtabular}
\end{table*}

\begin{table*}[htp]
\caption{
Expectation values $\langle H\rangle$, $\langle H_0\rangle$,
$\langle V\rangle$, and three-body binding energies $E_3$
for the Malfliet--Tjon potentials in the present 3D calculations.
The rows labeled $l_{\rm max}=\infty$ use the full vector-variable
interaction. Historical results from Ref.~\cite{Elster98b} are
identified explicitly as partial-wave (PW) or three-dimensional
(3D) calculations. Their eigenvalues are taken from Tables~3
and 2 of that reference, respectively, while their expectation
values and residuals are taken from Table~5.
The historical values are reproduced at their original
reported precision.}
\label{tab:mt_3d_convergence}
\begin{ruledtabular}
\setlength{\tabcolsep}{0pt}
\begin{tabular*}{\textwidth}{
    @{\extracolsep{\fill}}
    l
    S[table-format=2.0, group-digits=false]
    S[table-format=2.7, group-digits=false]
    S[table-format=-3.7, group-digits=false]
    S[table-format=-2.7, group-digits=false]
    S[table-format=-2.7, group-digits=false]
    S[table-format=1.7, group-digits=false]
}
\multicolumn{1}{c}{Potential} &
\multicolumn{1}{c}{$l_{\rm max}$} &
\multicolumn{1}{c}{$\langle H_0\rangle$} &
\multicolumn{1}{c}{$\langle V\rangle$} &
\multicolumn{1}{c}{$\langle H\rangle$} &
\multicolumn{1}{c}{$E_{3}$} &
\multicolumn{1}{c}{$|E_{3}-\langle H\rangle|$} \\
& &
\multicolumn{1}{c}{[MeV]} &
\multicolumn{1}{c}{[MeV]} &
\multicolumn{1}{c}{[MeV]} &
\multicolumn{1}{c}{[MeV]} &
\multicolumn{1}{c}{[MeV]} \\
\hline\\[-8pt]
MT-IV &  0 & 76.820184 & -101.682010 & -24.861826 & -24.861907 & 0.000080 \\
MT-IV &  2 & 77.245198 & -102.291906 & -25.046708 & -25.046784 & 0.000077 \\
MT-IV &  4 & 77.265184 & -102.320609 & -25.055425 & -25.055480 & 0.000055 \\
MT-IV &  6 & 77.267537 & -102.324046 & -25.056509 & -25.056542 & 0.000033 \\
MT-IV &  8 & 77.267926 & -102.324663 & -25.056737 & -25.056749 & 0.000013 \\
MT-IV & 10 & 77.267945 & -102.324747 & -25.056802 & -25.056796 & 0.000006 \\
MT-IV & 12 & 77.267995 & -102.324820 & -25.056825 & -25.056817 & 0.000009 \\
MT-IV & {$\infty$} & 77.267335 & -102.324189 & -25.056854 & -25.056768 & 0.000087 \\
\hline\\[-8pt]
MT-IV (PW)~\cite{Elster98b} & 12     & 77.2653 & -102.3207 & -25.0554 & -25.0565 & 0.0011 \\
MT-IV (3D)~\cite{Elster98b} & {$\infty$} & 77.2055 & -102.2550 & -25.0495 & -25.0499 & 0.0004 \\
\hline\\[-8pt]
MT-V  &  0 & 29.0115589 &  -36.5513174 &  -7.5397585 &  -7.5397651 & 0.0000067 \\
MT-V  &  2 & 29.7092519 &  -37.4239578 &  -7.7147058 &  -7.7147155 & 0.0000097 \\
MT-V  &  4 & 29.7715359 &  -37.5053761 &  -7.7338402 &  -7.7338498 & 0.0000096 \\
MT-V  &  6 & 29.7769459 &  -37.5130768 &  -7.7361308 &  -7.7361403 & 0.0000095 \\
MT-V  &  8 & 29.7775116 &  -37.5140038 &  -7.7364922 &  -7.7365013 & 0.0000091 \\
MT-V  & 10 & 29.7775791 &  -37.5141463 &  -7.7365672 &  -7.7365761 & 0.0000089 \\
MT-V  & 12 & 29.7775867 &  -37.5141737 &  -7.7365870 &  -7.7365957 & 0.0000087 \\
MT-V  & {$\infty$} & 29.7775830 &  -37.5141815 &  -7.7365985 &  -7.7366067 & 0.0000082 \\
\hline\\[-8pt]
MT-V (PW)~\cite{Elster98b} & 12 & 29.7776 & -37.5139 & -7.73634 & -7.73658 & 0.00024 \\
MT-V (3D)~\cite{Elster98b} & {$\infty$} & 29.77706 & -37.51340 & -7.73634 & -7.73650 & 0.00016 \\
\hline\\[-8pt]
\end{tabular*}
\end{ruledtabular}
\end{table*}

Table~\ref{tab:summary_t_vs_v} compares the corresponding converged
$t$-driven and $V$-driven 2D kernels for the MT-V potential. The two
eigenvalue energies differ by only $3.5\times10^{-6}~\mathrm{MeV}$ (with
detailed step-by-step convergence shown in Appendix~A,
Table~\ref{tab:mesh_mt5_t_vs_v_2d}). Thus the agreement observed for the
separable YAMA-23 benchmark also persists for a local interaction with
stronger short-range structure.

\subsection{\boldmath Three-Dimensional Vector Formulation for Local
Interactions}
\label{sec:res_mt_3d}

The local interactions are also tested in the full 3D
vector-variable formulation.
Table~\ref{tab:cutoff_convergence_MT53D_N384} shows the cutoff
convergence for the MT-V potential using the truncated
interaction space with $l_{\rm max}=12$. The convergence pattern is consistent
with the 2D calculation: increasing the cutoff from $60~\mathrm{fm}^{-1}$ to
$200~\mathrm{fm}^{-1}$ reduces the internal residual from
$4.6\times10^{-5}~\mathrm{MeV}$ to $8.9\times10^{-6}~\mathrm{MeV}$, while a
further increase to $300~\mathrm{fm}^{-1}$ changes the residual only
slightly.

The $q_{\rm cut}=200~\mathrm{fm}^{-1}$ entry in this cutoff study
was obtained with the custom subspace solver. The corresponding
$t$-driven entries at $l_{\rm max}=12$ and $N_p=N_q=384$ in
Tables~\ref{tab:mt_3d_convergence} and
\ref{tab:mesh_mt5_t_vs_v_3d} were obtained in a separate ARPACK run
with identical input grids. The eigenvalues differ by
$2.2\times10^{-7}$~MeV, while $\langle H\rangle$ agrees at the
displayed precision.

Table~\ref{tab:mt_3d_convergence} presents the corresponding
3D vector-variable results for the MT-IV and MT-V potentials.
For the truncated interaction spaces, the partial-wave
convergence pattern follows that of the 2D calculation.
Increasing $l_{\rm max}$ from 10 to 12 changes the binding
energy by $2.1\times10^{-5}$~MeV for MT-IV and
$1.96\times10^{-5}$~MeV for MT-V. These changes are comparable
to those in the 2D formulation and remain visible at the
displayed precision.

For MT-V at $l_{\rm max}=12$, the 3D calculation gives
$E_3=-7.7365957~\mathrm{MeV}$ with
$\Delta E=8.7\times10^{-6}~\mathrm{MeV}$.
Its binding energy differs from the corresponding 2D result,
$E_3=-7.7365935~\mathrm{MeV}$, by
$2.2\times10^{-6}~\mathrm{MeV}$.
For MT-IV, the 3D result is
$E_3=-25.056817~\mathrm{MeV}$, compared with
$E_3=-25.056870~\mathrm{MeV}$ in 2D, giving a larger
difference of $5.3\times10^{-5}~\mathrm{MeV}$.
The achieved inter-formulation agreement is therefore
potential-dependent.

\begin{table*}[!htbp]
\caption{
\linespread{1.0}\selectfont
Partial-wave norm contributions and energy expectation values
in momentum and coordinate space for the 3D solutions.
YAMA-23 uses the interaction truncated at $l_{\rm max}=12$;
MT-V uses the full vector-variable interaction.
The coordinate-space results are obtained by projecting the
3D wave functions onto partial waves through $l=12$ and
applying the double spherical Bessel transform.
The normalization is $3\langle\psi|\Psi\rangle=1$. The individual
norm contributions $P_l$ are given in percent.
}
\label{tab:combined_observables_3d}
\begin{ruledtabular}
\setlength{\tabcolsep}{0pt}
\begin{tabular*}{\textwidth}{
    @{\extracolsep{\fill}}
    l
    S[table-format=-3.8, group-digits=false]
    S[table-format=-3.8, group-digits=false]
    S[table-format=-3.8, group-digits=false]
    S[table-format=-3.8, group-digits=false]
}
\multicolumn{1}{c}{} &
\multicolumn{2}{c}{YAMA-23} &
\multicolumn{2}{c}{MT-V} \\
\cline{2-3} \cline{4-5} \\[-8pt]
\multicolumn{1}{c}{$l$} &
\multicolumn{1}{c}{Momentum space} &
\multicolumn{1}{c}{Coordinate space} &
\multicolumn{1}{c}{Momentum space} &
\multicolumn{1}{c}{Coordinate space} \\
\hline\\[-8pt]
\quad 0  & 99.15395139 & 99.15396651 & 99.08519745 & 99.08523217 \\
\quad 2  &  0.78070495 &  0.78070513 &  0.74823273 &  0.74823133 \\
\quad 4  &  0.05445755 &  0.05445752 &  0.11591912 &  0.11591863 \\
\quad 6  &  0.00808827 &  0.00808825 &  0.03304888 &  0.03304942 \\
\quad 8  &  0.00185160 &  0.00185158 &  0.01088165 &  0.01088270 \\
\quad 10 &  0.00055906 &  0.00055905 &  0.00394051 &  0.00394164 \\
\quad 12 &  0.00020407 &  0.00020406 &  0.00154594 &  0.00154690 \\
\hline\\[-8pt]
$\langle\Psi | \Psi\rangle$ [\%]
& 99.99981689 & 99.99983210 & 99.99876628 & 99.99880277 \\
\hline\\[-8pt]
$\langle H_0 \rangle$ [MeV]
& 67.1491712 & 67.1494527 & 29.7775830 & 29.7768789 \\
$\langle V \rangle$ [MeV]
& -91.6877378 & -91.6877642 & -37.5141815 & -37.5141942 \\
$\langle H \rangle$ [MeV]
& -24.5385667 & -24.5383116 & -7.7365985 & -7.7373154 \\
\end{tabular*}
\end{ruledtabular}
\end{table*}

The rows labeled $l_{\rm max}=\infty$ use the full
vector-variable Malfliet--Tjon interaction rather than an
embedded finite partial-wave truncation. These calculations
provide a direct comparison with the earlier full-3D results
of Ref.~\cite{Elster98b}. The absolute binding-energy
differences are approximately $1.1\times10^{-4}$~MeV
for MT-V and $6.9\times10^{-3}$~MeV for MT-IV.
The origin of the larger MT-IV discrepancy has not been
isolated in the present comparison.

Finally, Table~\ref{tab:summary_t_vs_v} demonstrates that the excellent
agreement between the $t$-driven and $V$-driven kernels holds for the MT-V
potential in the full 3D continuous space. The available $V$-driven
calculations agree with the corresponding $t$-driven results within the
displayed numerical accuracy (see Tables~\ref{tab:mesh_mt5_t_vs_v_3d} and
\ref{tab:mesh_mt5nopartial_t_vs_v_3d} in Appendix~A for full convergence
details). This shows that the potential-driven equation preserves the same
permutation geometry as the standard $t$-driven formulation also in the full
vector-variable setting, within the displayed numerical accuracy.

\subsection{Wave Function Analysis and Coordinate-Space Transformation}
\label{sec:res_wavefunctions}

To investigate the spatial structure of the bound state, we use the
Fourier-transform framework developed in Sec.~\ref{sec:observables} to map
the converged momentum-space wave functions into coordinate space. This
provides an independent check of the wave-function normalization, the
partial-wave content, and the expectation values evaluated in the two
representations.

Table~\ref{tab:combined_observables} compares the partial-wave norm
contributions obtained in momentum space with those extracted after the
coordinate-space transformation for the YAMA-23 and MT-V potentials in the
2D formulation. The agreement of the individual partial-wave contributions
shows that the Fourier--Bessel transformation preserves the angular
decomposition of the wave function at high numerical accuracy. For the MT-V
potential, we also include previous partial-wave norm calculations from
Ref.~\cite{Elster98b}. The agreement is consistent with the precision of the
earlier results, while the denser grids used here provide additional
stabilization of the norm and expectation-value diagnostics. The summed norms
remain close to the momentum-space normalization, with the small deviations
reflecting the combined effects of finite momentum cutoffs, finite
coordinate-space grids, and oscillatory quadrature.

\begin{figure*}[!tpb]

\par\addvspace{\intextsep}
\noindent
\begin{minipage}{\textwidth}
  \vspace{-15mm}
  \includegraphics[width=0.48\textwidth]{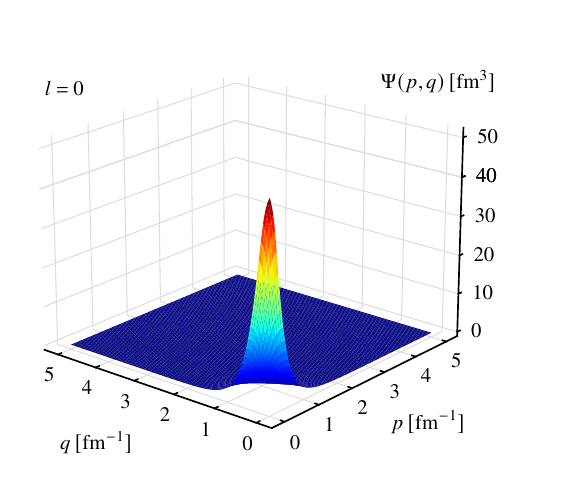}
  \hfill
  \includegraphics[width=0.48\textwidth]{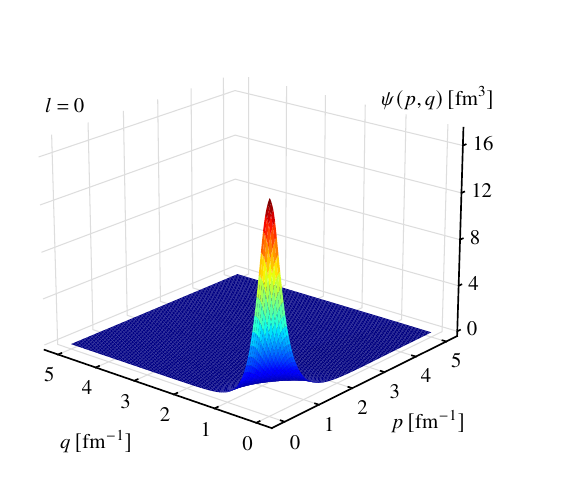}
\end{minipage}
\vspace{-4mm}

\begin{minipage}{\textwidth}
  \includegraphics[width=0.48\textwidth]{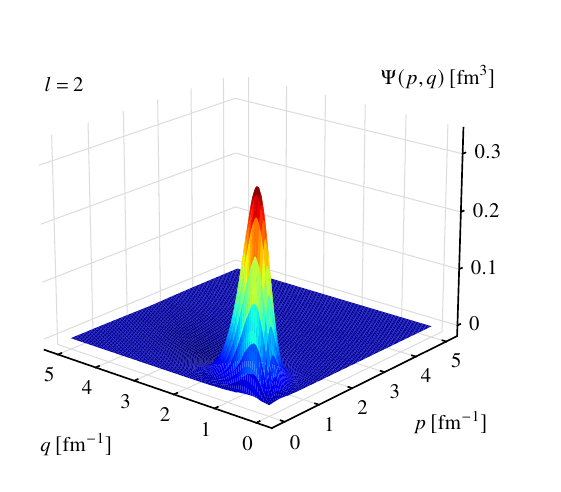}
  \hfill
  \includegraphics[width=0.48\textwidth]{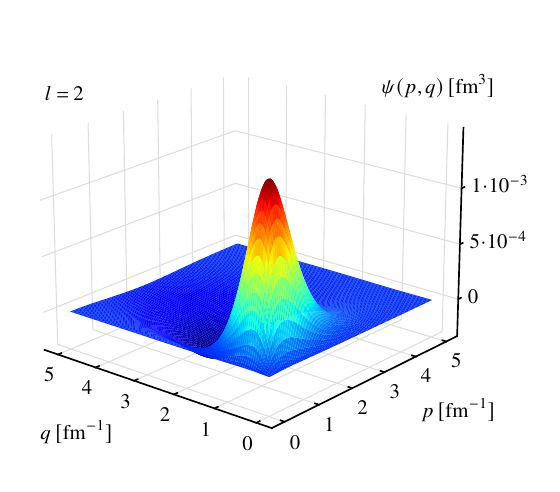}
\end{minipage}
\vspace{-4mm}

\begin{minipage}{\textwidth}
  \includegraphics[width=0.48\textwidth]{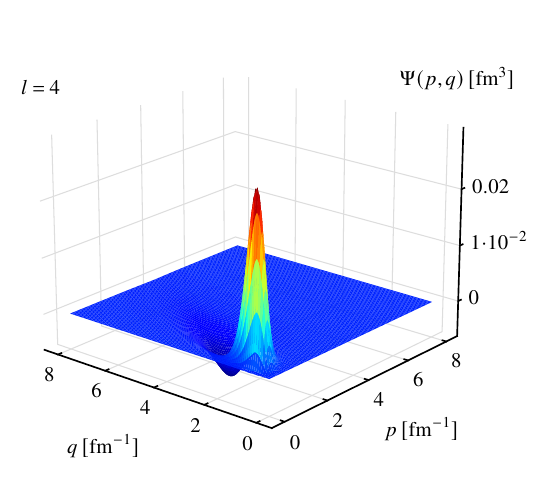}
  \hfill
  \includegraphics[width=0.48\textwidth]{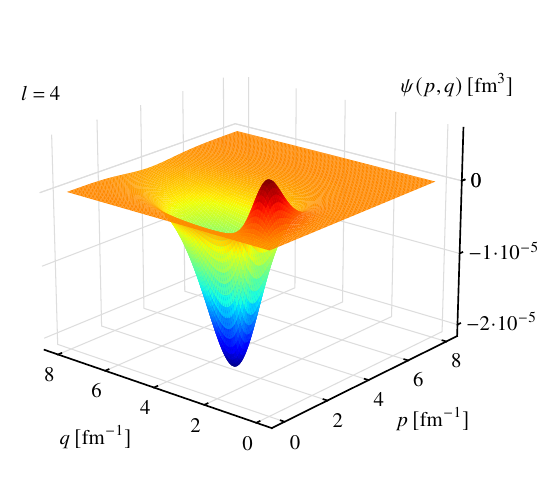}
  \captionof{figure}{\label{fig:tritwave_mt_low_l} Three-boson wave functions
  $\Psi_l(p,q)$ (left)
  and the Faddeev components $\psi_l(p,q)$ (right) with the MT-V potential
  and the angular momentum $l = 0, 2, 4$ in
  a two-dimensional approach (2D).}
\end{minipage}
\end{figure*}

\begin{figure*}[!tpb]
\noindent
\begin{minipage}{\textwidth}
  \vspace{-15mm}
  \includegraphics[width=0.48\textwidth]{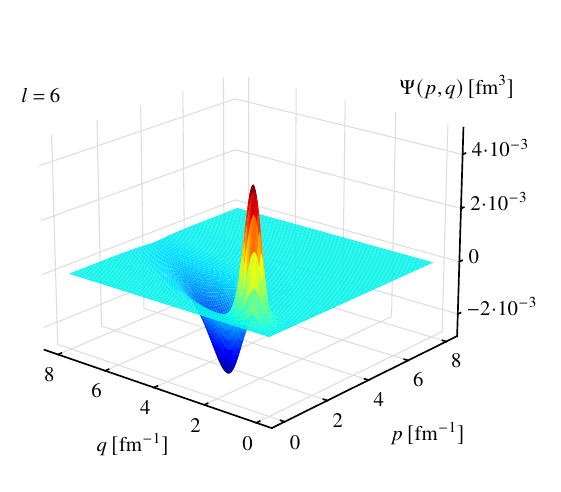}
  \hfill
  \includegraphics[width=0.48\textwidth]{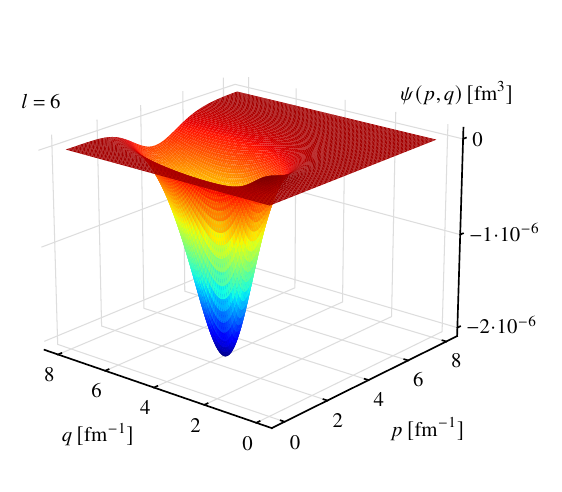}
\end{minipage}
\vspace{-4mm}

\noindent
\begin{minipage}{\textwidth}
  \includegraphics[width=0.48\textwidth]{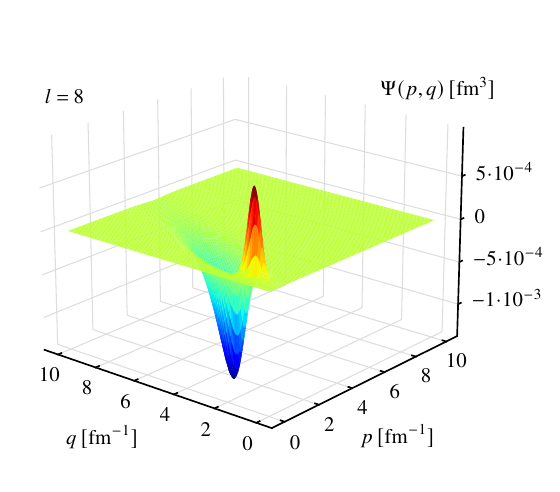}
  \hfill
  \includegraphics[width=0.48\textwidth]{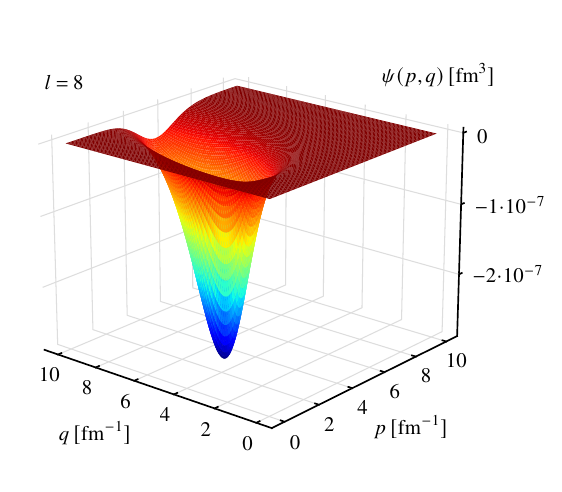}
\end{minipage}
\vspace{-4mm}

\noindent
\begin{minipage}{\textwidth}
  \includegraphics[width=0.48\textwidth]{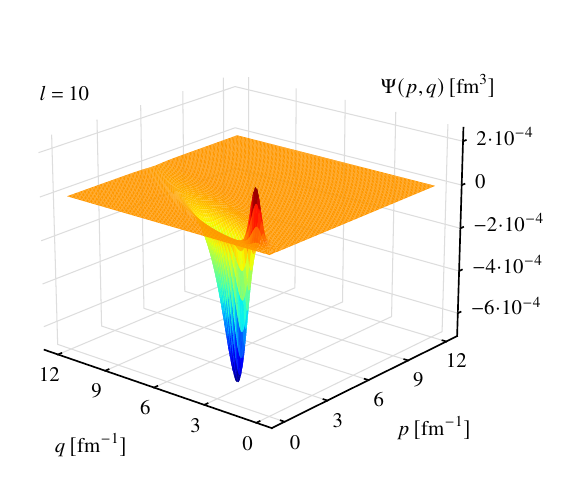}
  \hfill
  \includegraphics[width=0.48\textwidth]{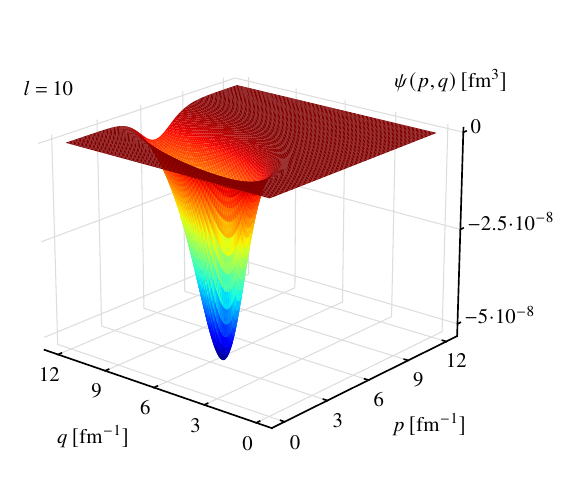}
  \captionof{figure}{\label{fig:tritwave_mt_high_l} Three-boson wave functions
   $\Psi_l(p,q)$ (left)
  and the Faddeev components $\psi_l(p,q)$ (right) with the MT-V potential
  and the angular momentum $l = 6, 8, 10$ in
  a two-dimensional approach (2D).}
\end{minipage}
\end{figure*}

\begin{figure*}[!thp]
\noindent
\begin{minipage}{\textwidth}
  \vspace{-4mm}
  \includegraphics[width=0.48\textwidth]{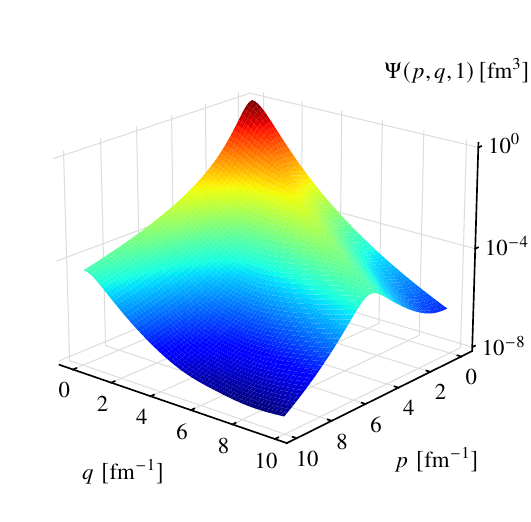}
  \hfill
  \includegraphics[width=0.48\textwidth]{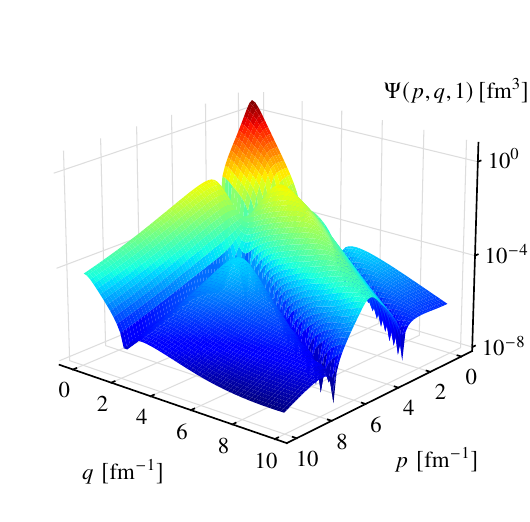}
  \vspace{-2mm}
\captionof{figure}{\label{fig:tritwave_yamamt_3dl}
Magnitudes of the full three-boson wave functions
$|\Psi(p,q,1)|$ for the YAMA-23 (left) and MT-V (right)
potentials, obtained in the 3D vector-variable formulation.
The angular variable is fixed at $x=1$, and the vertical
axis is logarithmic.}
  \vspace{2mm}
\end{minipage}
\end{figure*}

Table~\ref{tab:combined_observables_3d} presents the corresponding
comparison for the 3D momentum-space solutions. Here the
coordinate-space results are obtained by first projecting the
3D wave functions onto partial waves through $l=12$ and then
applying the double spherical Bessel transform. The summed
partial-wave norm contributions change by approximately
$1.5\times10^{-5}$ percentage points for YAMA-23 and
$3.6\times10^{-5}$ percentage points for MT-V.
The Hamiltonian expectation values are more sensitive:
the differences between the momentum- and coordinate-space
evaluations are approximately $2.6\times10^{-4}$~MeV and
$7.2\times10^{-4}$~MeV, respectively. These differences
characterize the combined numerical effects of projection
and coordinate-space evaluation for the selected calculations.

\begin{figure*}[!t]
\hspace{-0.2cm}\includegraphics[width=8.2cm, trim=0cm 0cm 0cm 0cm, clip]
  {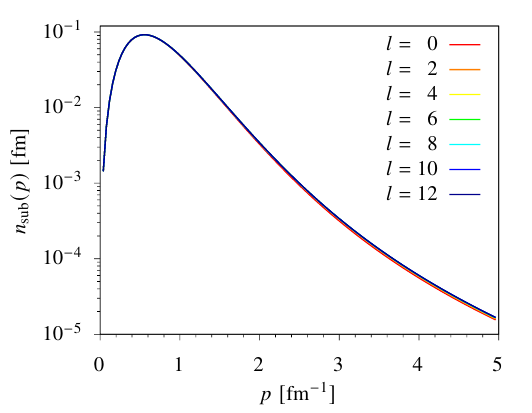}
  \includegraphics[width=8.2cm, trim=0cm 0cm 0cm 0cm, clip]
  {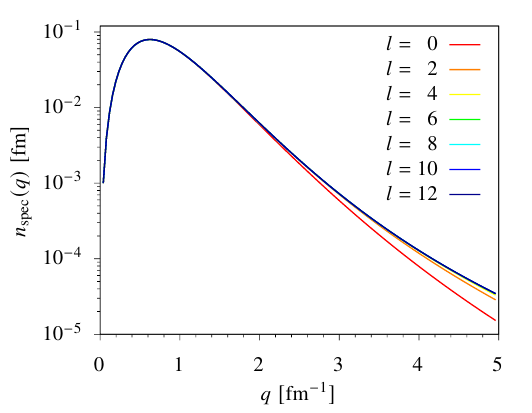} \\[-5mm]
  \caption{\label{fig:mom_dist_yama}The momentum distributions
  $n_{\mathrm{sub}}(p)$ and $n_{\mathrm{spec}}(q)$ with the YAMA-23 potential in
  a two-dimensional approach (2D). The curves represent the monotonically
  increasing partial-wave sums for $l=0,2,4,6,8,10$, and $12$.}
\end{figure*}

\begin{figure*}[!t]
  \hspace{-0.2cm}\includegraphics[width=8.2cm, trim=0cm 0cm 0cm 0cm, clip]
  {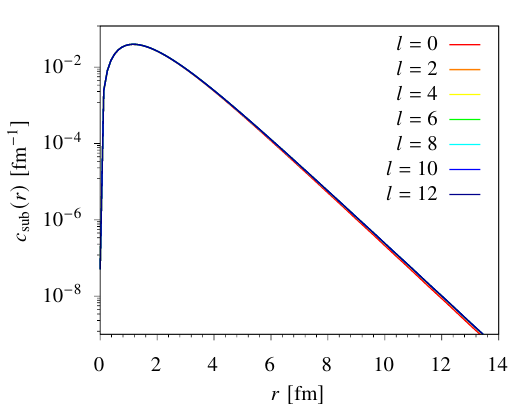}
  \includegraphics[width=8.2cm, trim=0cm 0cm 0cm 0cm, clip]
  {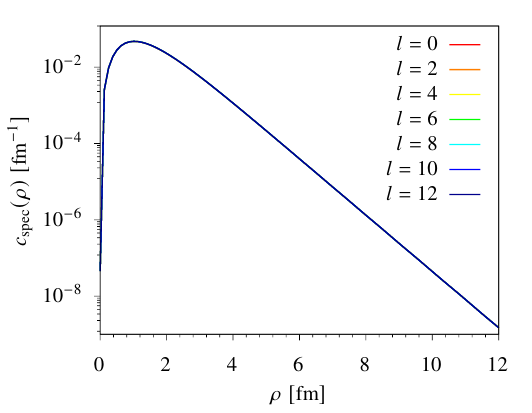} \\[-5mm]
  \caption{\label{fig:correlation_dist_yama}The spatial pair correlation
  function $c_{\mathrm{sub}}(r)$ and spectator distribution
  $c_{\mathrm{spec}}(\rho)$ with the YAMA-23 potential in a two-dimensional
  approach (2D). The curves represent the monotonically increasing partial-wave
  sums for $l=0,2,4,6,8,10$, and $12$.}
\end{figure*}

\begin{figure*}[htbp]
  \hspace{-0.2cm}\includegraphics[width=8.2cm, trim=0cm 0cm 0cm 0cm, clip]
  {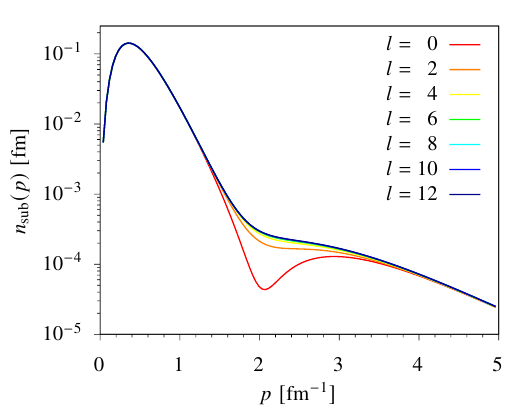}
  \includegraphics[width=8.2cm, trim=0cm 0cm 0cm 0cm, clip]
  {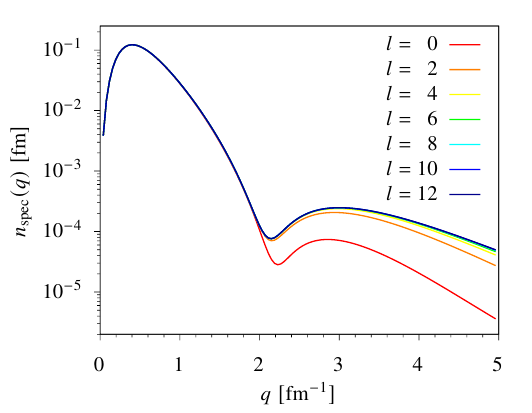} \\[-5mm]
  \caption{\label{fig:mom_dist_mt}The momentum distributions
  $n_{\mathrm{sub}}(p)$ and $n_{\mathrm{spec}}(q)$ with the MT-V potential in
  a two-dimensional approach (2D). The curves represent the monotonically
  increasing partial-wave sums for $l=0,2,4,6,8,10$, and $12$.}
\end{figure*}

\begin{figure*}[htbp]
  \hspace{-0.2cm}\includegraphics[width=8.2cm, trim=0cm 0cm 0cm 0cm, clip]
  {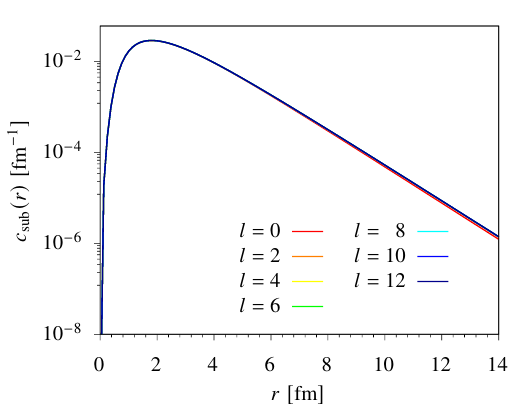}
  \includegraphics[width=8.2cm, trim=0cm 0cm 0cm 0cm, clip]
  {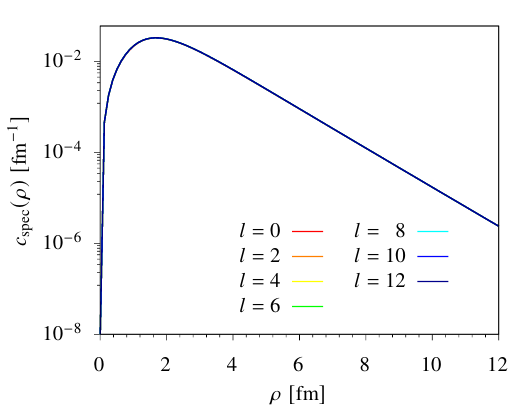} \\[-5mm]
  \caption{\label{fig:correlation_dist_mt}The spatial pair correlation
  function $c_{\mathrm{sub}}(r)$ and spectator distribution
  $c_{\mathrm{spec}}(\rho)$ with the MT-V potential in a two-dimensional
  approach (2D). The curves represent the monotonically increasing partial-wave
  sums for $l=0,2,4,6,8,10$, and $12$.}
\end{figure*}

Figures~\ref{fig:mom_dist_yama} and \ref{fig:mom_dist_mt} show the momentum
distributions $n_{\mathrm{sub}}(p)$ and $n_{\mathrm{spec}}(q)$ for the
YAMA-23 and MT-V potentials. Compared with the soft separable Yamaguchi
interaction, the local MT-V potential generates a more structured
intermediate-momentum region. In particular, the two-body momentum
distribution exhibits a pronounced suppression around
$p\approx 2~\mathrm{fm}^{-1}$, consistent with the interplay between the
attractive long-range and repulsive short-range parts of the interaction.
The corresponding 2D partial-wave components are shown in
Figs.~\ref{fig:tritwave_mt_low_l} and
\ref{fig:tritwave_mt_high_l}.
Figure~\ref{fig:tritwave_yamamt_3dl} displays the magnitudes
of the full 3D vector-variable wave functions at $x=1$.

The corresponding coordinate-space information is shown in
Figs.~\ref{fig:correlation_dist_yama} and
\ref{fig:correlation_dist_mt}. These figures display the pair correlation
function $c_{\mathrm{sub}}(r)$ and the spectator distribution
$c_{\mathrm{spec}}(\rho)$ for YAMA-23 and MT-V, respectively. For the MT-V
potential, the short-range repulsive component suppresses the pair
correlation at small distances. Resolving this depletion requires higher
partial waves, in agreement with the slower partial-wave convergence
observed for the local interactions in Table~\ref{tab:mt_2d_convergence}.

\begin{table*}[!htbp]
\caption{
\linespread{1.0}\selectfont
Spatial expectation values and geometric diagnostics for the MT-V and
YAMA-23 interactions. The first part compares the mean Jacobi-coordinate
separations in the 2D and 3D formulations. The second part gives the RMS
separations and the permutation-symmetry residual $\epsilon_{\rm sym}$
defined in Eq.~(\ref{eq:rms_symmetry_residual}). For MT-V, both the
interaction truncated at $l_{\rm max}=12$ and the full vector-variable
interaction are shown in the 3D calculation. The potential labels
specify the interaction truncation; the spatial analysis retains
partial-wave components through $l=12$ in all cases.
The second moments used to form the RMS separations are normalized
by the corresponding direct coordinate-space norm.
Ratios and residuals are evaluated from the unrounded numerical
results. The equilateral reference ratio is
$2/\sqrt{3}\approx1.1547$.
}
\label{tab:geometric_comparison}
\begin{ruledtabular}
\setlength{\tabcolsep}{0pt}
\begin{tabular*}{\textwidth}{@{\extracolsep\fill}l c c c c}
Potential & $\langle r \rangle$ [fm] & $\langle \rho \rangle$ [fm] &
$\langle r \rangle / \langle \rho \rangle$ & $\delta$ [\%] \\
\hline\\[-8pt]
\multicolumn{5}{c}{Mean separations} \\
\hline\\[-8pt]
\multicolumn{5}{c}{2D Approach} \\
\hline\\[-8pt]
MT-V ($l_{\rm max}=12$)
& 2.620617 & 2.270469 & 1.154219 & $-0.041734$ \\
YAMA-23 ($l_{\rm max}=12$)
& 1.658895 & 1.442613 & 1.149923 & $-0.413706$ \\
\hline\\[-8pt]
\multicolumn{5}{c}{3D Approach} \\
\hline\\[-8pt]
MT-V ($l_{\rm max}=\infty$)
& 2.620616 & 2.270468 & 1.154218 & $-0.041749$ \\
MT-V ($l_{\rm max}=12$)
& 2.620616 & 2.270468 & 1.154218 & $-0.041748$ \\
YAMA-23 ($l_{\rm max}=12$)
& 1.658895 & 1.442614 & 1.149923 & $-0.413716$ \\
\hline\\[-8pt]
\multicolumn{5}{c}{RMS separations and symmetry check} \\
\hline\\[-8pt]
Potential &
$r_{\rm rms}$ [fm] &
$\rho_{\rm rms}$ [fm] &
$r_{\rm rms}/\rho_{\rm rms}$ &
$\epsilon_{\rm sym}$ \\
\hline\\[-8pt]
\multicolumn{5}{c}{2D Approach} \\
\hline\\[-8pt]
MT-V ($l_{\rm max}=12$)
& 2.960779 & 2.564157 & 1.154679 & $3.73\times10^{-5}$ \\
YAMA-23 ($l_{\rm max}=12$)
& 1.896992 & 1.642847 & 1.154698 & $4.95\times10^{-6}$ \\
\hline\\[-8pt]
\multicolumn{5}{c}{3D Approach} \\
\hline\\[-8pt]
MT-V ($l_{\rm max}=\infty$)
& 2.960777 & 2.564156 & 1.154679 & $3.76\times10^{-5}$ \\
MT-V ($l_{\rm max}=12$)
& 2.960778 & 2.564157 & 1.154679 & $3.76\times10^{-5}$ \\
YAMA-23 ($l_{\rm max}=12$)
& 1.896992 & 1.642848 & 1.154697 & $5.28\times10^{-6}$ \\
\end{tabular*}
\end{ruledtabular}
\end{table*}

\subsection{Spatial Geometry and the Equilateral Ideal}
\label{sec:res_geometry}

The expectation values of the Jacobi coordinates provide a compact measure of
the average spatial geometry of the three-body bound state. In the convention
used here, $\langle r\rangle$ represents the average pair separation, while
$\langle\rho\rangle$ measures the distance of the spectator particle from the
center of mass of the interacting pair. Their ratio can therefore be compared
with the value $2/\sqrt{3}$ expected for the corresponding Jacobi-coordinate
representation of an equilateral triangle.

For the root-mean-square (RMS) separations,
\begin{equation}
r_{\rm rms}=\sqrt{\langle r^2\rangle},
\qquad
\rho_{\rm rms}=\sqrt{\langle\rho^2\rangle},
\end{equation}
an additional exact relation follows from permutation symmetry. With particle
coordinates $\mathbf R_i$,
\begin{equation}
\mathbf r=\mathbf R_2-\mathbf R_3,
\qquad
\boldsymbol{\rho}=
\mathbf R_1-\frac{\mathbf R_2+\mathbf R_3}{2},
\end{equation}
and hence
\begin{equation}
\rho^2=
\frac{1}{2}\left(r_{12}^2+r_{13}^2\right)
-\frac{1}{4}r_{23}^2 .
\end{equation}
For three identical bosons, permutation symmetry implies
$\langle r_{12}^2\rangle=\langle r_{13}^2\rangle
=\langle r_{23}^2\rangle=\langle r^2\rangle$, and therefore
\begin{equation}
\langle\rho^2\rangle=\frac{3}{4}\langle r^2\rangle,
\qquad
\frac{r_{\rm rms}}{\rho_{\rm rms}}=\frac{2}{\sqrt{3}} .
\label{eq:rms_symmetry}
\end{equation}
We use the corresponding dimensionless residual
\begin{equation}
\epsilon_{\rm sym}
=
\left|
\frac{4\langle\rho^2\rangle}
     {3\langle r^2\rangle}-1
\right|
\label{eq:rms_symmetry_residual}
\end{equation}
as an additional numerical consistency check.

Table~\ref{tab:geometric_comparison} shows that both interactions produce an
average Jacobi geometry close to the equilateral reference value, with
deviations well below one percent. For MT-V, replacing the interaction
truncated at $l_{\rm max}=12$ by the full vector-variable interaction changes
neither $\langle r\rangle$ nor $\langle\rho\rangle$ at the displayed
precision. The independently calculated 2D and 3D first moments also agree
to within approximately $10^{-6}$~fm.

The absolute sizes of the spatial expectation values reflect the markedly
different spatial extent of the two bound states. The weakly bound MT-V
state has $\langle r\rangle\approx2.62$~fm, compared with
$\langle r\rangle\approx1.66$~fm for the more deeply bound YAMA-23 state.
The corresponding RMS pair separations are approximately $2.96$~fm and
$1.90$~fm, respectively.

The first-moment ratio is not fixed by permutation symmetry and therefore
provides a compact diagnostic of the average spatial correlations.
For MT-V, $\delta\simeq-0.04\%$, whereas for YAMA-23
$\delta\simeq-0.41\%$. These values should not be interpreted as a
pointwise geometric constraint on the wave function.

In contrast, the RMS ratio obeys the exact permutation-symmetry relation
of Eq.~(\ref{eq:rms_symmetry}) for a fully converged three-boson state.
The independently evaluated second moments satisfy this relation at high
accuracy in both formulations. For MT-V at $l_{\rm max}=12$,
$\epsilon_{\rm sym}=3.73\times10^{-5}$ in the 2D calculation and
$3.76\times10^{-5}$ in the 3D calculation; the full-interaction 3D result
is also $3.76\times10^{-5}$. For YAMA-23, the corresponding residuals are
$4.95\times10^{-6}$ and $5.28\times10^{-6}$ in the 2D and 3D
calculations, respectively.

The RMS separations themselves show similarly close agreement. At matched
$l_{\rm max}=12$, the 2D--3D differences are below
$10^{-6}$~fm for both MT-V RMS radii and below
$4\times10^{-7}$~fm for YAMA-23. These independent coordinate-space
results therefore provide an additional cross-formulation check of the
Fourier transformations and of the reconstructed three-body wave functions.

\subsection{Numerical Error Budget}
\label{sec:res_error_budget}

To consolidate the convergence properties discussed above, we quantify the
dominant numerical uncertainties of the calculation. The accuracy of the
three-body binding energies and expectation values is controlled mainly by
four sources:

\emph{1. Discretization and Cutoff Error:}
Mesh and momentum-cutoff effects are assessed at fixed
interaction content. The reported calculations use
$q_{\rm cut}=400~\mathrm{fm}^{-1}$
($p_{\rm cut}=600~\mathrm{fm}^{-1}$) for Yamaguchi and
$q_{\rm cut}=200~\mathrm{fm}^{-1}$
($p_{\rm cut}=300~\mathrm{fm}^{-1}$) for Malfliet--Tjon.
The convergence tables distinguish the stability of the
binding energy from that of the individual expectation
values $\langle H_0\rangle$ and $\langle V\rangle$,
which are more sensitive to the high-momentum tail.
The achieved numerical stability must therefore be assessed
for each interaction, formulation, and observable separately.

\emph{2. Interpolation and Kernel-Implementation Error:} The evaluation of
the shifted permutation arguments requires interpolation of the wave function
and, in the $t$-driven formulations, of the fully off-shell two-body
$t$-matrix. The comparison between the standard $t$-driven formulation and
the $V$-driven formulation provides a stringent combined check of the
interpolation, quadrature, and permutation-kernel implementation. For the
Yamaguchi benchmark, the two approaches agree at the $10^{-6}$~MeV level,
while for the local MT-V interaction the agreement is at the few-$10^{-6}$~MeV
level. These differences are comparable to or smaller than the
corresponding internal residuals.

\emph{3. Partial-Wave Truncation Effects:}
The local MT interactions exhibit slower partial-wave
convergence than the separable Yamaguchi interactions.
Comparing the 3D calculations with $l_{\rm max}=12$
and the full vector-variable interaction gives binding-energy
differences of approximately $1.1\times10^{-5}$~MeV
for MT-V and $4.9\times10^{-5}$~MeV for MT-IV.
These finite-mesh differences contain both the effect of
the interaction truncation and numerical differences between
the calculations. In particular, the MT-IV difference
should not be interpreted as a separately resolved
truncation error without further convergence tests.

\emph{4. Coordinate-Space Transformation Error:}
The coordinate-space results reported in
Table~\ref{tab:combined_observables_3d} are obtained by projecting
the 3D momentum-space wave functions onto partial waves through
$l=12$ and applying the double spherical Bessel transform.
The summed partial-wave norm contributions differ between momentum
and coordinate space by approximately $1.5\times10^{-5}$
percentage points for YAMA-23 and $3.6\times10^{-5}$ percentage
points for MT-V. The corresponding Hamiltonian expectation values
differ by approximately $2.6\times10^{-4}$~MeV and
$7.2\times10^{-4}$~MeV, respectively. These differences characterize
the combined numerical effects of projection and coordinate-space
evaluation for the reported calculations; they are not rigorous
error bounds on the binding energies.

The numerical assessment therefore distinguishes convergence
under mesh and cutoff refinement, agreement between
formulations, and the internal residual
$\Delta E=|E_3-\langle H\rangle|$.
For the matched local-interaction calculations at
$l_{\rm max}=12$, the 2D--3D binding-energy differences
are $2.2\times10^{-6}$~MeV for MT-V and
$5.3\times10^{-5}$~MeV for MT-IV.
The internal residual provides a complementary consistency
check of the reconstructed wave function and expectation-value
integrations; it is not used as a rigorous error bound
on the binding energy.

\section{Summary and Outlook}
\label{sec:outlooksummary}

We presented a momentum-space benchmark of the three-boson
bound state that builds on earlier comparisons of
vector-variable and partial-wave calculations
\cite{Elster98b} and of $t$-matrix-driven and potential-driven
Faddeev equations \cite{Golak2013a}.
The contribution is a combined numerical cross-validation:
identical finite partial-wave interaction spaces are used
across the formulations, separable interactions provide an
additional 1D spectator-amplitude reference, and the 2D--3D
comparisons are supplemented by checks using both kernel
forms and coordinate-space observables.
This construction allows numerical differences to be studied
without conflating them with differences in the retained
interaction content.

For the local Malfliet--Tjon interactions, increasing $l_{\rm max}$ from
10 to 12 still changes the binding energies by approximately $2\times10^{-5}$~MeV.
At $l_{\rm max}=12$, the 2D--3D binding-energy differences are
$2.2\times10^{-6}$~MeV for MT-V and $5.3\times10^{-5}$~MeV for MT-IV.
The momentum-space internal residuals also depend on the
interaction and formulation, reaching $8.7\times10^{-5}$~MeV for the
full-interaction 3D MT-IV calculation. The achieved agreement and internal
consistency are therefore reported separately for the individual cases.

A central numerical result is the high-precision cross-validation of the
vector-variable formulation within the matched interaction spaces considered
here. Comparing the standard $t$-matrix-driven Faddeev equation with an alternative
bare-potential-driven formulation, we showed that these structurally distinct
kernels yield binding energies agreeing at the $10^{-6}$~MeV level for the
Yamaguchi benchmark and few-$10^{-6}$~MeV level for the MT-V interaction. This
 provides a stringent validation of the permutation geometry and
multidimensional integration machinery, showing that the continuous angular
dependence can be accurately resolved without an explicit partial-wave expansion.

We also examined the coordinate-space representations of the
momentum-space solutions. For the 3D solutions, the results reported
in Table~\ref{tab:combined_observables_3d} are obtained by projection
onto partial waves through $l=12$, followed by double spherical
Bessel transforms. The norm decompositions provide a sensitive
consistency check, while the momentum- and coordinate-space
Hamiltonian expectation values differ by approximately
$2.6\times10^{-4}$~MeV for YAMA-23 and
$7.2\times10^{-4}$~MeV for MT-V. The spatial expectation values
show close 2D--3D agreement, and the independently evaluated
second moments satisfy the permutation-symmetry relation with
$\epsilon_{\rm sym}<4\times10^{-5}$ in all cases reported in
Table~\ref{tab:geometric_comparison}.

The present results establish a numerical reference for three identical
spinless bosons interacting through the specified Yamaguchi and
Malfliet--Tjon pair potentials, without three-body forces. Future work
will extend this matched-interaction benchmark to elastic and breakup
scattering, where the treatment of singular kernels and the convergence
of scattering observables require additional validation. Related
theoretical developments include the 3D formulation of nucleon--deuteron
breakup by Ramazani-Sharifabadi and Ziaeian
\cite{RamazaniSharifabadi2026a}, which includes spin and isospin degrees
of freedom, omits three-nucleon forces, and proposes a treatment of
moving singularities.

Further applications may include the structural properties and
low-energy dynamics of weakly bound atomic trimers, such as noble-gas
clusters. These extensions require separate convergence and
cross-formulation studies with the corresponding interactions;
the numerical accuracy established for the present model systems
should not be assumed to carry over unchanged.

\section{Data availability}
All data generated or analysed during this study are included in this
published article and its Supplementary Information. The Supplementary
Information contains the input files and computation logs for the reported
calculations, together with the scripts used to extract the tabulated
results.

\section{Acknowledgments}
M.R.H. acknowledges support from the National Science Foundation under
Grant Nos. NSF-OIA-2430293 and NSF-EES-2436204 at Central State University.
\vspace{3mm}


\appendix

\begingroup
\makeatletter
\renewenvironment{widetext}{%
  \par\ignorespaces
  \onecolumngrid
  \vskip10\p@
  \vskip6\p@
  \prep@math@patch
}{%
  \par
  \vskip6\p@
  \vskip8.5\p@
  \twocolumngrid\global\@ignoretrue
  \@endpetrue
}
\makeatother

\begin{widetext}

\section{Mesh Convergence of the $t$- and $V$-Driven Formulations}

This appendix collects the momentum-mesh convergence tests
for the $t$-matrix-driven and potential-driven Faddeev
equations. Tables~\ref{tab:mesh_yama_t_vs_v_2d} and
\ref{tab:mesh_yama_t_vs_v_3d} give the YAMA-23 results
in the 2D and 3D formulations.
The corresponding MT-V comparisons with the interaction
truncated at $l_{\rm max}=12$ are presented in
Tables~\ref{tab:mesh_mt5_t_vs_v_2d} and
\ref{tab:mesh_mt5_t_vs_v_3d}.
Table~\ref{tab:mesh_mt5nopartial_t_vs_v_3d} gives the
3D comparison using the full vector-variable MT-V interaction.
In each table, the pair and spectator momentum meshes are
refined together, with $N=N_p=N_q$, at the stated momentum
cutoffs.

\begin{table}[htbp]
\caption{
\linespread{1.0}\selectfont
Comparison of the $t$-matrix-driven and potential-driven ($V$-driven) 2D
formulations for the YAMA-23 potential with $l_{\rm max} = 12$ and
$q_{\rm cut} = 400$~fm$^{-1}$.
Expectation values $\langle H_0\rangle$, $\langle V\rangle$,
the three-body binding energy $E_3$, and the residual
$|E_{3}-\langle H\rangle|$ are shown. Here $N=N_p=N_q$ denotes the number
of grid points in each momentum variable.}
\label{tab:mesh_yama_t_vs_v_2d}
\begin{ruledtabular}
\setlength{\tabcolsep}{0pt}
\begin{tabular*}{\textwidth}{
    @{\extracolsep{\fill}}
    S[table-format=3.0, group-digits=false]
    S[table-format=2.7, group-digits=false]
    S[table-format=-2.7, group-digits=false]
    S[table-format=-2.7, group-digits=false]
    S[table-format=1.7, group-digits=false]
    S[table-format=2.7, group-digits=false]
    S[table-format=-2.7, group-digits=false]
    S[table-format=-2.7, group-digits=false]
    S[table-format=1.7, group-digits=false]
}
& \multicolumn{4}{c}{$t$-driven approach} &
  \multicolumn{4}{c}{$V$-driven approach} \\
\cline{2-5} \cline{6-9}\\[-8pt]
\multicolumn{1}{c}{$N$} &
\multicolumn{1}{c}{$\langle H_0\rangle$} &
\multicolumn{1}{c}{$\langle V\rangle$} &
\multicolumn{1}{c}{$E_3$} &
\multicolumn{1}{c}{$|E_{3}-\langle H\rangle|$} &
\multicolumn{1}{c}{$\langle H_0\rangle$} &
\multicolumn{1}{c}{$\langle V\rangle$} &
\multicolumn{1}{c}{$E_3$} &
\multicolumn{1}{c}{$|E_{3}-\langle H\rangle|$} \\
&
\multicolumn{1}{c}{[MeV]} &
\multicolumn{1}{c}{[MeV]} &
\multicolumn{1}{c}{[MeV]} &
\multicolumn{1}{c}{[MeV]} &
\multicolumn{1}{c}{[MeV]} &
\multicolumn{1}{c}{[MeV]} &
\multicolumn{1}{c}{[MeV]} &
\multicolumn{1}{c}{[MeV]} \\
\hline\\[-8pt]
128 & 67.1487913 & -91.6869525 & -24.5385237 & 0.0003625 & 67.1488427 & -91.6870039 & -24.5385284 & 0.0003672 \\
192 & 67.1492014 & -91.6876902 & -24.5385618 & 0.0000729 & 67.1492076 & -91.6876965 & -24.5385608 & 0.0000719 \\
256 & 67.1492081 & -91.6877513 & -24.5385634 & 0.0000202 & 67.1492680 & -91.6878113 & -24.5385660 & 0.0000228 \\
384 & 67.1492713 & -91.6878349 & -24.5385690 & 0.0000054 & 67.1492907 & -91.6878543 & -24.5385680 & 0.0000044 \\
512 & 67.1492758 & -91.6878427 & -24.5385691 & 0.0000022 & 67.1492943 & -91.6878612 & -24.5385683 & 0.0000014 \\
768 & 67.1492787 & -91.6878468 & -24.5385697 & 0.0000015 & 67.1492958 & -91.6878639 & -24.5385684 & 0.0000003 \\
\end{tabular*}
\end{ruledtabular}
\end{table}

\begin{table}[htbp]
\caption{
\linespread{1.0}\selectfont
Comparison of the $t$-matrix-driven and potential-driven ($V$-driven) 3D
formulations for the YAMA-23 potential with $l_{\rm max} = 12$ and
$q_{\rm cut} = 400$~fm$^{-1}$.
Expectation values $\langle H_0\rangle$, $\langle V\rangle$, the
three-body binding energy $E_3$, and the residual
$|E_3-\langle H\rangle|$ are shown. Here $N=N_p=N_q$ denotes the number
of grid points in each momentum variable.}
\label{tab:mesh_yama_t_vs_v_3d}
\begin{ruledtabular}
\setlength{\tabcolsep}{0pt}
\begin{tabular*}{\textwidth}{
    @{\extracolsep{\fill}}
    S[table-format=3.0, group-digits=false]
    S[table-format=2.7, group-digits=false]
    S[table-format=-2.7, group-digits=false]
    S[table-format=-2.7, group-digits=false]
    S[table-format=1.7, group-digits=false]
    S[table-format=2.7, group-digits=false]
    S[table-format=-2.7, group-digits=false]
    S[table-format=-2.7, group-digits=false]
    S[table-format=1.7, group-digits=false]
}
& \multicolumn{4}{c}{$t$-driven approach} &
  \multicolumn{4}{c}{$V$-driven approach} \\
\cline{2-5} \cline{6-9}\\[-8pt]
\multicolumn{1}{c}{$N$} &
\multicolumn{1}{c}{$\langle H_0\rangle$} &
\multicolumn{1}{c}{$\langle V\rangle$} &
\multicolumn{1}{c}{$E_3$} &
\multicolumn{1}{c}{$|E_{3}-\langle H\rangle|$} &
\multicolumn{1}{c}{$\langle H_0\rangle$} &
\multicolumn{1}{c}{$\langle V\rangle$} &
\multicolumn{1}{c}{$E_3$} &
\multicolumn{1}{c}{$|E_{3}-\langle H\rangle|$} \\
&
\multicolumn{1}{c}{[MeV]} &
\multicolumn{1}{c}{[MeV]} &
\multicolumn{1}{c}{[MeV]} &
\multicolumn{1}{c}{[MeV]} &
\multicolumn{1}{c}{[MeV]} &
\multicolumn{1}{c}{[MeV]} &
\multicolumn{1}{c}{[MeV]} &
\multicolumn{1}{c}{[MeV]} \\
\hline\\[-8pt]
128 & 67.1480852 & -91.6862463 & -24.5384474 & 0.0002863 & 67.1480739 & -91.6862351 & -24.5384351 & 0.0002739 \\
192 & 67.1490228 & -91.6875116 & -24.5385495 & 0.0000608 & 67.1490418 & -91.6875307 & -24.5385371 & 0.0000482 \\
256 & 67.1491213 & -91.6876644 & -24.5385671 & 0.0000240 & 67.1492396 & -91.6877828 & -24.5385656 & 0.0000224 \\
384 & 67.1491625 & -91.6877259 & -24.5385689 & 0.0000056 & 67.1492840 & -91.6878476 & -24.5385675 & 0.0000039 \\
512 & 67.1491712 & -91.6877378 & -24.5385697 & 0.0000030 & 67.1492929 & -91.6878598 & -24.5385682 & 0.0000014 \\
\end{tabular*}
\end{ruledtabular}
\end{table}

\begin{table}[htbp]
\caption{
\linespread{1.0}\selectfont
Comparison of the $t$-matrix-driven and potential-driven ($V$-driven) 2D
formulations for the MT-V potential with $l_{\rm max} = 12$ and
$q_{\rm cut} = 200$~fm$^{-1}$.
Expectation values $\langle H_0\rangle$, $\langle V\rangle$, the
three-body binding energy $E_3$, and the residual
$|E_3-\langle H\rangle|$ are shown. Here $N=N_p=N_q$ denotes the number
of grid points in each momentum variable.}
\label{tab:mesh_mt5_t_vs_v_2d}
\begin{ruledtabular}
\setlength{\tabcolsep}{0pt}
\begin{tabular*}{\textwidth}{
    @{\extracolsep{\fill}}
    S[table-format=3.0, group-digits=false]
    S[table-format=2.7, group-digits=false]
    S[table-format=-2.7, group-digits=false]
    S[table-format=-2.7, group-digits=false]
    S[table-format=1.7, group-digits=false]
    S[table-format=2.7, group-digits=false]
    S[table-format=-2.7, group-digits=false]
    S[table-format=-2.7, group-digits=false]
    S[table-format=1.7, group-digits=false]
}
& \multicolumn{4}{c}{$t$-driven approach} &
  \multicolumn{4}{c}{$V$-driven approach} \\
\cline{2-5} \cline{6-9}\\[-8pt]
\multicolumn{1}{c}{$N$} &
\multicolumn{1}{c}{$\langle H_0\rangle$} &
\multicolumn{1}{c}{$\langle V\rangle$} &
\multicolumn{1}{c}{$E_3$} &
\multicolumn{1}{c}{$|E_{3}-\langle H\rangle|$} &
\multicolumn{1}{c}{$\langle H_0\rangle$} &
\multicolumn{1}{c}{$\langle V\rangle$} &
\multicolumn{1}{c}{$E_3$} &
\multicolumn{1}{c}{$|E_{3}-\langle H\rangle|$} \\
&
\multicolumn{1}{c}{[MeV]} &
\multicolumn{1}{c}{[MeV]} &
\multicolumn{1}{c}{[MeV]} &
\multicolumn{1}{c}{[MeV]} &
\multicolumn{1}{c}{[MeV]} &
\multicolumn{1}{c}{[MeV]} &
\multicolumn{1}{c}{[MeV]} &
\multicolumn{1}{c}{[MeV]} \\
\hline\\[-8pt]
128 & 29.7775183 & -37.5139908 & -7.7365960 & 0.0001235 & 29.7775337 & -37.5140062 & -7.7366010 & 0.0001285 \\
192 & 29.7775796 & -37.5141411 & -7.7365942 & 0.0000327 & 29.7776093 & -37.5141528 & -7.7365915 & 0.0000480 \\
256 & 29.7776002 & -37.5141769 & -7.7365936 & 0.0000169 & 29.7776092 & -37.5141776 & -7.7365903 & 0.0000219 \\
384 & 29.7776037 & -37.5141861 & -7.7365935 & 0.0000111 & 29.7776049 & -37.5141873 & -7.7365900 & 0.0000075 \\
\end{tabular*}
\end{ruledtabular}
\end{table}

\begin{table}[htbp]
\caption{
\linespread{1.0}\selectfont
Comparison of the $t$-matrix-driven and potential-driven ($V$-driven) 3D
formulations for the MT-V potential with $l_{\rm max} = 12$ and
$q_{\rm cut} = 200$~fm$^{-1}$.
Expectation values $\langle H_0\rangle$, $\langle V\rangle$, the
three-body binding energy $E_3$, and the residual
$|E_3-\langle H\rangle|$ are shown. Here $N=N_p=N_q$ denotes the number
of grid points in each momentum variable.}
\label{tab:mesh_mt5_t_vs_v_3d}
\begin{ruledtabular}
\setlength{\tabcolsep}{0pt}
\begin{tabular*}{\textwidth}{
   @{\extracolsep{\fill}}
   S[table-format=3.0, group-digits=false]
   S[table-format=2.7, group-digits=false]
   S[table-format=-2.7, group-digits=false]
   S[table-format=-2.7, group-digits=false]
   S[table-format=1.7, group-digits=false]
   S[table-format=2.7, group-digits=false]
   S[table-format=-2.7, group-digits=false]
   S[table-format=-2.7, group-digits=false]
   S[table-format=1.7, group-digits=false]
}
& \multicolumn{4}{c}{$t$-driven approach} &
 \multicolumn{4}{c}{$V$-driven approach} \\
\cline{2-5} \cline{6-9}\\[-8pt]
\multicolumn{1}{c}{$N$} &
\multicolumn{1}{c}{$\langle H_0\rangle$} &
\multicolumn{1}{c}{$\langle V\rangle$} &
\multicolumn{1}{c}{$E_3$} &
\multicolumn{1}{c}{$|E_{3}-\langle H\rangle|$} &
\multicolumn{1}{c}{$\langle H_0\rangle$} &
\multicolumn{1}{c}{$\langle V\rangle$} &
\multicolumn{1}{c}{$E_3$} &
\multicolumn{1}{c}{$|E_{3}-\langle H\rangle|$} \\
&
\multicolumn{1}{c}{[MeV]} &
\multicolumn{1}{c}{[MeV]} &
\multicolumn{1}{c}{[MeV]} &
\multicolumn{1}{c}{[MeV]} &
\multicolumn{1}{c}{[MeV]} &
\multicolumn{1}{c}{[MeV]} &
\multicolumn{1}{c}{[MeV]} &
\multicolumn{1}{c}{[MeV]} \\
\hline\\[-8pt]
128 & 29.7777583 & -37.5142357 & -7.7366735 & 0.0001961 &
      29.7776559 & -37.5141334 & -7.7366706 & 0.0001931 \\
192 & 29.7776117 & -37.5141778 & -7.7366096 & 0.0000434 &
      29.7776174 & -37.5141836 & -7.7366044 & 0.0000382 \\
256 & 29.7775720 & -37.5141533 & -7.7365988 & 0.0000175 &
      29.7776097 & -37.5141911 & -7.7365933 & 0.0000119 \\
384 & 29.7775867 & -37.5141737 & -7.7365957 & 0.0000087 &
      29.7776091 & -37.5141962 & -7.7365900 & 0.0000029 \\
\end{tabular*}
\end{ruledtabular}
\end{table}
\begin{table}[htbp]
\caption{
\linespread{1.0}\selectfont
Comparison of the $t$-matrix-driven and potential-driven ($V$-driven) 3D
formulations for the MT-V potential using the full vector-variable
interaction rather than a truncated partial-wave representation, with
$q_{\rm cut} = 200$~fm$^{-1}$.
Expectation values $\langle H_0\rangle$, $\langle V\rangle$, the
three-body binding energy $E_3$, and the residual
$|E_3-\langle H\rangle|$ are shown. Here $N=N_p=N_q$ denotes the number
of grid points in each momentum variable.}
\label{tab:mesh_mt5nopartial_t_vs_v_3d}
\begin{ruledtabular}
\setlength{\tabcolsep}{0pt}
\begin{tabular*}{\textwidth}{
    @{\extracolsep{\fill}}
    S[table-format=3.0, group-digits=false]
    S[table-format=2.7, group-digits=false]
    S[table-format=-2.7, group-digits=false]
    S[table-format=-2.7, group-digits=false]
    S[table-format=1.7, group-digits=false]
    S[table-format=2.7, group-digits=false]
    S[table-format=-2.7, group-digits=false]
    S[table-format=-2.7, group-digits=false]
    S[table-format=1.7, group-digits=false]
}
& \multicolumn{4}{c}{$t$-driven approach} &
  \multicolumn{4}{c}{$V$-driven approach} \\
\cline{2-5} \cline{6-9}\\[-8pt]
\multicolumn{1}{c}{$N$} &
\multicolumn{1}{c}{$\langle H_0\rangle$} &
\multicolumn{1}{c}{$\langle V\rangle$} &
\multicolumn{1}{c}{$E_3$} &
\multicolumn{1}{c}{$|E_{3}-\langle H\rangle|$} &
\multicolumn{1}{c}{$\langle H_0\rangle$} &
\multicolumn{1}{c}{$\langle V\rangle$} &
\multicolumn{1}{c}{$E_3$} &
\multicolumn{1}{c}{$|E_{3}-\langle H\rangle|$} \\
&
\multicolumn{1}{c}{[MeV]} &
\multicolumn{1}{c}{[MeV]} &
\multicolumn{1}{c}{[MeV]} &
\multicolumn{1}{c}{[MeV]} &
\multicolumn{1}{c}{[MeV]} &
\multicolumn{1}{c}{[MeV]} &
\multicolumn{1}{c}{[MeV]} &
\multicolumn{1}{c}{[MeV]} \\
\hline\\[-8pt]
128 & 29.7777546 & -37.5142440 & -7.7366844 & 0.0001950 &
      29.7776553 & -37.5141447 & -7.7366818 & 0.0001924 \\
192 & 29.7776080 & -37.5141857 & -7.7366206 & 0.0000429 &
      29.7776165 & -37.5141943 & -7.7366156 & 0.0000378 \\
256 & 29.7775682 & -37.5141611 & -7.7366098 & 0.0000170 &
      29.7776089 & -37.5142017 & -7.7366044 & 0.0000116 \\
384 & 29.7775830 & -37.5141815 & -7.7366067 & 0.0000082 &
      29.7776082 & -37.5142068 & -7.7366011 & 0.0000026 \\
\end{tabular*}
\end{ruledtabular}
\end{table}

\clearpage

\section{Explicit Representation of the partial wave projected
Permutation Operator}
\label{sec:appendixperm}

This appendix contains the quantities $G$ and $\tilde G$ related to
the permutation operators  \cite{Elster98b,Gloeckle83} for general
total angular momentum ${\cal L}$.
The function $G_{lL \, l'L\!'  {\cal L}}(q,q',x)$ from Eq.~(\ref{eq:faddeev_2d_t}) is a
combination of Legendre polynomials $P_k(x)$
\begin{align}
G_{l L \, l' L\!'  {\cal L}} (q,q',x) &= \sum_k P_k(x) \sum_ {\mu_1+\mu_2=l} \sum_
{\nu_1+\nu_2=l'} \ q'^{\mu_1+\nu_2} \ q^{\mu_2+\nu_1} \
g_{lL \, l'L\!'{\cal L} }^{k \mu_1 \nu_1  \mu_2 \nu_2}.
\end{align}

For general ${\cal L}$ the geometrical coefficient is given as
\begin{align}
g_{lL \, l'L\!' {\cal L}}^{k \mu_1 \nu_1 \mu_2 \nu_2} &= \sum_{gg'} \hat k \,
     \sqrt{ \hat l \ms \hat L \ms \hat l' \ms \hat L\!'} \,
     \sqrt{ \frac{\hat l ! \, \hat l ' !}{(2 \mu_1)! \, (2 \mu_2)! \,
           (2 \nu_1)! \, (2 \nu_2)! } }
           \; (-)^{l'} \left( {\textstyle{\frac{1}{2}}}
     \right)^{\mu_2+\nu_2}
    \cgk{\mu_2 }{ L    }{ g } {0}{0}{0}  \\[1pt]
& \quad \times
    \cgk{\nu_2 }{ L\!' }{ g'} {0}{0}{0}
    \cgk{k     }{ \nu_1}{ g } {0}{0}{0}
    \cgk{k     }{ \mu_1}{ g'} {0}{0}{0}
     \left\{ \begin{array}{ccc} \mu_1   & \mu_2 & l \cr
                            L &  {\cal L}    & g \end{array}
     \right\}
     \left\{ \begin{array}{ccc} \nu_1   & \nu_2 & l' \cr
                            L\!'&  {\cal L}  & g' \end{array}
     \right\}
     \left\{ \begin{array}{ccc} \mu_1   & g & {\cal L} \cr
                            \nu_1   & g'& k  \end{array}
     \right\} . \nonumber
\end{align}

\noindent
Here $l$ and $L$ are the orbital angular momenta associated with
$p$ and $q$, respectively; $\hat j\equiv 2j+1$ and
$\hat j!\equiv(2j+1)!$.
For the three-boson ground state considered here, ${\cal L}=0$,
which requires $L=l$ and $L'=l'$. The pair orbital angular
momenta $l$ and $l'$ are even. Under these restrictions, the
geometrical coefficient reduces to
\begin{align}
g_{ll'}^{k \mu_1 \nu_1  \mu_2 \nu_2} &=
      \sqrt{\frac{\hat l \ms \hat l ' }{\hat \mu_1 \ \hat \nu_1 }}  \
     \sqrt{ \frac{\hat l ! \, \hat l ' !}{(2 \mu_1)! \, (2 \mu_2)! \,
           (2 \nu_1)! \, (2 \nu_2)! } } \nonumber \\[1pt]
& \quad \times \, \left( {\textstyle{\frac{1}{2}}} \right)^{\mu_2+\nu_2} \ (-)
    ^ { \mu_1 + \nu_1}
     \cgk{ \mu_2 }{ l }{ \mu_1}{0}{0}{0}
     \cgk{ \nu_2 }{ l'}{ \nu_1}{0}{0}{0}
     \cgk{ \mu_1 }{ \nu_1 }{ k }{0}{0}{0}^{\!2}
\end{align}

\noindent
and $G_{lL \, l'L\!' {\cal L}}$ reduces to $G_{l l'}$.

The quantity $\tilde G_{lL \, l'L\!' {\cal L}}(p,q,x)$ occurs in Eq.~(\ref{eq:partialwave}).
In the case of general ${\cal L}$ the function $\tilde G$ reads
\begin{align}
\tilde G_{lL \, l'L\!'{\cal L} } (p,q,x)  &= \sum_k P_k(x) \sum_ {\mu_1+\mu_2=l'} \sum_
{\nu_1+\nu_2=L '} \ p^{\mu_1+\nu_1} \ q^{\mu_2+\nu_2} \
\tilde g_{lL \, l'L\!'{\cal L}}^{k \mu_1 \nu_1  \mu_2 \nu_2}
\end{align}
with the geometrical factor
\begin{align}
\tilde g_{lL \, l'L\!' {\cal L}}^{k \mu_1 \nu_1  \mu_2 \nu_2} &= \sum_{gg'}
     \hat k  \,
     \sqrt{ \hat l ' \ms  \hat L\!' \ms \hat g \ms \hat g '} \,
     \sqrt{ \frac{\hat l ' !  \, \hat L\!' !}{(2 \mu_1)! \, (2 \mu_2)!  \,
           (2 \nu_1)! \, (2 \nu_2)! } }  \; (-)^{g+{\cal L}+\nu_2+l+L'}
     \left( {\textstyle{\frac{1}{2}}} \right)^{\mu_1+\nu_2}
     \left( {\textstyle{\frac{3}{4}}} \right)^{\mu_2}  \\[1pt]
& \quad \times
\cgk{\mu_1 }{\nu_1 }{ g  }{0}{0}{0}
         \cgk{\mu_2 }{\nu_2 }{ g'  }{0}{0}{0}
     \cgk{ g  }{k }{ l  }{0}{0}{0}
     \cgk{ g' }{k }{ L  }{0}{0}{0}
     \left\{ \begin{array}{ccc} g'    &    g    & {\cal L}  \cr
                            l    & L & k  \end{array}
     \right\} \
     \left\{ \begin{array}{ccc} \mu_1   & \mu_2 & l ' \cr
                            \nu_1   & \nu_2 & L\!' \cr
                               g    & g'    & {\cal L}    \end{array}
     \right\}  . \nonumber
\end{align}

\noindent
Under the same ${\cal L}=0$ and even-partial-wave restrictions,
the coefficient $\tilde g$ reduces to
\begin{align}
\tilde g_{l l'}^{k \mu_1 \nu_1 \mu_2 \nu_2 } &= \sum_{g} \hat k   \,
     (-) ^ { \mu_2+g+k} \,
     \left( {\textstyle{\frac{1}{2}}} \right)^{\mu_1+\nu_2} \,
     \sqrt{\frac{\hat l'}{\hat l }} \,
     \frac{\hat l' !}{\sqrt{ (2 \mu_1)! \, (2 \mu_2)!  \,
           (2 \nu_1)!\, (2 \nu_2)! } } \nonumber \\[1pt]
& \quad \times \,
     \left( {\textstyle{\frac{3}{4}}} \right)^{\mu_2}
     \cgk{\mu_1 }{\nu_1 }{ g  }{0}{0}{0}
     \cgk{\mu_2 }{\nu_2 }{ g  }{0}{0}{0}
     \cgk{ g    }{  k   }{ l  }{0}{0}{0}^{\!2}
     \left\{ \begin{array}{ccc} \mu_1   & \mu_2 & l ' \cr
                            \nu_2   & \nu_1 & g   \end{array}
     \right\}
\end{align}

\noindent
and correspondingly $\tilde G_{lL \, l'L\!' {\cal L}}$ reduces to $\tilde G_{l l'}$.
\vspace{-5mm}
 \end{widetext}
 \endgroup


%

\end{document}